\documentstyle[12pt,psfig]{article}
\input xy
\xyoption{all}
\begin{document}

\hfill DUKE-CGTP-2001-03

\hfill hep-th/0102211

\vspace*{1.25in}

\begin{center}

{\large\bf String Orbifolds and Quotient Stacks}

\vspace{0.75in}

Eric Sharpe \\
Department of Physics \\
Box 90305 \\
Duke University \\
Durham, NC  27708 \\
{\tt ersharpe@cgtp.duke.edu} \\

$\,$ \\

\end{center}

In this note we observe that, contrary to the usual lore,
string orbifolds do {\it not} describe strings on quotient spaces,
but rather seem to describe strings on objects called
quotient stacks, a result that follows from simply unraveling
definitions, and is further justified by a number of results.
Quotient stacks are very closely related to quotient spaces;
for example, when the orbifold group acts freely,
the quotient space and the quotient stack are homeomorphic.
We explain how sigma models on quotient stacks naturally
have twisted sectors, and why a sigma model on a quotient stack
would be a nonsingular CFT even when the associated quotient space
is singular.  We also show how to understand twist fields in
this language, and outline the derivation of the orbifold
Euler characteristic purely in terms of stacks.
We also outline why there is a sense in which
one naturally finds $B \neq 0$ on exceptional divisors of resolutions.
These insights are not limited to merely understanding existing
string orbifolds:  we also point out how this technology enables
us to understand orbifolds in M-theory, as well as how this means that
string orbifolds provide the first example of an entirely new class
of string compactifications.
As quotient stacks are not a staple of the physics literature,
we include a lengthy tutorial on quotient stacks,
describing how one can perform differential geometry on stacks.

\begin{flushleft}
February 2001  
\end{flushleft}

\newpage

\tableofcontents

\newpage

\section{Introduction}

One often hears that string orbifolds \cite{strorb1,strorb2,dixonthes}
define strings propagating
on quotient spaces (with some `stringy' effects at singularities).
Of course, a string orbifold is not described in terms of maps into a 
quotient space, but rather is set up in terms of group actions
on covering spaces.

\begin{figure}
\centerline{\psfig{file=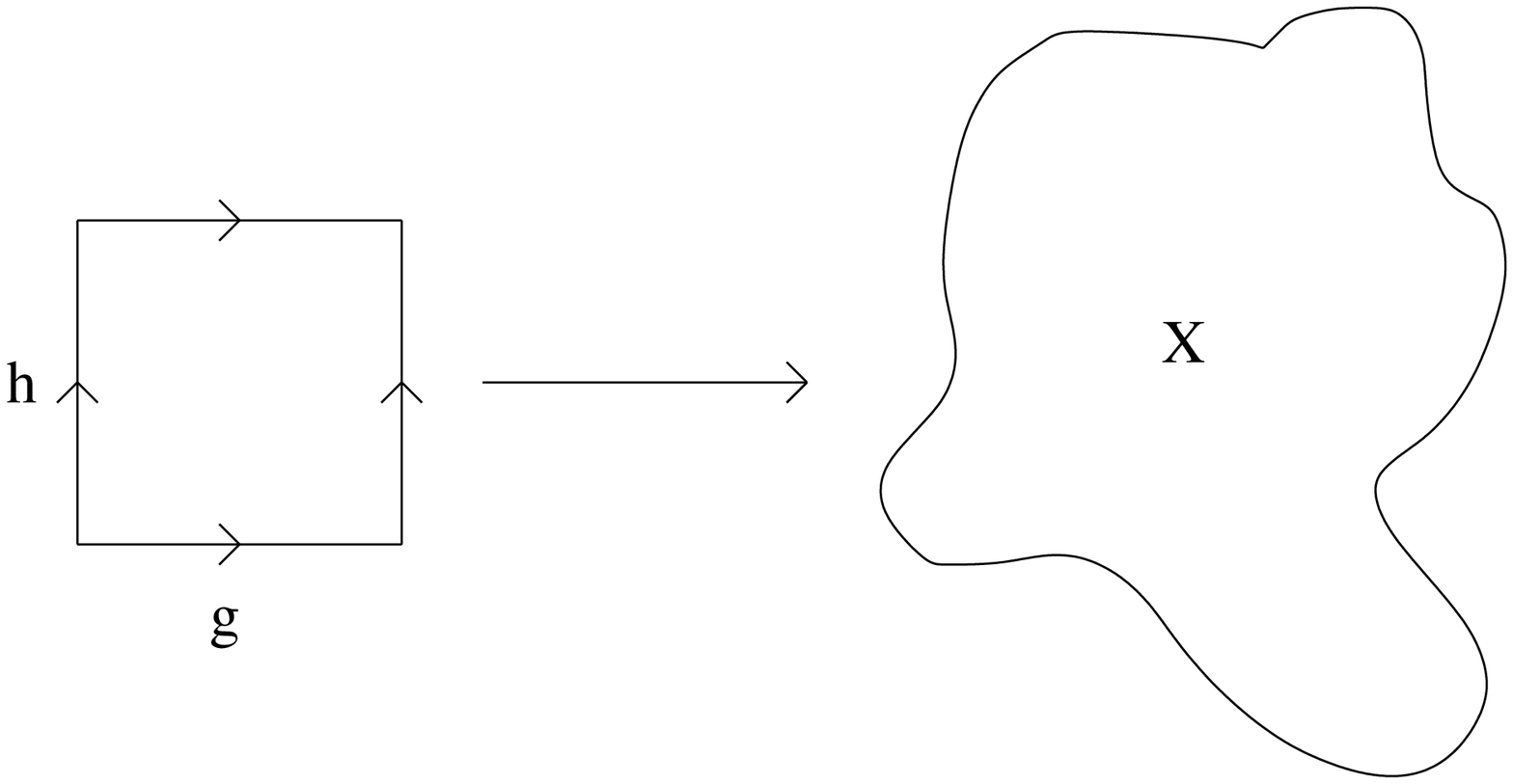,width=4in}}
\caption{\label{figsig1} Contribution to the $(g,h)$ twisted sector
of a string orbifold on $T^2$}
\end{figure}

For example, 
recall that the partition function for a string orbifold (at 1-loop, say)
is of the form
\begin{displaymath}
Z \: = \: \frac{ 1 }{ |G| } \sum_{g,h, gh=hg} Z_{g,h}
\end{displaymath}
where each $Z_{g,h}$ is the partition function of a sigma model of
maps from a square into the covering space $X$, such that the images of 
the sides
of the square are identified under the action of the orbifold group
$\Gamma$, as illustrated
in figure~\ref{figsig1}.
In particular, the partition function of the original, ``unorbifolded''
theory is just $Z_{1,1}$.  (From a Hamiltonian perspective, summing
over twisted sectors is equivalent to inserting projection operators
that only allow $\Gamma$-invariant states to propagate in loops.)
Clearly this is not a sigma model on the quotient space $X/\Gamma$,
if for no other reason than the fact that one does not sum over
maps into $X/\Gamma$, but rather is a sigma model on $X$
in which the $\Gamma$-action on $X$ has been gauged. 

It seems reasonable that a theory defined in this fashion
determines some physical theory living on the quotient space.
Let us take a moment to study this matter more closely. 

%First, the partition function $Z$ above is described in terms of
%the covering space, not the quotient.  A sigma model on a quotient
%space $X/\Gamma$ would involve a sum over maps from the worldsheet
%into $X/\Gamma$, but that is clearly not what we sum over in a string orbifold.
%At least naively, a string orbifold seems
%to be more nearly a $\Gamma$-equivariant sigma model on the covering
%space, not a sigma model on the quotient space.

%More precisely, a contribution to the sigma model seems to consist
%of a principal $\Gamma$-bundle over the worldsheet, together
%with a $\Gamma$-equivariant map from that bundle into the covering
%space $X$, judging from figure~\ref{figsig1}.

%Let us be a little more precise.  

Traditionally, contributions to
twisted-sector partition functions are described as maps from
polygons into $X$, such that inside $X$, the sides of the polygon
are identified by the group action.  One can set up a category
of continuous maps of this form, and we shall check in section~\ref{Gfree}
that this category is strictly not
the same as the category of continuous maps into the quotient
space $X/\Gamma$ when $\Gamma$ does not act freely.
(The problem is that lying over any given map into $X/\Gamma$, there 
can be several inequivalent maps of the form illustrated in 
figure~\ref{figsig1}.)
The category of continuous maps into a space completely
determines that space (see appendix~\ref{mapsgivespace} for more details),
so if we can think of a string orbifold as describing 
strings on {\it some} space, then that space cannot be the same as $X/\Gamma$,
but rather appears to be something containing more information at
singularities than $X/\Gamma$.

%we cannot think of a string orbifold as being literally
%a sigma model into a quotient space $X/\Gamma$.
%If anything, a string orbifold appears to be a sigma model into
%something containing more information at singularities than $X/\Gamma$.

%We have just argued that thinking of string orbifolds as strings
%on quotient spaces is not consistent.
%We shall spend a little more time listing other problems that
%crop up in trying to understand string orbifolds as describing
%strings on quotient spaces, and then we shall show that all of these
%problems are immediately resolved by recognizing that string orbifolds
%actually describe strings on something other than a quotient space.

Another way of thinking about the partition function 
$Z$ of a string orbifold is that we have gauged the
action of the orbifold group $\Gamma$ on the covering space.
However, the projection $X \rightarrow X/\Gamma$ is only a principal
$\Gamma$-bundle when $\Gamma$ acts freely on $X$.  Thus, again, if we
try to understand physical properties of string orbifolds
in terms of quotient spaces, we seem to have a minor difficulty.

%question to what extent $Z$ is the partition function of a sigma
%model on the quotient space $X/\Gamma$.

Another difficulty lies in interpreting twist fields.
Many physicists have believed that twist fields could be
understood in terms of some cohomology theory of the
quotient space\footnote{As opposed to merely a description in
terms of group actions on covers, as in, for example, \cite{ruan1}.}
At the moment, there does not seem to exist
any completely satisfactory calculation that reproduces twist fields 
directly in terms of some
cohomology of the quotient space.  Now, although many physicists
have this perspective, there does seem to be a bit of confusion
regarding twist fields in the community, and we shall
take a moment to speak to two classes of misconceptions that
have arisen in some places:
\begin{enumerate}
\item Equivariant cohomology:  Some physicists have
claimed that the twist fields can be reproduced using equivariant
cohomology.  A simple counterexample will convince the reader
that this is incorrect.  Consider the ${\bf Z}_2$ action on
${\bf C}^2$ that leaves the origin invariant.
Now,
\begin{displaymath}
H^*_{{\bf Z}_2}\left( {\bf C}^2, {\bf Z}\right) \: = \:
H^*_{{\bf Z}_2}\left( \mbox{point}, {\bf Z} \right) \: = \:
H^*( B{\bf Z}_2, {\bf Z}) \: = \: H^*( {\bf R}{\bf P}^{\infty}, {\bf Z})
\: = \: \{ {\bf Z}, 0, {\bf Z}_2, 0, {\bf Z}_2, \cdots \}
\end{displaymath}
% The last assertion is given in, for ex, Hatcher p. 171 (my copy)
On the other hand,
\begin{displaymath}
H^*\left( \widetilde{ {\bf C}^2/{\bf Z}_2 }, {\bf Z} \right) \: = \:
H^*\left( {\bf P}^1, {\bf Z} \right) \: = \: \{ {\bf Z}, 0,
{\bf Z}, 0, 0, 0, \cdots \}
\end{displaymath}
Clearly, equivariant cohomology does not compute twist fields,
as not even its freely-generated part is appropriate.
\item Resolutions.  Because string orbifolds behave as though
they described strings on smooth spaces, and because their massless
spectrum contains twist fields, which often couple to exceptional divisors
of resolutions, people sometimes speak loosely of string orbifolds
describing strings on resolutions of quotient spaces.
If that were literally true, then twist fields would be naturally
understood in terms of properties of resolutions.
However, not all singularities that one encounters can be resolved;
for example, ${\bf C}^4/{\bf Z}_2$ admits no Calabi-Yau resolution,
yet a ${\bf C}^4/{\bf Z}_2$ orbifold possesses a twist field.
\end{enumerate}

We shall propose in this paper that string orbifolds are literally
strings compactified on spaces known as quotient {\it stacks}
(denoted $[X/\Gamma]$), 
which are closely related to quotient spaces $X/\Gamma$.
Quotient stacks differ from quotient spaces by possessing
extra structure at singularities, so in effect, we are arguing
that the ``stringy'' structure believed to exist at singularities
in string orbifolds, actually has a geometric interpretation.
This picture of string orbifolds sheds new light on their
physical properties.

%How can these difficulties be resolved?
%As we noted earlier, string orbifolds do not,
%strictly speaking, define string propagation on the quotient
%space $X/\Gamma$, but instead describe string propagation on
%another, very closely related space.  This space is known as the quotient
%{\it stack}, and denoted $[X/\Gamma]$ (to be distinguished from
%the quotient space $X/\Gamma$).

At first blush, this may sound radical,
but actually it is quite conservative.
In particular, to a mathematician familiar with stacks,
this would be very natural.  Among other things, quotient stacks
are one way of describing group actions on covers -- the
term ``quotient stack'' could be used synonymously with
``orbifold'' -- and moreover, as stacks, they have the structure
of a ``generalized space,'' so that one can, for example,
perform differential geometry on a (quotient) stack.
(One way to think of stacks is as slight generalizations of spaces.)
In a nutshell, we are proposing that the extra structure of
a generalized space has physical relevance.

%, on the grounds that
%stacks appear elsewhere in physics as well, though their presence
%usually goes unremarked.  For example, in
%\cite{stacks} we observed that the twisted ``bundles'' on D-brane
%worldvolumes in the presence of nontrivial B fields are naturally
%understood as sheaves on stacks; more flippantly, D-branes
%naturally live on stacks, not spaces.
%Again, stacks have appeared implicitly in physics
%previously (though this fact usually goes unnoticed), 
%and we are merely pointing out another one of
%their appearances.

Quotient stacks have the following basic properties:
\begin{enumerate}
\item Maps from the string worldsheet into $[X/\Gamma]$ are described
by maps from twisted sectors into the cover $X$.  Put another
way, a string orbifold is literally a weighted sum over maps into
$[X/\Gamma]$, a smoking gun for an interpretation
as a sigma model on $[X/\Gamma]$. 
This is, in the author's mind, the single most important point --
by merely unraveling definitions, one finds the signature of an
interpretation of string orbifolds
in terms of a sigma model on a quotient stack $[X/\Gamma]$.
We also propose a classical action for a sigma model on a stack that
reproduces string orbifolds (down to $|G|^{-1}$ factors in
partition functions) for the special case of global quotient stack
targets.
%We shall support this result in a number of other ways; however,
%that does {\it not} mean
%we have made some wild
%guess that we are backing up with computations in a few examples.  Rather,
%we have found directly from unraveling definitions,
%and incidentally shall also give a wide array of additional support.
\item If $\Gamma$ acts freely, so that $X/\Gamma$ is nonsingular,
then $[X/\Gamma] \cong X/\Gamma$.
\item Regardless of whether or not $\Gamma$ acts
freely, there is a natural map $X \rightarrow [X/\Gamma]$ which describes
$X$ as a principal $\Gamma$-bundle over $[X/\Gamma]$.
\item The quotient stack $[X/\Gamma]$ is smooth if $X$ is smooth
and $\Gamma$ acts by diffeomorphisms,
regardless of whether or not $\Gamma$ acts freely.  Thus, the CFT
of a sigma model on $[X/\Gamma]$ is naturally well-behaved, regardless of
whether $X/\Gamma$ is singular.
In the past, people have claimed that
string orbifolds are nonsingular because ``string theory smooths out
singularities;'' we instead are claiming that string orbifolds are
nonsingular because they describe sigma models into smooth spaces --
nonsingularity has nothing to do with string theory.
The old lore about ``strings smoothing out singularities'' is just a result
of misinterpreting the underlying mathematical structure.
Later we shall argue that, moreover, quotient stacks are smooth
in precisely the sense relevant for physics.
\end{enumerate}

In addition,
\begin{enumerate}  \setcounter{enumi}{4}
\item We can now naturally interpret twist fields,
in terms of cohomology of a stack associated to a quotient stack
rather than a quotient space. 
More precisely, we shall
argue later that the low-energy spectrum of a sigma model on $[X/\Gamma]$
is determined not by a cohomology of $[X/\Gamma]$ (a subtlety
of generalized spaces), but rather by a cohomology of the associated inertia
group stack $I_{[X/\Gamma]}$.  This stack has the pleasant property that
\cite{toen}
\begin{displaymath}
I_{[X/\Gamma]} \: \cong \: \coprod_{[g]} \, \left[ \, X^g / C(g) \, \right]
\end{displaymath}
which is the precise form of the Hirzebruch-H\"ofer \cite{hirzhofer}
expression for the orbifold Euler
characteristic
\begin{displaymath}
\chi_{orb}(X, \Gamma) \: = \: \sum_{[g]} \, \chi( X^g / C(g) )
\end{displaymath}
For physicists who have been expecting a description of twist fields
in terms of a cohomology of a quotient space, this is rather new.
However, this result is not new to some mathematicians,
who may recognize it as an overcomplicated way of describing
twist fields in terms of group actions on covers.
In essence, we are arguing that a description of twist fields
in terms of group actions on covers is the best one can expect.
\item The lore that string orbifolds have ``nonzero B fields at
quotient singularities'' also has a natural understanding in this
framework, which we shall explore in detail.  Very briefly,
if we think of a quotient stack as looking like a quotient space
with some sort of ``extra structure'' at the singularities, then
that extra structure is precisely a gerbe over the singularities,
precisely right to duplicate the existing lore.
\item The role of equivariance in string orbifolds is now clarified.
We shall show that a bundle (or sheaf) on the quotient stack
$[X/\Gamma]$ is the same thing as a $\Gamma$-equivariant bundle
(or sheaf) on $X$.  Thus, the standard intuition that one defines
an orbifold group action by specifying an equivariant structure,
is clarified.
\end{enumerate} 

%We would like to emphasize at this point that our plan in this
%paper is to unravel definitions to show that a string orbifold
%is a sigma model on a quotient stack.  In particular, we are not
%making some wild guess and attempting to motivate it by checking
%a handful of examples.  Rather, we are unraveling elementary
%physics definitions to get correct mathematical understanding.
%We also provide a wide variety of additional evidence to support
%our conclusion, but again, the results in this paper follow
%from unraveling definitions, not wild guesses.

So far we have merely put a new perspective on existing physics.
However, there are some important new directions we can use
this technology to pursue:
\begin{enumerate}
\item This now gives us an entirely new family of spaces to compactify on.
Quotient stacks are examples of ``generalized'' spaces, and cannot
be understood within the usual notions of point-set topology.
Since we now have an example of strings on generalized spaces
(provided by string orbifolds), we see that strings can indeed
consistently propagate on these generalized spaces, and so these can
be used for compactification in general.
\item We can now understand orbifolds in M-theory (in the sense of,
quantum theory underlying eleven-dimensional supergravity).
Previously, one might have argued that understanding orbifolds
in M-theory was unclear, as twisted sectors seem to be purely stringy.
However, now that we properly understand the underlying mathematics,
we see that twisted sectors are not in any way stringy,
and can now make sense of the notion of an orbifold in M-theory,
as M-theory compactified on a quotient stack $[X/\Gamma]$.
\item As a further minor note, this appears to
give an alternative way of
understanding the computation of B fields at quotient singularities.
%as well as finally giving a mathematical understanding of
%twist fields and related concepts.
\end{enumerate}

Unfortunately, the picture is not completely set.
One of the immediate consequences of a description of string
orbifolds as sigma models on quotient stacks, is that
deformation theory of the string orbifold CFT must be understood
in terms of deformation theory of the quotient stack,
as opposed to the quotient space.  We shall give indirect evidence
that this may coincide with the usual picture,
in terms of a quotient space ``with B fields,'' but much work remains
to be done to check whether this is actually consistent.

Phrased another way, although we describe sigma models on stacks
classically, we have not checked whether there are any obstructions
to quantization, which would be due to global effects
(see for example \cite{mmn} for a discussion of this matter in the context
of sigma models on spaces).  By performing consistency checks
(such as checks of the deformation theory, as above),
one can hope to gain some insight into whether this picture
is consistent.  In any event, much work remains to be done.

We begin in section~\ref{genlgpoid} with an overview of groupoids
and stacks, the technology that will be central to this paper.
In section~\ref{qstacks} we then give a lengthy pedagogical tutorial
on the basics of quotient stacks.  In section~\ref{sigmamodel} we consider
sigma models on quotient stacks, and argue that immediately from the
definitions, a sum over maps into a quotient stack
duplicates both the twisted sector sum and the functional integrals
within each twisted sector, a smoking gun for an interpretation
of string orbifolds in terms of sigma models on stacks.
In section~\ref{Gfree} we examine quotient stacks when the
orbifold group action is free, and argue that in this special
case, $[X/\Gamma] \cong X/\Gamma$.  In section~\ref{Xprinbundle} we
point out that $X$ is always the total space of a principal $\Gamma$-bundle
over $[X/\Gamma]$, regardless of whether or not $\Gamma$ acts freely,
a fact we will use often.  In section~\ref{equivlore} we examine the
lore on equivariance in string orbifolds, and show how a sheaf on
$[X/\Gamma]$ is equivalent to a $\Gamma$-equivariant sheaf on $X$.
In section~\ref{smooth}
we show that quotient stacks are always smooth (so, as we argue later,
a sigma model on a quotient stack should be well-behaved, as indeed
string orbifold CFT's are).  
In section~\ref{redux} we complete the picture begun in
section~\ref{sigmamodel} by writing down a proposal for a 
classical action for a sigma model on a stack, which generalizes
both ordinary sigma models and string orbifolds.
In section~\ref{wellbehave} we briefly outline what conditions
would naively be necessary for a sigma model on a stack
to yield a well-behaved CFT, and argue that smoothness of the
target stack are the conditions needed -- so, in other words,
the notion of `smoothness' for stacks has physical relevance
in this context.
In section~\ref{twistfield} we examine
the low-energy spectrum of strings propagating on a quotient stack,
and argue that it is computed by the cohomology of an auxiliary
stack $I_{[X/\Gamma]}$, a result that, although known to some
mathematicians, has a very different form from what many
physicists have expected.  
In section~\ref{noneff} we talk about non-effective orbifold
group actions.  
In section~\ref{Bfield} we describe how the old lore about B fields and
string orbifolds seems to fit into this picture.  
In section~\ref{Mthy} we talk about what our results
implies about understanding orbifolds in M theory.  
In section~\ref{newcomp}
we very briefly speak about the possibility of getting new string
compactifications from stacks.  
In section~\ref{ncg}, on a slightly different note,
we discuss some mathematical lore concerning
the relationship between stacks and noncommutative geometry.
Finally we conclude in 
section~\ref{conclusions} with a list of followup projects that we
encourage readers to pursue.
We have also included several appendices on more technical aspects
of quotient stacks.

%Before we proceed with the rest of this paper, a few comments are
%in order.  First, the insight into string orbifolds we present
%in this paper does not in any way alter existing physics calculations
%(although, as just outlined, it does extend our knowledge to new
%realms, and provides a depth of insight previously impossible).
%Even most of the physics lore is still intact.  The only change
%we are making in the existing lore
%is the interpretation of the string orbifold itself;
%all string orbifold CFT calculations people have done are
%unchanged. 

%Second, we should comment on our quotient stacks.  Historically
%quotient stacks have occurred only in algebraic geometry; we do not
%know of any writeup of any topological or smooth quotient
%stacks.  To really describe
%all string orbifolds, not just those appearing in an algebro-geometric
%context, in terms of quotient stacks, it is necessary to properly
%define and study topological and smooth quotient stacks,
%in addition to the algebraic quotient stacks already well-studied.
%We have outlined the basics of such quotient stacks in this paper,
%but more work remains to be done to put our topological
%constructions on firm ground.  

Before we proceed with the rest of this paper,
a few comments are in order.  First,
we should comment on our quotient stacks.
Historically, stacks have been used primarily by algebraic
geometers, so many discussions of stacks refer to schemes.  
However, stacks have nothing to do {\it per se} with schemes,
and can be described purely topologically (see 
\cite{stacks,dt2,brylinski,breen,fantechi,teleman} for examples).
Now, although stacks in general are not specific to algebraic
geometry, all previous discussions of quotient stacks we
have seen are written in terms of schemes.
To make them more useful for physics (and accessible for physicists),
we have written up a description of quotient stacks from 
purely topological and differential-geometric perspectives.
The details and analysis of such ``topological quotient stacks''
is virtually identical to that of traditionally-defined quotient stacks.
In particular, in this paper we really only use very basic properties
of quotient stacks, and these properties all go over without modification
to our ``topological quotient stacks'' (essentially because the
underlying analysis is identical between the cases).
Put another way, because our ``topological quotient stacks''
are in principle very slightly different from traditional quotient stacks,
one has to check that everything works -- but after checking,
one finds that everything basic works just as usual, and essentially
for the same reasons.  Because these ``topological quotient stacks''
are virtually identical to traditional quotient stacks,
we have opted to refer to them as quotient stacks; dignifying them
with a new name seems excessive.
Furthermore, although our description
of and perspective on quotient stacks is very slightly novel, the results
on quotient stacks that we use are very well-entrenched in the existing
mathematics literature.

%The reader might then be concerned at the possibility of mathematical
%errors in this paper.  Not to worry!
%Such a reader should rest easy, as 
%the results about quotient stacks we use in this paper are all very basic
%properties, quoted in any review of quotient stacks, and really
%have nothing to do with algebra or algebraic geometry, but rather
%have a much broader understanding.  Although our description
%of and perspective on quotient stacks is slightly novel, the results
%on quotient stacks that we use are very well-entrenched in the existing
%mathematics literature.

It should also be said that
there exists a group of mathematicians who already use
quotient stacks to describe string orbifolds.
Indeed, as described above, quotient stacks are an overcomplicated way
to describe group actions on covering spaces, and also possess
extra structure that gives them an interpreation as a sort of
`generalized space,' so that, for example, one can make sense of
differential geometry on a quotient stack.  However, as far as the
author has been able to determine, the mathematicians in question
have not done any of the work required to justify making the claim
that a string orbifold CFT coincides with the CFT for a string 
compactified
on a quotient stack, or even to justify the claim that the notion of
string compactification on a stack makes sense.  They do not seem to
have attempted to study sigma models on stacks, and do not even realize
why this is relevant.  They also do not seem to be aware of even the most
elementary physical implications of such a claim, such as the fact that,
to be consistent, any deformations of a string orbifold CFT would have
to be interpreted in terms of deformations of the quotient stack,
rather than the quotient space.  Thus, mathematicians reading this paper
should interpret our work as the beginnings of a program to fill in
the logical steps that they seem to
have omitted.  In fact, this paper was written
for a physics audience; mathematicians are encouraged to instead read
lecture notes \cite{wisc} we shall publish shortly.

Finally, in our discussion of quotient stacks as generalized
topological spaces, we assume that $\Gamma$ is discrete\footnote{This
condition can be dropped in general, but it simplifies certain technical
aspects, and is sufficient for our purposes.} and acts by homeomorphisms.
When we discuss quotient stacks as manifolds, we assume
that $X$ is a smooth manifold and that $\Gamma$ acts by diffeomorphisms.

\section{Generalities on groupoids and stacks}    \label{genlgpoid}

We shall begin with a brief overview of groupoids and stacks,
which we shall be manipulating for the rest of this paper.
The original reference for the material present is
\cite[section 4]{dm}.  For algebraic stacks in general,
and quotient stacks in particular,
\cite{gomez} is an excellent resource, and we also recommend
\cite[appendix A]{vistoli}.

For alternative perspectives on stacks, less useful in the present
context but more useful for other things, see for example
\cite{stacks,dt2,brylinski,breen,fantechi,teleman}.

\subsection{Groupoid basics}

Let ${\cal C}$ be any category.  Let ${\cal F}$ be another category
and $p: {\cal F} \rightarrow {\cal C}$ a functor.
We say that ${\cal F}$ is a groupoid over ${\cal C}$ 
\cite[section 4]{dm} if it obeys
the following two axioms:
\begin{enumerate}
\item If $\rho: U \rightarrow V$ is a morphism in ${\cal C}$ and
$\eta \in \mbox{Ob } {\cal F}$ with $p(\eta) = V$, then there exists an 
object $\xi \in \mbox{Ob } {\cal F}$ and a morphism $f: \xi \rightarrow 
\eta$ in ${\cal F}$ such that $p(\xi) = U$ and $p(f) = \rho$.
\item If $\phi: \xi \rightarrow \zeta$ and $\psi: \eta \rightarrow
\zeta$ are morphisms in ${\cal F}$ and $h: p(\xi) \rightarrow
p(\eta)$ is a morphism in ${\cal C}$ such that
$p(\psi) \circ h = p(\phi)$, then there exists a unique morphism
$\chi: \xi \rightarrow \eta$ such that $\psi \circ \chi = \phi$
and $p(\chi) = h$.
\end{enumerate}

It is straightforward to check that the object $\xi$ determined in the
first axiom is unique up to unique isomorphism, from the second axiom.
Since $\xi$ is determined uniquely (up to unique isomorphism),
$\xi$ is commonly denoted $\rho^* \eta$ (remembering that it comes
with a morphism $\rho^* \eta \rightarrow \eta$).  

It is also straightforward to check that if $\alpha$ is a morphism in ${\cal F}$
such that $p(\alpha) = \mbox{Id}$, then $\alpha$ is invertible.

Any object of ${\cal C}$ naturally defines a groupoid over ${\cal C}$.
Specifically, let $X \in \mbox{Ob }{\cal C}$, and define
${\cal F} = \mbox{Hom}_{{\cal C}}(-, X)$, i.e., the category whose
\begin{enumerate}
\item Objects are morphisms $Y \rightarrow X$ in ${\cal C}$, for
any other object $Y \in \mbox{Ob } {\cal C}$, and
\item Morphisms $(Y \rightarrow X) \longrightarrow (Z \rightarrow X)$
are morphisms $Y \rightarrow Z$ such that the diagram
\begin{displaymath}
\xymatrix{
Y \ar[rr] \ar[dr] & & Z \ar[dl] \\
 & X &
}
\end{displaymath}
commutes.
\end{enumerate}
Later in this paper, we shall often use the notation $(X)_{map}$
in place of $\mbox{Hom}(-,X)$, in an attempt to give readers
inexperienced in these matters a better grip.
We can define a functor $p: \mbox{Hom}(-, X) \rightarrow {\cal C}$,
that maps an object $( Y \rightarrow X)$ to the object $Y \in
\mbox{Ob } {\cal C}$, and acts in the obvious trivial way upon morphisms.
It is straightforward to check that with these definitions,
$\mbox{Hom}(-, X)$ is a groupoid over ${\cal C}$.

A morphism of groupoids ${\cal F} \rightarrow {\cal G}$ over ${\cal C}$
is a functor between the categories such that the following diagram
commutes:
\begin{displaymath}
\xymatrix{
{\cal F} \ar[rr] \ar[dr] & & {\cal G} \ar[dl] \\
& {\cal C} & 
}
\end{displaymath}

Note that since we have defined groupoids associated to objects
$X \in \mbox{Ob }{\cal C}$, we can now talk about maps $X \rightarrow
{\cal F}$, i.e., maps from objects of ${\cal C}$ into groupoids ${\cal F}$
over ${\cal C}$.  Such a map is simply a functor 
\begin{displaymath}
\mbox{Hom}_{{\cal C}}\left( -, X \right) \: \longrightarrow \: 
{\cal F}
\end{displaymath}
compatible with the projections to ${\cal C}$.

We can now talk about representable groupoids.
We say that a groupoid ${\cal F}$ is representable, 
or represented by $X \in \mbox{Ob } {\cal C}$, if there exists
an object $X \in \mbox{Ob }{\cal C}$ and an equivalence of categories
$F: {\cal F} \longrightarrow \mbox{Hom}(-,X)$ (compatible with the
projection maps to ${\cal C}$).  

Now that we have described groupoids for general categories
${\cal C}$, what categories does one actually use in practice?
Sometimes, especially if one wants to think of a groupoid
as being a ``presheaf of categories'' on a space,
one takes ${\cal C}$ to be the category of open sets on
that space (see for example \cite{stacks,dt2,brylinski,breen}.)

However, when describing quotient stacks and other algebraic
stacks, one typically wants ${\cal C}$ to be ${\it Sch/S}$
(the category of schemes relative to a fixed scheme), whose
topological equivalent is ${\it Top}$, the category of topological
spaces. 

Now, in this paper we want to describe stacks from a topological and
differential-geometric perspective, so we shall refer to groupoids
over ${\it Top}$, instead of ${\it Sch/S}$.  Also, we shall be
specifically concerned with quotient stacks, which can also
be understood as groupoids over quotient spaces.  This fact
is often obscure in usual discussions of quotient stacks
(see \cite{vistoli} for a recent discussion), but as it adds
a great deal of geometric insight, we shall emphasize this
connection in this paper.

\subsection{Stack basics}

A stack is a special kind of groupoid, one that satisfies certain
gluing axioms (just as a sheaf is a special kind of presheaf, satisfying
certain gluing axioms).  The relation to sheaves is more than
just an analogy -- stacks are often thought of as, essentially,
sheaves of categories.  One demands that objects and morphisms satisfy
certain gluing axioms.  Of course, to make sense of those gluing
axioms, one must define a notion of covering, which requires
putting what is known as a Grothendieck topology on the categories
in question. 

In this paper, although we will be manipulating groupoids a very
great deal, nothing we shall be doing will require us to work
with stacks specifically.  Put another way, although groupoids
are very important in this paper, stacks {\it per se} have no role
at all.  Although we will be describing quotient ``stacks,''
we will only be manipulating them at the level of the underlying
groupoid.

As stacks {\it per se} (as opposed to groupoids) play such
a marginal role in this paper, we shall not discuss them
further.

For more information on stacks specifically,
see for example \cite{stacks,dt2,gomez,vistoli,dm,brylinski,breen}.

%**** Should I mention the canonical exs from algebraic stacks: $Sch/S$,
%in \'etale and fppf topologies ??

\section{Quotient stacks}   \label{qstacks}

In this section we shall give a basic description of 
the quotient stack
$[X/\Gamma]$.  
Our presentation is ordered pedagogically.  We begin with some
generalities on generalized spaces, then discuss
the points of a quotient stack, and then outline how one could 
understand the open sets of a quotient stack.  Then, finally,
we come to a thorough technically useful description, in terms
of the continuous maps into a quotient stack.

In passing, for the benefit of the expert on stacks,
we shall say a few things about our presentation.  First, since we are
interested in describing stacks from a topological and
differential-geometric perspective, our groupoids will be defined
over ${\it Top}$, the category of topological spaces,
rather than ${\it Sch/S}$, the category of schemes relative to
some fixed scheme $S$.  Second, we are primarily interested in
quotient stacks in this paper, which can be understood
as groupoids over quotient spaces as well as ${\it Top}$
(see \cite{vistoli} for a recent discussion of this matter).
In order to help the reader gain some geometric insight into
quotient stacks, we shall primarily refer to them in terms
of groupoids over quotient spaces.  
In particular, we shall talk about the category of continuous
maps $[X/\Gamma]_{map}$ into a quotient stack, as a groupoid
over the category of continuous maps $(X/\Gamma)_{map}$
into the quotient space.  For pedagogical reasons 
we shall also set up the category
of points of $[X/\Gamma]$, as a groupoid over the  
discrete\footnote{A discrete category is the same thing as a set -- 
it is a category in which the only morphisms are the identity morphisms.} 
category of points of $X/\Gamma$.
% and to further aid the reader,
%we have included an ansatz for the category of ``open sets'' of
%$[X/\Gamma]$, also described as a groupoid over the category of
%open sets of $X/\Gamma$.

The general intuition for quotient stacks is that a quotient
stack $[X/\Gamma]$ looks essentially like a quotient space
$X/\Gamma$, but with extra structure at the singularities.
That extra structure makes quotient stacks far better behaved than
quotient spaces.  The technical description of quotient stacks
rapidly obscures this intuition; the reader may want to keep this
picture in mind.

\subsection{Generalized spaces}

Quotient stacks are examples of ``generalized spaces,'' which cannot
be understood within the usual notions of point-set topology.
In a nutshell, in generalized spaces, all occurrences of
``set'' in point-set topology should be replaced by ``category.''
For example, instead of having a {\it set} of points, a generalized space
has a category of points.  In other words, distinct points can be
isomorphic, and a single point can have nontrivial automorphisms.

Ordinary spaces can also be thought of as generalized spaces.
For example, since sets are discrete categories, we can think of
the set of points of an ordinary space as a (discrete) category of
points.

Generalized spaces are perhaps best defined in terms of the
category of maps into them.  After all, certainly ordinary
topological spaces are completely determined by the categories
of maps into them (see appendix~\ref{mapsgivespace}), so this
is a very natural approach.  One can then recover, for example,
the category of points as the category of maps from a fixed point
into the space.  This is somewhat analogous to noncommutative geometry,
where spaces are defined by the rings of functions on them.
Instead of altering the product structure of the ring to describe
new spaces, we alter the category structure of maps.
This may sound somewhat cumbersome, but it is actually an ideal
setup for string sigma models.

In passing, we should also mention that the idea of defining
spaces in terms of the maps into them is a commonly-used setup
in algebraic geometry (see discussions of ``Grothendieck's functor
of points'' in, for example, \cite[section II.6]{mumred} or
\cite[section VI]{eisenbudharris}).
For example, Hilbert schemes are defined by the maps into them;
the defining statement \cite[chapter 1.1]{nakajima}
is that a map from a scheme $U$ into a Hilbert scheme of ideals over $X$
(i.e., an object of the category)
is a closed subscheme of $U \times X$, flat over $U$, of fixed Hilbert
polynomial.

We shall introduce quotient stacks in two successive stages.
First, we shall discuss the category of points of a quotient stack.
%Second, we shall discuss an ansatz for the category of open sets
%on a quotient stack.  Now, in practice, a notion of ``open set''
%for a quotient stack does not seem very useful at all; however,
%readers new to this material will no doubt wonder what an open
%set should be, and so our ansatz will hopefully help them
%adjust to these spaces.  
The category of points
%nor our ansatz for open sets, however, 
does not completely characterize
quotient stacks in a useful form, however.  Completely characterizing
quotient stacks is the business of the second stage, where
we discuss the continuous maps into a quotient stack.

For an introduction to generalized spaces, see for example
\cite{stacks,mumford1,artingt,milne,johnstone,maclane}.
% Used to also cite phsites

\subsection{Points}

\subsubsection{Definition and intuition}

The category $[X/\Gamma]_{pt}$ of points of 
$[X/\Gamma]$ has the following objects
and morphisms:
\begin{enumerate}
\item Objects are $\Gamma$-equivariant maps $\Gamma \rightarrow X$,
i.e., maps $f: \Gamma \rightarrow X$ such that $f(g h) = g f(h)$
for $g, h \in \Gamma$.
\item Morphisms $f_1 \rightarrow f_2$ are $\Gamma$-equivariant
bijections $\lambda: \Gamma 
\rightarrow \Gamma$ such that $f_2 \circ \lambda = f_1$
\end{enumerate}

Loosely speaking, the category $[X/\Gamma]_{pt}$ 
of points of $[X/\Gamma]$ consists
of the orbits of points of $X$ under $\Gamma$.

Note that there is a natural projection functor $[X/\Gamma]_{pt} \rightarrow
(X/\Gamma)_{pt}$ (where $(X/\Gamma)_{pt}$ denotes the discrete category
(i.e., set) of points of $X/\Gamma$):  
any object $(f: \Gamma \rightarrow X)$ maps to the
image of $f(e)$ in $X/\Gamma$, and any morphism maps to the identity
morphism.  It is straightforward to check that, with this projection
functor, $[X/\Gamma]_{pt}$ is a groupoid over $(X/\Gamma)_{pt}$. 

%To gain a little intuition for quotient stacks,
%let us examine some special cases.

Let us pause to study these definitions.
For any one orbit of a point on $X$,
the various elements of the corresponding isomorphism class of
points on $[X/\Gamma]$ consists of the various ways to map points of
$\Gamma$ into the points of the orbit.
Thus, 
%if $\Gamma$ acts freely, then at the level of points,
isomorphism classes of points of $[X/\Gamma]$ are in one-to-one
correspondence with points of $X/\Gamma$.  In fact,
when $\Gamma$ acts freely, we shall see later that $[X/\Gamma]$ and
$X/\Gamma$ are homeomorphic\footnote{For example, it is
straightforward to check that if
$\Gamma$ acts freely, then the category of points of $[X/\Gamma]$
is equivalent to the discrete category of points of $X/\Gamma$.}, 
so away from fixed points of $\Gamma$,
$[X/\Gamma]$ and $X/\Gamma$ look
the same.

What happens at fixed points of $\Gamma$?
To gain a little insight, first consider the case that $\Gamma = {\bf Z}_2$.
At a fixed point, there will be a single corresponding point of $[X/\Gamma]$,
but this point will have a nontrivial automorphism in the category
of points.

More generally, it is straightforward to check that if $f_1$, $f_2$
are two objects in the category of points, then the set of morphisms
$f_1 \rightarrow f_2$ will be 
\begin{enumerate}
\item empty if $f_1$ and $f_2$ do not have the same image on $X$, or
\item nonempty if they do have the same image on $X$, and containing
as many elements as the number of elements of the isotropy group of
a point in the orbit
\end{enumerate}

As mentioned earlier, the intuitive picture of quotient stacks
is that $[X/\Gamma]$ looks like $X/\Gamma$ except at singularities,
where the quotient stack has extra structure.  Here, we have seen
this explicitly, at the level of points of the stack.

\subsubsection{The canonical functor $(X)_{pt} \rightarrow
[X/\Gamma]_{pt}$}

In general, the projection map $X \rightarrow X/\Gamma$ only defines
$X$ as a principal $\Gamma$-bundle over $X/\Gamma$ if $\Gamma$ acts
freely.  However, there is a natural map $X \rightarrow [X/\Gamma]$.

We can understand this map at the level of points as
a functor
$\pi: (X)_{pt} \rightarrow [X/\Gamma]_{pt}$, defined 
as follows.  For any point $x \in X$ (i.e., any object of the 
discrete category $(X)_{pt}$),
define $f_x: \Gamma \rightarrow X$ by, $f_x(g) = g \cdot x$.
(The only morphisms in $(X)_{pt}$ are the identity morphisms, which
are mapped to identity morphisms.)
We shall show later that this map always defines $X$ as the total space
of a principal $\Gamma$-bundle over $[X/\Gamma]$, regardless of
whether or not $\Gamma$ acts freely.

Before continuing, we shall make a technical observation concerning
this functor, that will be useful later.  In principle, one could define other functors $\pi_g:
X \rightarrow [X/\Gamma]$, one for each $g \in \Gamma$, as the
composition
\begin{displaymath}
X \: \stackrel{g}{\longrightarrow} \: X \:
\stackrel{\pi}{\longrightarrow} \: [X/\Gamma]
\end{displaymath}
However, these maps are all equivalent, in the sense that the
corresponding functors are always isomorphic.
(Indeed, the reader might have expected this, since the analogous
maps into the quotient space $X/\Gamma$ are all equivalent.)
Specifically, at the level of points, we can explicitly define a
natural transformation $\omega: \pi \Rightarrow \pi_g$ as follows:
for each $x \in X$, define $\omega(x): \Gamma \rightarrow \Gamma$
to be the $\Gamma$-equivariant map defined by $\omega(x)(e) = g$.
It is straightforward to check that this is an invertible natural
transformation, and so, at least at the level of points,
$\pi \cong \pi_g$.  We shall see this is also the case in general later.

\subsection{Continuous maps}

As mentioned earlier, quotient stacks $[X/\Gamma]$ are defined
by the category $[X/\Gamma]_{map}$ of maps into them.
In this section we shall describe that category in detail.

Much of this section is rather technical.  The basic idea
is motivated in the next subsection; a reader working through this
material for the first time might want to read only the next subsection
and leave the rest for later study.

\subsubsection{Intuition}

In this subsection we shall merely motivate the definition of
a continuous map into a quotient stack $[X/\Gamma]$, rather than
try to give an intrinsic definition of continuity for generalized spaces.

Recall points of $[X/\Gamma]$ are $\Gamma$-equivariant maps
$\Gamma \rightarrow X$.  What should a ``continuous'' map
$f: Y \rightarrow [X/\Gamma]$ look like?

Well, for each point $y \in Y$, one would have a copy of $\Gamma$,
together with a $\Gamma$-equivariant map $f_y: \Gamma \rightarrow X$.
So, naively, it looks as though continuous maps $f: Y \rightarrow [X/\Gamma]$
should be described by $\Gamma$-equivariant maps
$Y \times \Gamma \rightarrow X$.

However, this description of the possible maps is slightly naive.
Ordinarily, when we think of a function $f: Y_1 \rightarrow Y_2$
between two (ordinary) spaces $Y_1$, $Y_2$,
the function should assign a unique point of $Y_2$ to each point 
of $Y_1$.  That is what it means for a function to be well-defined, 
after all.  However, for generalized spaces, matters are somewhat
more interesting.  For a function $f: Y \rightarrow [X/\Gamma]$
to be well-defined, we do not need to assign a unique point
of $[X/\Gamma]$ to each point of $Y$, but rather only a unique
equivalence class of points to each point of $Y$ -- it does not matter
if $f$ does not assign a well-defined point to each point of $Y$,
so long as the possible points of $[X/\Gamma]$ assigned to any one
point of $Y$ are isomorphic.

Thus, in general, a continuous map $Y \rightarrow [X/\Gamma]$
is defined by a principal $\Gamma$-bundle $E \rightarrow Y$,
together with a $\Gamma$-equivariant map $f: E \rightarrow X$.

\subsubsection{Technical definition}

Now that we have developed some intuition for continuous maps into
$[X/\Gamma]$, we shall give the complete technical definition of the
category of continuous maps into $[X/\Gamma]$.  First, however,
we shall define the category of continuous maps $(Y)_{map}$ into
a topological space $Y$, both for purposes of comparison, and for
later use.

The remainder of this section is rather technical; the reader
may wish to skip it upon a first reading.

Given any topological space $Y$, we can define a category $(Y)_{map}$
(also denoted $\mbox{Hom}(-,Y)$)
consisting of all the continuous maps into $Y$:
\begin{enumerate}
\item Objects are continuous maps $f: W \rightarrow Y$,
where $W$ is any topological space
\item Morphisms $( W_1 \stackrel{ f_1}{\longrightarrow} Y )
\longrightarrow ( W_2 \stackrel{f_2}{\longrightarrow} Y )$
are maps $\lambda: W_1 \rightarrow W_2$ such that the following
diagram commutes:
\begin{displaymath}
\xymatrix{
W_1 \ar[rr]^{\lambda} \ar[dr]_{f_1} & & W_2 \ar[dl]^{f_2} \\
& Y & 
}
\end{displaymath}
\end{enumerate}
It is straightforward to check that, for any topological space $Y$,
the category $(Y)_{map}$ is a groupoid over the category ${\it Top}$
of topological spaces, whose objects are topological spaces and morphisms
are continuous maps.  (The projection functor is defined by
sending an object $(W \rightarrow Y)$ of $(Y)_{map}$ to the object
$W \in \mbox{Ob } {\it Top}$, and in the obvious fashion on morphisms.)

Now, we shall finally define the category of continuous maps into
a quotient stack.
Define the category $[X/\Gamma]_{map}$ of continuous maps into 
the quotient stack $[X/\Gamma]$ as follows:
\begin{enumerate}
\item Objects are pairs $( E \longrightarrow Y, E \stackrel{f}{\longrightarrow}
X )$, where 
\begin{tabbing}
abcdefg \= yuck \kill  
 \> $Y$ is any topological space, \\
 \> $E \rightarrow Y$ is
a principal $\Gamma$-bundle, and \\
 \> $f: E \rightarrow X$ is a continuous $\Gamma$-equivariant map.
\end{tabbing}
\item Morphisms
\begin{displaymath}
\left( \, E_1 \: \longrightarrow \: Y_1, \: E_1 \: 
\stackrel{f_1}{\longrightarrow} \: X \, \right)
\: \longrightarrow \:
\left( \, E_2 \: \longrightarrow \: Y_2, \: E_2 \:
\stackrel{f_2}{\longrightarrow} \: X \, \right)
\end{displaymath}
are pairs $(\rho, \lambda)$,
where $\rho: Y_1 \rightarrow Y_2$ is a continuous map, and
$\lambda: E_1 \rightarrow E_2$ is a bundle morphism, meaning
\begin{displaymath}
\xymatrix{
E_1 \ar[r]^{\lambda} \ar[d] & E_2 \ar[d] \\
Y_1 \ar[r]^{\rho} & Y_2 
}
\end{displaymath}
commutes, and is constrained to make the diagram
\begin{displaymath}
\xymatrix{
E_1 \ar[rr]^{\lambda} \ar[dr]_{f_1} & & E_2 \ar[dl]^{f_2} \\
& X & 
}
\end{displaymath}
commute.
\end{enumerate}
It is straightforward to check that $[X/\Gamma]_{map}$ is a groupoid
over the category ${\it Top}$ of topological spaces.
(The projection functor sends pairs $( E \rightarrow Y, E \rightarrow X)$
to $Y \in \mbox{Ob } {\it Top}$, and sends a morphism $(\rho, \lambda)$
to $\rho$.)
Now, we shall argue later that
$[X/\Gamma]_{map}$ is also a groupoid over the category
$(X/\Gamma)_{map}$ of continuous maps into the quotient space
$X/\Gamma$, and it is this second groupoid structure that we
shall emphasize in this paper.

In passing, note that we can now recover the points of $[X/\Gamma]$
that we described earlier, simply as the maps from a fixed point
into $[X/\Gamma]$.  More generally, if one knows the category of
maps into some space, then the points of that space can be identified
with maps from a fixed point into the space.

\subsubsection{Philosophy}

If the reader pauses to think carefully about what we have
described so far, they may be somewhat confused about our
description of continuous maps into a quotient stack.
Given two spaces $X$ and $Y$, a continuous map $f: X \rightarrow Y$
should be equivalent to a functor between the categories
$(X)_{map} \rightarrow (Y)_{map}$.  (After all, given any $h: Z \rightarrow X$,
one can compose with $f$ to form $f \circ h: Z \rightarrow Y$.
One could recover $f$ itself as the image of the identity map $X \rightarrow
X$.)
Thus, a continuous map from any space $Y$ into the quotient stack
$[X/\Gamma]$ should be equivalent to a functor
$(Y)_{map} \rightarrow [X/\Gamma]_{map}$.

On the other hand, these functors should somehow be defined by
objects of $[X/\Gamma]_{map}$ -- after all, we defined $[X/\Gamma]_{map}$
to be the category of maps into $[X/\Gamma]$.

In fact, this is precisely the case, thanks to the Yoneda
lemma (see for example \cite{stacks,gomez,vistoli}), which says that
a functor $(Y)_{map} \rightarrow {\cal F}$, for any groupoid ${\cal F}$,
is determined (up to equivalence) by the image of the identity
map $\mbox{Id}: Y \rightarrow Y$.

In the present case, this means the following.
Let 
\begin{displaymath}
\left( \, E \: \longrightarrow \: Y, \: E \:
\stackrel{f}{\longrightarrow} \: X \, \right)
\end{displaymath}
be an object of $[X/\Gamma]_{map}$.  Given this object, define
a functor $(Y)_{map} \rightarrow [X/\Gamma]_{map}$ as follows:
\begin{enumerate}
\item Objects:
Let $( h: Z \rightarrow Y ) \in
\mbox{Ob } (Y)_{map}$.  The image of this object under the functor
we are defining is the pair
\begin{displaymath}
\left( \, h^* E \: \longrightarrow \: Z, \:
h^* E \: \stackrel{canonical}{\longrightarrow} \: E \:
\stackrel{f}{\longrightarrow} \: X \, \right)
\end{displaymath}
\item Morphisms:  Let 
\begin{displaymath}
\lambda: \: \left( \, h_1: \: Z_1 \: \longrightarrow \: Y \, \right)
\: \longrightarrow \: \left( \,
h_2: \: Z_2 \: \longrightarrow \: Y \, \right)
\end{displaymath}
be a morphism in $(Y)_{map}$, meaning that $\lambda: Z_1 \rightarrow Z_2$
is a map such that the diagram
\begin{displaymath}
\xymatrix{
Z_1 \ar[rr]^{\lambda} \ar[dr]_{h_1} & & Z_2 \ar[dl]^{h_2} \\
& Y &
}
\end{displaymath}
commutes.  The map $\lambda$ induces a canonical map
\begin{displaymath}
\lambda^{\#}: \: h_1^* E \, \left( \, = \: \lambda^* h_2^* E \, \right)
\: \longrightarrow \: h_2^* E
\end{displaymath}
so the image of the morphism $\lambda$ under the functor
$(Y)_{map} \rightarrow [X/\Gamma]_{map}$ is the morphism
$(\lambda, \lambda^{\#})$.
\end{enumerate}

Conversely, given a functor $(Y)_{map} \rightarrow [X/\Gamma]_{map}$,
the object of $[X/\Gamma]_{map}$ corresponding to this functor
is the image of the object $(\mbox{Id}: Y \rightarrow Y) \in
\mbox{Ob } (Y)_{map}$.

For example, there is a canonical map $X \rightarrow [X/\Gamma]$
defined by the object
\begin{displaymath}
\left( \, X \times \Gamma \: \stackrel{p_1}{\longrightarrow} \: X, \:
X \times \Gamma \: \stackrel{eval}{\longrightarrow} \: X \, \right)
\end{displaymath}
of $[X/\Gamma]_{map}$.  This map plays the same role for quotient
stacks that the canonical projection $X \rightarrow X/\Gamma$ plays
for quotient spaces.
In subsection~\ref{canproj} we shall discuss this example
in more detail.

\subsubsection{ $[X/\Gamma]_{map}$ is a groupoid over $( X/\Gamma )_{map}$}

It is straightforward to check that the category
$[X/\Gamma]_{map}$ of maps into the quotient stack
$[X/\Gamma]$ is a groupoid over
the category $(X/\Gamma)_{map}$ of maps into the quotient space
$X/\Gamma$.

Define a projection functor $p: [X/\Gamma]_{map} \rightarrow
(X/\Gamma)_{map}$ as follows:
\begin{enumerate}
\item Objects:  Let $( E \stackrel{\pi}{\longrightarrow} Y, \:
E \stackrel{f}{\longrightarrow} X )$ be an object of $[X/\Gamma]_{map}$.
Given this object, define a map $g: Y \rightarrow X/\Gamma$ as follows:
for any $y \in Y$, let $e \in \pi^{-1}(y)$, and define
$g(y) = ( \pi_0 \circ f)(e)$, where $\pi_0: X \rightarrow X/\Gamma$
is the canonical map.  It is straightforward to check that this
map is well-defined and continuous.\footnote{To show that this map
is continuous, we have used the assumption that $\Gamma$ acts by
homeomorphisms.}

So, define 
\begin{displaymath}
p\left( \, \left( \, E \: \stackrel{\pi}{\longrightarrow} \: Y, \:
E \: \stackrel{f}{\longrightarrow} \: X \, \right) \, \right) \: = \:
\left( \, Y \: \stackrel{g}{\longrightarrow} \: X/\Gamma \, \right)
\end{displaymath}
\item Morphisms:  Let
\begin{displaymath}
( \rho, \lambda ): \:
\left( \, E_1 \: \stackrel{\pi_1}{\longrightarrow} \: Y_1, \:
E_1 \: \stackrel{f_1}{\longrightarrow} \: X \, \right) \: \longrightarrow \:
\left( \, E_2 \: \stackrel{\pi_2}{\longrightarrow} \: Y_2, \:
E_2 \: \stackrel{f_2}{\longrightarrow} \: X \, \right)
\end{displaymath}
be a morphism.  Define $p\left( (\rho, \lambda) \right) = \rho$.
\end{enumerate}

It is straightforward to check that this definition makes
$p$ a well-defined functor 
\begin{displaymath}
[X/\Gamma]_{map} \rightarrow
(X/\Gamma)_{map}
\end{displaymath}
Furthermore, it is also straightforward to check that 
$[X/\Gamma]_{map}$ is a groupoid over $(X/\Gamma)_{map}$.

\subsubsection{The canonical functor $(X)_{map} \rightarrow
[X/\Gamma]_{map}$}     \label{canproj}

At the level of points, recall we discussed a map
$X \rightarrow [X/\Gamma]$, given by mapping any point $x \in X$
to the function $f_x: \Gamma \rightarrow X$ defined by $f_x(g) = g \cdot x$.

We can now define this map more thoroughly.

At the level of maps, the functor $\pi: (X)_{map} \rightarrow
[X/\Gamma]_{map}$ is defined by the pair
\begin{displaymath}
\left( \, X \times \Gamma \: \stackrel{p_1}{\longrightarrow} \: X, \:
X \times \Gamma \: \stackrel{eval}{\longrightarrow} \: X \, \right)
\end{displaymath}
In other words, at the level of maps, this functor sends the object
$( Y \stackrel{f}{\longrightarrow} X )$ in $(X)_{map}$ to the object
\begin{displaymath}
\left( \, f^* ( X \times \Gamma \rightarrow X), \: f^* ( X \times \Gamma)
\: \longrightarrow \: X \times \Gamma \: \stackrel{eval}{\longrightarrow}
\: X \, \right)
\end{displaymath}

Now, for later use, note we could also define other functors $\pi_g: (X)_{map}
\rightarrow [X/\Gamma]_{map}$, for any $g \in \Gamma$, by the
composition
\begin{displaymath}
(X)_{map} \: \stackrel{g}{\longrightarrow} \: (X)_{map} \:
\stackrel{\pi}{\longrightarrow} \: [X/\Gamma]_{map}
\end{displaymath}
However, these functors are all equivalent.  Specifically,
define an invertible natural transformation $\omega: \pi_g \Rightarrow
\pi$ as follows.  For any object $( Y \stackrel{f}{\longrightarrow} X)$,
define $\omega\left( ( Y \stackrel{f}{\longrightarrow} X) \right)$
to be the pair $(\mbox{id}_Y, 1 \times g)$, where
$\mbox{id}_Y$ is the identity map $Y \rightarrow Y$, and $1 \times g:
Y \times \Gamma \rightarrow Y \times \Gamma$ acts as,
$(1 \times g)(y,h) = (y, hg)$.  It is straightforward to check that
this is a well-defined and invertible morphism in $[X/\Gamma]_{map}$, and
moreover that such $\omega$ define an invertible natural transformation
$\pi_g \Rightarrow \pi$.  Thus, the functors $\pi_g$ and $\pi$ are
isomorphic.

\subsubsection{General properties of continuous maps}   \label{mapprops}

Continuous maps between ordinary topological spaces
are often classified by certain extra properties they may possess.
For example, they may be open, closed, surjective, injective, 
local homeomorphisms, and so forth.

Closely analogous notions exist for certain continuous maps into stacks.
In order to understand how to phrase notions such as ``open''
and ``surjective'' in language that makes sense for stacks,
we need to reexamine these notions for ordinary topological spaces.
Now, although properties such as open and surjective do not, as ordinarily
formulated, make sense for stacks, the notion of a fiber product does make
sense for stacks.  With this in mind, let us rewrite such notions  
in terms of fiber products, so as to
be potentially useful for stacks.

Define a property ``P'' of continuous maps between
(ordinary) topological spaces to be local if
it is true that, a map $f: Y \rightarrow X$ has property ``P'' if and only
if for all continuous maps $g: Z \rightarrow X$,
the first projection map
$Z \times_X Y \stackrel{p_1}{\longrightarrow} Z$ has property ``P''
also.

It is straightforward to check that the properties open, surjective, injective,
and local homeomorphism are all examples of local properties.
In other words, for example, a continuous map $f: Y \rightarrow X$ between
ordinary topological spaces is open if and only if for all
continuous maps $g: Z \rightarrow X$, the projection
$Z \times_X Y \stackrel{p_1}{\longrightarrow} Z$ is also open.

In order to be able to apply this to stacks, we still need to relate
fiber products of stacks to ordinary spaces.
Now, many morphisms $f: Y \rightarrow {\cal F}$ 
into a stack ${\cal F}$ 
have the useful property
that for all continuous maps 
$g: Z \rightarrow {\cal F}$, the fiber product $Z \times_{{\cal F}} Y$
is an ordinary space, not another stack.  
We call a morphism $f: Y \rightarrow {\cal F}$ representable if it has
this property.  Intuitively, for a morphism to be representable
means that the fibers of the morphism are ordinary spaces, not stacks.
(For a useful theorem on representable morphisms, see
appendix~\ref{represent}.)

At least for
the representable morphisms, 
we can now formulate for stacks the local properties of
morphisms.
Specifically, 
%for a stack ${\cal F}$ such that fiber products
%$Z \times_{{\cal F}} Y$ are always representable by ordinary spaces,
if ``P'' is a local property of morphisms between ordinary spaces,
then we say that a representable map $f: Y \rightarrow {\cal F}$ has
property ``P'' precisely when for all continuous maps $g:
Z \rightarrow {\cal F}$, the projection map $Z \times_{{\cal F}} Y
\stackrel{p_1}{\longrightarrow} Z$ has property ``P.'' 
Since, by assumption, $Z \times_{{\cal F}} Y$ is an ordinary space,
it is sensible to talk about properties of the continuous map
$Z \times_{{\cal F}} Y \rightarrow Z$ in the ordinary sense.

In this paper, essentially the only map into a stack for which
we shall be concerned with local properties is the canonical
map $X \rightarrow [X/\Gamma]$, and we shall show
in section~\ref{Xprinbundle} that this map is representable.
In fact, we shall show that 
if $Y \rightarrow [X/\Gamma]$ is defined by the pair
\begin{displaymath}
\left( \, E \: \stackrel{\pi}{\longrightarrow} \: Y, \:
E \: \stackrel{f}{\longrightarrow} \: X \, \right)
\end{displaymath}
then $Y \times_{[X/\Gamma]} X \cong E$, and the projection map
$Y \times_{[X/\Gamma]} X \stackrel{p_1}{\longrightarrow} Y$
is isomorphic to the
projection $E \stackrel{\pi}{\longrightarrow} Y$.

Let us put our description of properties of morphisms into
practice.
For example, we can now ask whether the canonical (and
representable) map $X \rightarrow
[X/\Gamma]$ is open.   Let $Y \rightarrow [X/\Gamma]$ be any
continuous map, defined by some pair
\begin{displaymath}
\left( \, E \: \stackrel{\pi}{\longrightarrow} \: Y, \:
E \: \stackrel{f}{\longrightarrow} \: X \, \right)
\end{displaymath}
As described above, the projection $Y \times_{[X/\Gamma]} X \longrightarrow Y$
is isomorphic to the projection $\pi: E \rightarrow Y$.
The projection map $\pi$ is always open, for all principal $\Gamma$-bundles
$E$ over any $Y$, hence the canonical map $X \rightarrow [X/\Gamma]$ is open.

Similarly, if $\Gamma$ is discrete, then the projection map
$\pi: E \rightarrow Y$ is always a local homeomorphism,
hence if $\Gamma$ is discrete, the canonical map $X \rightarrow
[X/\Gamma]$ is a local homeomorphism.

\section{Sigma models on $[X/\Gamma]$ -- first pass}    \label{sigmamodel}

What would a sigma model on a quotient stack look like?
We shall argue that, after unraveling definitions, a string orbifold
is precisely a sigma model on a quotient stack.

How does this work?  A sigma model on $[X/\Gamma]$ is a weighted
sum over maps into $[X/\Gamma]$.  Now, we have just discussed
continuous maps $Y \rightarrow [X/\Gamma]$ -- these are the same as
pairs
\begin{displaymath}
\left( \, E \: \stackrel{\pi}{\longrightarrow} \: Y, \:
E \: \stackrel{f}{\longrightarrow} \: X \, \right)
\end{displaymath}
where $E$ is a principal $\Gamma$-bundle over $Y$, and $f$ is
$\Gamma$-equivariant.

\begin{figure}
\centerline{\psfig{file=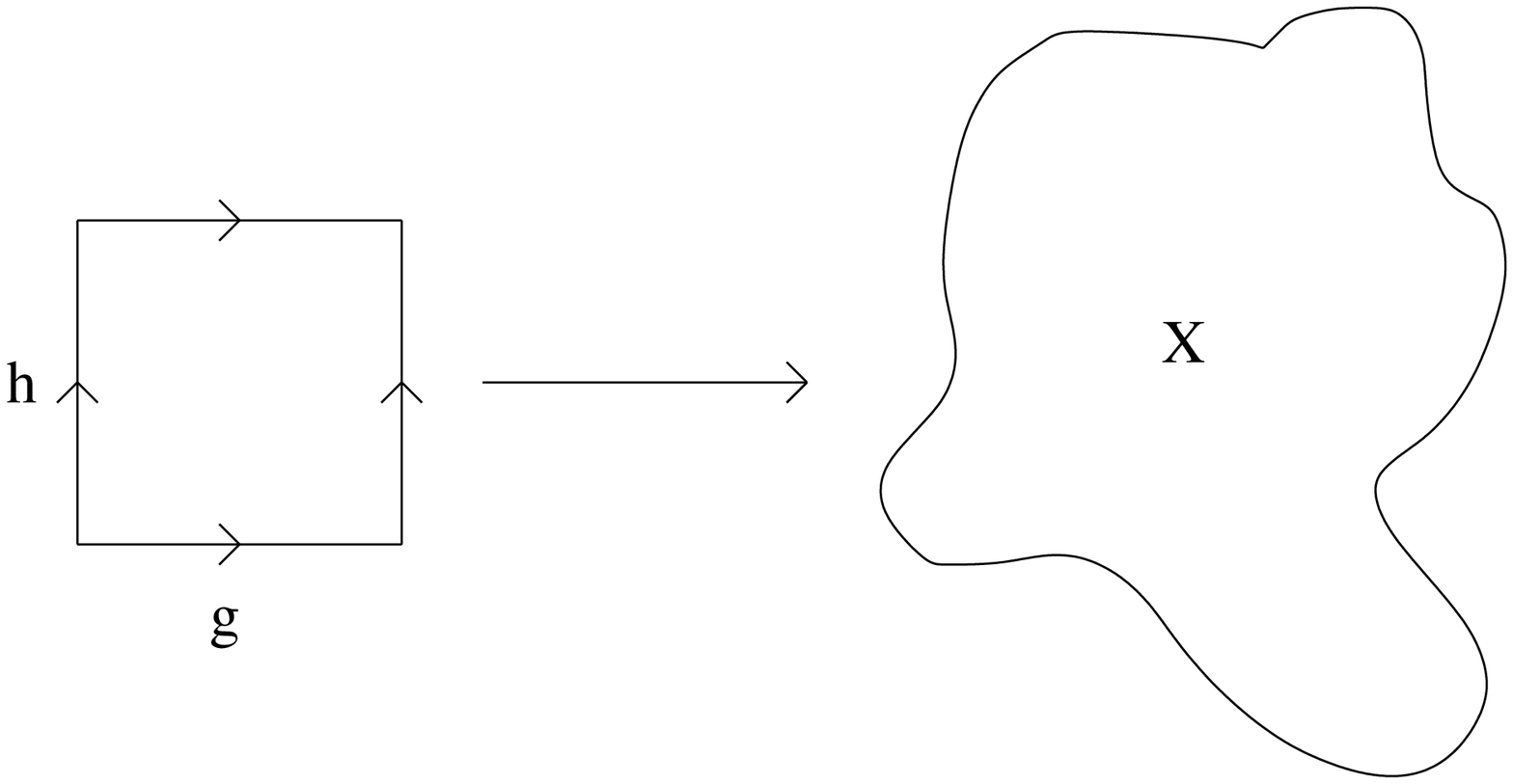,width=4in}}
\caption{\label{figsig2} Contribution to the $(g,h)$ twisted sector
of a string orbifold on $T^2$}
\end{figure}

Now, it is straightforward to check that this data is the same
thing as a map from a twisted sector into $X$, as illustrated
in figure~\ref{figsig2}.  Over any Riemann surface, restrict
to a maximally large contractible open subset (the interior of
the square in figure~\ref{figsig2}).  Over that open subset,
$E$ is trivializable, and so admits a section.  That section is precisely
a twisted sector -- the choice of boundary conditions dictates the extent
to which the section fails to extend globally, and also
dictates the equivalence class\footnote{
As a check, it is straightforward to compute that
the classification of twisted sectors agrees with the classification
of principal $\Gamma$-bundles on the worldsheet (given by
$H^1(\Sigma, C^{\infty}(\Gamma))$, for $\Sigma$ the worldsheet).
For example, for $\Sigma = T^2$, we can calculate
\begin{eqnarray*}
H^1\left( T^2, C^{\infty}(\Gamma) \right) & = & H^1\left( T^2, \Gamma \right)
\mbox{ for $\Gamma$ discrete} \\
 & = & \mbox{Hom}\left( H_1(T^2), \Gamma \right) \\
 & = & \mbox{Hom}\left( {\bf Z}^2, \Gamma \right) \\
 & = & \left\{ (g,h) \, | \, g,h \in \Gamma, gh = hg \right\}
\end{eqnarray*}
} of the bundle $E$.
Now, a map from that local section into $X$ is the same thing
as a $\Gamma$-equivariant map from the entire bundle $E$ into $X$.
(Although the map $E \rightarrow X$ may naively appear to contain
more information, $\Gamma$-equivariance dictates that the action on a choice
of a local section of the form above completely determines the map.)

%If we describe twisted sectors in terms of principal $\Gamma$-bundles
%on the worldsheet, then a map into $X$ defining a contribution to
%the sigma-model partition function is a $\Gamma$-equivariant
%map from the total space of the principal $\Gamma$-bundle into
%$X$. 

In other words, the maps one sums over in twisted sectors,
as illustrated in figure~\ref{figsig2}, are the same thing as pairs
\begin{displaymath}
\left( \, E \: \stackrel{\pi}{\longrightarrow} \: Y, \:
E \: \stackrel{f}{\longrightarrow} \: X \, \right)
\end{displaymath}
and hence the same thing as maps into $[X/\Gamma]$.

So far so good, but why cannot we understand these maps as maps
into the quotient space?  Given a pair
$(E \rightarrow Y, E \rightarrow X)$, we can certainly
define a map $Y \rightarrow X/\Gamma$, as we shall study in
more detail later.  However, the category of continuous
maps of the form summed over in twisted sectors (the
category of continuous maps into $[X/\Gamma]$) is only equivalent
to the category of continuous maps into the quotient space $X/\Gamma$
in the special case that $\Gamma$ acts freely.  If $\Gamma$ does not
act freely, then these categories are {\it not} equivalent
(essentially because there are more twisted-sector-type maps than corresponding
maps into $X/\Gamma$).  Since a topological space is completely
determined by the maps into it (see, for example,
appendix~\ref{mapsgivespace}), we see
a string orbifold can {\it not}
be understood as a sigma model on a quotient space $X/\Gamma$ in general.
We shall study this further in the next section.

So far, we have argued that after unraveling definitions, the maps
one sums over in a string orbifold are naturally understood as
maps into the quotient stack $[X/\Gamma]$, and in general not as
maps into the quotient space $X/\Gamma$. 
Put another way, twisted sectors are not naturally understood in
terms of quotient spaces, but are naturally understood in terms
of quotient stacks.  Put another way still, we have argued that, after
unraveling definitions, a string orbifold is a sum over maps into 
the quotient stack $[X/\Gamma]$, a `smoking gun' for an interpretation
as a sigma model.

We would like to emphasize that our analysis really does proceed
by unraveling definitions.  In particular, we are not making some
wild guess and trying to justify it by checking a handful of examples.

Now, in order to positively identify string orbifolds with
sigma models on quotient stacks, there is at least one more thing we need
to do.  Namely, we need to identify the action weighting the
contributions to the string orbifold path integral with the action
weighting the contributions to the sigma model partition function.
In section~\ref{redux} we shall describe a proposal for a classical
action for a sigma model on a stack, which generalizes both
ordinary sigma models and string orbifolds.
Of course, beyond classical actions one also needs to check whether
such a sigma model can be consistently quantized, which involves
thinking about global properties of the target; this we shall not do
in this paper.  We are really just trying to set up the basics of a program,
and such considerations are beyond what we hope to accomplish in this
particular paper.

%In fact, this is easy.  As a rule of thumb, any object xyz on 
%a quotient stack $[X/\Gamma]$ is the same thing as a $\Gamma$-invariant
%or -equivariant object xyz on the cover $X$.  (We shall
%examine this in more detail in section~\ref{equivlore}.)
%Sheaves, differential
%forms, {\it et cetera} on $[X/\Gamma]$ are the same as $\Gamma$-equivariant
%sheaves, $\Gamma$-invariant differential forms, {\it et cetera} on $X$.
%Thus, an action for a sigma model on $[X/\Gamma]$ would look like
%an action for a sigma model on $X$ involving $\Gamma$-invariant and
%-equivariant differential forms, metrics, spinors, and so forth.
%In other words, an action for a sigma model on $[X/\Gamma]$
%would look like a $\Gamma$-invariant action for a sigma model on $X$,
%which is what crops up in string orbifolds.
%We shall not speak further about sigma model actions.

%The reader might still be uncomfortable with the notion of
%compactifying a string theory on a stack -- after all, this is not
%something that has been widely considered in the past.
%However, other examples exist implicitly within the literature.
%For example, we showed in \cite{stacks} that the twisted ``bundle''
%on a D-brane worldvolume in the presence of a nontrivial B field
%could be naturally understood as a bundle on a stack (a stack naturally
%associated to the B field).  Thus, one way of thinking about D-branes
%is that they live on stacks, not spaces.  So, clearly, stacks have
%been present implicitly within the literature in the past.

Given that a string orbifold is a sigma model on a quotient stack
$[X/\Gamma]$, what can we now understand that was unclear before?
Many things!  For example, if $X$ is smooth, then $[X/\Gamma]$ is smooth,
regardless of whether $\Gamma$ acts freely; hence, a string orbifold
should always be well-behaved, regardless of whether the quotient space
$X/\Gamma$ is singular, and indeed this is a well-known property.
The story that the B field is somehow nonzero (in simple examples)
in string orbifolds can also be naturally understood.
Twist fields and orbifold Euler characteristics also have a natural
explanation.

In the rest of this paper, we shall describe why quotient stacks have
these nice properties.  We begin in the next section with an analysis
of how $[X/\Gamma]$ is related to $X/\Gamma$ in the special case that
$\Gamma$ acts freely.

\section{The case that $\Gamma$ acts freely}    \label{Gfree}

One can check that when $\Gamma$ acts freely, 
$[X/\Gamma]$ and $X/\Gamma$ are homeomorphic.
As a result of the computations below, one often thinks of the quotient stack
$[X/\Gamma]$ heuristically as roughly being the quotient space
$X/\Gamma$ with some sort of extra structure at the singularities.

What does it mean for $[X/\Gamma]$ to be homeomorphic to $X/\Gamma$?
Simply that there is an equivalence of categories, compatible
with the projection map, that respects coverings defining the
Grothendieck topology on $[X/\Gamma]$.  For the most part
we shall continue to ignore
questions regarding Grothendieck topologies on $[X/\Gamma]$, as
our manipulations really only involve the underlying groupoid structure.

We shall check this at three levels:  in terms of points,
in terms of our ansatz for open sets, and in terms of the continuous maps.
Of course, the only case that matters is the continuous maps case;
the other two cases are presented for pedagogical purposes.

\subsection{Points}

In this section we shall show that the functor
$p: [X/\Gamma]_{pt} \rightarrow (X/\Gamma)_{pt}$ defines an
equivalence of categories in the case that
$\Gamma$ acts freely on $X$.  To do this, we shall construct a functor
$q: (X/\Gamma)_{pt} \rightarrow [X/\Gamma]_{pt}$, and show that
when $\Gamma$ acts freely, then $p \circ q \cong \mbox{Id}$ and
$q \circ p \cong \mbox{Id}$.

Define a functor $q: (X/\Gamma)_{pt} \rightarrow [X/\Gamma]_{pt}$
as follows.
For any $y \in \mbox{Ob } (X/\Gamma)_{pt}$, let $y(x) \in
\pi_0^{-1}(x) \subset X$, and define a $\Gamma$-equivariant
map $f: \Gamma \rightarrow X$ by, $f(1) = y(x)$.
Then, define $q$ to map $y$ to $(f: \Gamma \rightarrow X)$.
(As the only morphisms in $(X/\Gamma)_{pt}$ are the identity
morphisms, defining the action of $q$ on them is trivial.)

It should be clear that $p \circ q = \mbox{Id}$.
In the case that $\Gamma$ acts freely on $X$, we can also
show that $q \circ p \cong \mbox{Id}$.  Define a natural
transformation $q \circ p \Rightarrow \mbox{Id}$ as follows.
For any object $(f: \Gamma \rightarrow X) \in \mbox{Ob } [X/\Gamma]_{pt}$,
let $\eta: \Gamma \rightarrow \Gamma$ be the unique $\Gamma$-equivariant
map making the diagram
\begin{displaymath}
\xymatrix{
\Gamma \ar[rr]^{\eta} \ar[dr]_{  (q \circ p)(f) } & & \Gamma
\ar[dl]^{ f } \\
& X &
}
\end{displaymath}
commute.  (Uniqueness follows from the fact that $\Gamma$ acts
freely on $X$.)  This is clearly a morphism in $[X/\Gamma]_{pt}$,
and it is also straightforward to check that $\eta$ is a natural
transformation.

Thus, there exists a functor $q: (X/\Gamma)_{pt} \rightarrow
[X/\Gamma]_{pt}$ such that $p \circ q = \mbox{Id}$ and $q \circ p
\cong \mbox{Id}$, so $p$ defines an equivalence of categories.

Next we shall give a more thorough (and more technical) arguments
that when $\Gamma$ acts freely, $[X/\Gamma] \cong X/\Gamma$.

\subsection{Continuous maps}

In this subsection we shall check that when $\Gamma$ acts freely,
$[X/\Gamma]_{map} \cong (X/\Gamma)_{map}$, i.e., $p$ is an equivalence
of categories.  In general terms, the idea is that if $g: Y \rightarrow
X/\Gamma$ is any continuous map and $\Gamma$ acts freely,
then we can construct a bundle over $Y$ as $g^*( X \rightarrow X/\Gamma)$,
and this construction provides an inverse to the projection
$[X/\Gamma]_{map} \rightarrow (X/\Gamma)_{map}$.

The remainder of this section is rather technical; the reader
may wish to skip it on a first reading.

To prove the claim, we need to find a functor $q: (X/\Gamma)_{map} \rightarrow
[X/\Gamma]_{map}$ such that $p \circ q \cong \mbox{Id}$ and
$q \circ p \cong \mbox{Id}$.

Define a functor $q: (X/\Gamma)_{map} \rightarrow [X/\Gamma]_{map}$
as follows:
\begin{enumerate}
\item Objects:  Let $( Y \stackrel{g}{\longrightarrow} X/\Gamma )$
be an object of $(X/\Gamma)_{map}$, i.e., let $g: Y \rightarrow X/\Gamma$
be a continuous map.

Define a principal $\Gamma$-bundle $E$ over $Y$ by,
$E = g^* ( X \rightarrow X/\Gamma )$.  (Since $\Gamma$ acts freely,
$X$ is itself a principal $\Gamma$-bundle over $X/\Gamma$.)

Since we defined $E$ in terms of a pullback by $g$,
there is a canonical bundle map $f: E \rightarrow X$
that makes the following diagram commute:
\begin{displaymath}
\xymatrix{
E \ar[r]^{f} \ar[d] & X \ar[d] \\
Y \ar[r]^{g} & X/\Gamma
}
\end{displaymath}

So, we can now consistently define
\begin{displaymath}
q: \: \left( \, Y \: \stackrel{g}{\longrightarrow} \: X/\Gamma \, \right)
\: \mapsto \: \left( \, E \: \longrightarrow \: Y, \:
E \: \stackrel{f}{\longrightarrow} \: X \, \right)
\end{displaymath}

\item Morphisms:  Let 
\begin{displaymath}
\rho: \: \left( \, Y_1 \: \stackrel{g_1}{\longrightarrow} \: X/\Gamma \, 
\right) \: \longrightarrow \: \left( \, Y_2 \:
\stackrel{g_2}{\longrightarrow} \: X/\Gamma \, \right)
\end{displaymath}
be a morphism in $(X/\Gamma)_{map}$, i.e., the diagram
\begin{displaymath}
\xymatrix{
Y_1 \ar[rr]^{\rho} \ar[dr]_{g_1} & & Y_2 \ar[dl]^{g_2} \\
& X/\Gamma &
}
\end{displaymath}
commutes.
Define $E_1 = g_1^* ( X \rightarrow X/\Gamma)$ and $E_2 = g_2^*( X
\rightarrow X/\Gamma)$.  Since $g_1 = g_2 \circ \rho$,
there is a canonical map $\lambda: E_1 ( = \rho^* E_2) \rightarrow E_2$
such that the diagram
\begin{displaymath}
\xymatrix{
E_1 \ar[r]^{\lambda} \ar[d] & E_2 \ar[d] \\
Y_1 \ar[r]^{\rho} & Y_2
}
\end{displaymath}
commutes.  Also, defining $f_i: E_i \rightarrow X$ to be the canonical
maps, we also have that
\begin{displaymath}
\xymatrix{
E_1 \ar[rr]^{\lambda} \ar[dr]_{f_1} & & E_2 \ar[dl]^{f_2} \\
& X &
}
\end{displaymath}
\end{enumerate}

With this definition, $q: (X/\Gamma)_{map} \rightarrow
[X/\Gamma]_{map}$ is a well-defined functor.

Finally, we must show that $p$ and $q$ are inverses.
It is immediately clear that $p \circ q = \mbox{Id}_{(X/\Gamma)_{map}}$.
Define a natural transformation $\psi: \mbox{Id}_{[X/\Gamma]_{map}}
\Rightarrow q \circ p$ as follows.

Let $\left( E \stackrel{\pi}{\longrightarrow} Y, \: 
E \stackrel{f}{\longrightarrow} X \right)$ be an object of
$[X/\Gamma]_{map}$, and let $\left( Y \stackrel{g}{\longrightarrow}
X/\Gamma \right)$ be its image under $p$.
Define $E' = g^*( X \rightarrow X/\Gamma)$, and $f': E' \rightarrow X$
the canonical map.

Define $\psi: E \rightarrow E'$ by,
\begin{displaymath}
\psi(e) \: = \: \left( \, \pi(e), \: f(e) \, \right)
\end{displaymath}
It is straightforward to check that this definition makes
$\psi: E \rightarrow E'$ a well-defined bundle isomorphism, such that
\begin{displaymath}
\xymatrix{
E \ar[rr]^{\psi} \ar[dr]_{f} & & E' \ar[dl]^{f'} \\
& X &
}
\end{displaymath}
commutes.  In other words, $\psi$ defines a morphism in 
$[X/\Gamma]_{map}$.  Furthermore, one can check that such $\psi$'s
define an invertible natural transformation 
$\mbox{Id}_{[X/\Gamma]_{map}}
\Rightarrow q \circ p$.

Thus, $p$ and $q$ are inverses, and so, for freely-acting $\Gamma$,
\begin{displaymath}
[X/\Gamma]_{map} \: \cong \: (X/\Gamma)_{map}
\end{displaymath}

In passing, when $\Gamma$ does not act freely, it is easy to check
that $[X/\Gamma]_{map}$ and $(X/\Gamma)_{map}$ are not equivalent
as categories, because $[X/\Gamma]_{map}$ has extra non-isomorphic
objects that all project down to the same object of $(X/\Gamma)_{map}$.  
For example, suppose $X$ has a point that is fixed under all
of $\Gamma$, for simplicity.  Then for any principal $\Gamma$-bundle
$E$ over a space $Y$, there is a $\Gamma$-equivariant map $f: E \rightarrow
X$ that sends all of $E$ to that fixed point.  Each of these objects
projects down to the same object of $(X/\Gamma)_{map}$, namely the
map that sends all of $Y$ to the image of this fixed point on 
$X/\Gamma$; however, for general $Y$, there can be objects
with non-isomorphic bundles $E$.  Thus, in this case,
$[X/\Gamma]_{map}$ contains non-isomorphic objects that project
down to the same object of $(X/\Gamma)_{map}$, and so these categories
cannot be the same.  Put another way, when $\Gamma$ does not act freely,
the quotient stack $[X/\Gamma]$ and the quotient space $X/\Gamma$
can not be homeomorphic.

\section{$X \rightarrow [X/\Gamma]$ is a principal $\Gamma$-bundle}
\label{Xprinbundle}

\subsection{Generalities}

Earlier in this paper we claimed that the natural map
$X \rightarrow [X/\Gamma]$ described $X$ as the total space of
a principal $\Gamma$-bundle over $[X/\Gamma]$, regardless of 
whether or not $\Gamma$ acted freely.  In this section we shall
explain this comment.

%First, heuristically:  in what sense does 
%$\pi: X \rightarrow [X/\Gamma]$ define a principal
%$\Gamma$-bundle?  To begin to gain some intution, note that
%for orbits on which $\Gamma$ acts freely,
%there are $|\Gamma|$ points in $X$ which $\pi$ maps into the
%isomorphism class.  For orbits on which $\Gamma$ does not act freely,
%there are fewer than $|\Gamma|$ points on $X$ which map to the
%isomorphism class, but that isomorphism class has nontrivial
%automorphisms.

In order to have a precise understanding of this claim,
we must explain how to make sense out of the notion of
bundle when the base space is a generalized space, and the 
projection map into the base does not carry open sets to
objects in the groupoid (i.e., open sets in the generalized space).

In particular, 
we need to find a way of defining bundles that involves no more
than fibered products.  Given such a definition for bundles
over ordinary spaces, we can then define bundles over
generalized spaces.  As a first step, note that 
if $\pi: E \rightarrow X$ is a 
fiber bundle over $X$, an ordinary topological space,
then for all open $U \subseteq X$,
$U \times_X E = \pi^{-1}(U)$, so we can say that $\pi: E \rightarrow X$
is a fiber bundle if, for all open $U \subseteq X$, 
$U \times_X E \rightarrow U$ is a fiber bundle over $U$.

It is straightforward to check that
this definition is equivalent to the following:
$\pi: E \rightarrow X$ is a fiber bundle if, for all spaces $Y$
and continuous maps $f: Y \rightarrow X$,
$Y \times_X E \rightarrow Y$ is a fiber bundle over $Y$,
with fixed ($Y$-independent) fiber.
This new definition has the useful property that we can make sense
out of it when the base space is a generalized space.

So, we say that $X \rightarrow [X/\Gamma]$ is a principal
$\Gamma$-bundle if, for all $Y \rightarrow [X/\Gamma]$,
$Y \times_{[X/\Gamma]} X \rightarrow Y$ is a principal
$\Gamma$-bundle, where the map $X \rightarrow [X/\Gamma]$ is the canonical
map discussed earlier.  Although this definition may look somewhat
unusual, it agrees with the usual notion of bundle in the special
case that the base space
is an ordinary space.  We shall first show how this works at the level
of points, where one can gain some insight with only a little work,
and then show how this is proven in general.

\subsection{Points}

Let $Y$ be any topological space, and $\psi: Y \rightarrow [X/\Gamma]$ a map.
It is straightforward to check that, at the level of points,
$Y \times_{[X/\Gamma]} X \rightarrow Y$ is a principal $\Gamma$-bundle.
After all, a point in $Y \times_{[X/\Gamma]} X$ is a triple 
$(y, x, \lambda)$, $y \in Y$, $x \in X$, and $\lambda: \psi(y) \rightarrow
h_X(x)$ is an isomorphism in the category of points of $[X/\Gamma]$
($h_X: X \rightarrow [X/\Gamma]$ is canonical).
If $\Gamma$ acts freely on the orbit of $x$, then
for each $y \in Y$, there are $|\Gamma|$ points in $X$ and for each,
a unique $\lambda$.  Thus, one has a principal 
$\Gamma$-bundle.  If $\Gamma$ does not act freely, then there
are fewer points $x \in X$, but their images in $[X/\Gamma]$ have
extra automorphisms, hence there are extra maps $\lambda$, so that
one still has a principal $\Gamma$-bundle.

\subsection{Continuous maps}

Now that we have seen how $Y \times_{[X/\Gamma]} X$ is a bundle
at the level of points, let us examine the matter rigorously
by considering the category of maps into $Y \times_{[X/\Gamma]} X$.

Let $Y$ be any topological space, and $Y \rightarrow [X/\Gamma]$
a continuous map.  As remarked earlier, specifying such a continuous
map is equivalent to specifying an object of $[X/\Gamma]_{map}$,
i.e., a pair
\begin{displaymath}
\left( \, E \: \stackrel{\pi}{\longrightarrow} \: Y, \:
E \: \stackrel{f}{\longrightarrow} \: X \, \right)
\end{displaymath}

A lengthy definition chase reveals that $Y \times_{[X/\Gamma]} X \cong E$,
and the projection maps $Y \times_{[X/\Gamma]} X \rightarrow Y, X$
are equivalent to the maps $E \rightarrow Y, X$.
Technically this is shown by proving that the categories
\begin{displaymath}
\left( Y \times_{[X/\Gamma]} X \right)_{map} \: \cong \: 
(E)_{map}
\end{displaymath}
are equivalent, and furthermore the functors
\begin{eqnarray*}
\left( \, Y \times_{[X/\Gamma]} X \: \stackrel{p_1}{\longrightarrow} \:
Y \, \right) & \cong & \left( \, E \: \stackrel{\pi}{\longrightarrow} \: Y 
\, \right) \\
\left( \, Y \times_{[X/\Gamma]} X \: \stackrel{p_2}{\longrightarrow} \:
X \, \right) & \cong & \left( \, E \, \stackrel{f}{\longrightarrow} \:
X \, \right)
\end{eqnarray*}
are also equivalent.  Because of their technical nature and length,
we have banished the formal proofs of these assertions to an
appendix.  However, it is now clear that, as claimed,
$Y \times_{[X/\Gamma]} X$ is indeed a principal $\Gamma$-bundle
over $Y$, and in fact is isomorphic to the bundle $E$
defining the map $Y \rightarrow [X/\Gamma]$.

Thus, $X \rightarrow [X/\Gamma]$ is a principal $\Gamma$-bundle.

\section{The lore on equivariance}   \label{equivlore}

In studying string orbifolds mathematically, one often
uses the notion of equivariance.  For example, one often
speaks of defining an orbifold group action on a bundle
by putting a $\Gamma$-equivariant structure on it.
Although this is a very natural thing to do, a very
heuristically reasonable thing to do, and there are no
other reasonable alternatives, one might still ask if there
is a 
deeper reason why this must work.  Put another way, why must
one use honest equivariant structures, and not, say,
projectivized\footnote{Projectivized equivariant structures appear
in \cite{doug1} in describing discrete torsion for D-branes;
however, it was argued in \cite{dt3} that this projectivized equivariant
structure on the D-brane bundle should really be understood
as an honest equivariant structure on a twisted ``bundle.''} 
equivariant structures?

Now that we have a deeper understanding of string orbifolds
(i.e., as sigma models on quotient stacks), we can now shed
some light on this matter.  In a nutshell, a $\Gamma$-equivariant
or -invariant object on $X$ is the same thing as a corresponding
object on the quotient stack $[X/\Gamma]$.  (This is one of
the famous properties of quotient stacks.)
For example, a sheaf on $[X/\Gamma]$ is equivalent to a 
$\Gamma$-equivariant sheaf on $X$.  So, by putting $\Gamma$-equivariant
structures on bundles, sheaves, {\it et cetera} on $X$, we are really specifying
bundles, sheaves, {\it et cetera} on $[X/\Gamma]$.  Put another way,
the fundamental thing to do is to construct a bundle, sheaf, whatever
on $[X/\Gamma]$, and then the fact that this corresponds to a bundle,
sheaf, whatever on $X$ with an honest equivariant structure
is dictated by the geometry of $[X/\Gamma]$.

In this section, we will show specifically that a sheaf on
the quotient stack $[X/\Gamma]$ is the same thing, after unraveling
definitions, as a $\Gamma$-equivariant sheaf on $X$,
another standard fact about quotient stacks (see for
example \cite{gomez,vistoli}).  From this fact, one can quickly
deduce that many other objects on $[X/\Gamma]$ are $\Gamma$-equivariant
or -invariant objects on $X$ -- for example, bundles are just special
kinds of sheaves, spinors and differential forms can be understood
as sections of sheaves, and so forth.  However, technically we shall
only explicitly consider the general construction for sheaves. 

Since understanding how to define a sheaf on a stack is a rather
technical matter, we begin our discussion with a much simpler
prototype:  functions on stacks.  Just as a sheaf on the
quotient stack $[X/\Gamma]$ is a $\Gamma$-equivariant sheaf
on $X$, it is also true that a function on the
quotient stack $[X/\Gamma]$ is the same as a $\Gamma$-invariant
function on $X$.  Also, the analysis for functions is not only
a prototype of the analysis for sheaves, but it is technically
much simpler, so for pedagogical purposes, we begin with functions.
After having described functions on stacks, 
we shall motivate the relevant definition of a sheaf
on a stack, then we shall show the equivalence between
sheaves on $[X/\Gamma]$ and $\Gamma$-equivariant sheaves on 
$X$.

As this section is rather technical, the reader may wish
to skip this section on a first reading.

\subsection{Functions on stacks}   \label{funcstack}

The ordinary definition of a function on a space does not
quite make sense for generalized spaces, a problem similar
to that we encountered when trying to understand in what
sense $X \rightarrow [X/\Gamma]$ could be a principal
$\Gamma$-bundle.  In order to make sense of functions in this
context, we need a somewhat different definition, that makes
sense for stacks.  The answer is to define functions in
terms of their pullbacks -- instead of trying to define
a function $f$ directly on the stack, we instead speak
of functions ``$\phi^* f$'' for each map $\phi: Y \rightarrow
{\cal F}$ into a stack ${\cal F}$.

Technically we define functions on stacks as follows.
For any stack ${\cal F}$,
a function $f: {\cal F} \rightarrow {\bf R}$ is defined by
associating, to each map $Y \rightarrow {\cal F}$,
a function $f_Y: Y \rightarrow {\bf R}$, such that for any
commuting diagram
\begin{equation}    \label{func1}
\xymatrix{
Y_1 \ar[rr]^{ g } \ar[dr] & & Y_2 \ar[dl] \\
 & {\cal F} & 
}
\end{equation}
one has $f_2 \circ g = f_1$.

Note that if ${\cal F}$ is itself an honest space,
then this definition uniquely specifies a function on ${\cal F}$
in the usual sense.

In the special case that ${\cal F} = [X/\Gamma]$,
it is straightforward to check that a function on $[X/\Gamma]$
is the same thing as a $\Gamma$-invariant function on $X$.

First, to show that a function on $[X/\Gamma]$ defines a 
$\Gamma$-invariant function on $X$, note that by virtue of the 
definition, one immediately has a function $f_X$ on $X$.
Moreover, because the diagram
\begin{displaymath}
\xymatrix{
X \ar[rr]^{ g } \ar[dr] & & X \ar[dl] \\
& [X/\Gamma] & 
}
\end{displaymath}
commutes for any $g \in \Gamma$, we see that
$f_X = f_X \circ g$, hence $f_X$ is $\Gamma$-invariant.

Next, let $f_X$ be a $\Gamma$-invariant function on $X$,
and let $Y \rightarrow [X/\Gamma]$ be any map.
We have the commuting diagram
\begin{displaymath}
\xymatrix{
Y \times_{ [X/\Gamma] } X \ar[r]^(0.7){p_2} \ar[d]_{p_1} & X \ar[d] \\
Y \ar[r] & [X/G]
}
\end{displaymath}
where $p_2$ is $\Gamma$-equivariant.
Now, $f_X \circ p_2$ is a function on ${\bf R}$,
and because $p_2$ is $\Gamma$-equivariant, we see that
$( f_X \circ p_2 )( g \cdot e ) = (f_X \circ p_2)(e)$
for all $g \in \Gamma$, $e \in Y \times_{ [X/\Gamma] } X$.
Hence, $f_X \circ p_2$ descends to a function on $Y$,
and it is also straightforward to check that 
diagram~(\ref{func1}) is satisfied.

\subsection{Sheaves on stacks}

The ordinary definition of a sheaf on a space does not
quite make sense for generalized spaces.  We had a similar problem
in understanding in what sense $X \rightarrow [X/\Gamma]$ could
be a principal $\Gamma$-bundle.  We shall proceed here in the same
fashion that we did there -- specifically, we shall find
a way to define sheaves on ordinary spaces that also makes sense
for generalized spaces.

We shall begin with the following observation
\cite[cor. I-14]{eisenbudharris}.
Let $\{ U_{\alpha} \}$ be an open cover of a space $X$.
If ${\cal S}_{\alpha}$ is a sheaf on $U_{\alpha}$ for each $\alpha$,
and if 
\begin{displaymath}
\varphi_{\alpha \beta}: {\cal S}_{\alpha}|_{U_{\alpha}
\cap U_{\beta}} \stackrel{\sim}{\longrightarrow} {\cal S}_{\beta}
|_{U_{\alpha} \cap U_{\beta}}
\end{displaymath}
are isomorphisms such that
\begin{displaymath}
\varphi_{\beta \gamma} \circ \varphi_{\alpha \beta} \: = \:
\varphi_{\alpha \gamma}
\end{displaymath}
on $U_{\alpha} \cap U_{\beta} \cap U_{\gamma}$, then 
\cite[cor. I-14]{eisenbudharris} there exists a unique
sheaf ${\cal S}$ on $X$ such that ${\cal S}|_{U_{\alpha}} = {\cal S}_{\alpha}$.

That definition is not quite useful for generalized spaces,
but we can step towards a more useful definition as follows.
It is straightforward to check that the definition above
is equivalent to the following data:
\begin{enumerate}
\item For all open $U \subseteq X$, and all maps $f_U: U \rightarrow X$,
a sheaf ${\cal S}_{U}$ on $U$.  
\item For all inclusion maps $\rho:
U \hookrightarrow V$ such that the following diagram commutes
\begin{displaymath}
\xymatrix{
U \ar@{^{(}->}[rr]^{\rho} \ar[dr]_{f_U} & & V \ar[dl]^{f_V} \\
& X &
}
\end{displaymath}
an isomorphism $\varphi_{\rho}: {\cal S}_U \stackrel{\sim}{\longrightarrow}
\rho^* {\cal S}_V$, such that for any triple
\begin{displaymath}
\xymatrix{
U \ar@{^{(}->}[r]^{\rho_1} \ar[dr]_{f_U} &
V \ar@{^{(}->}[r]^{\rho_2} \ar[d]^{f_V} &
W \ar[dl]^{f_W} \\
 & X &
}
\end{displaymath}
\end{enumerate}

In particular, to show that the first definition implies the second,
work with a cover containing $U$ and $V$ as elements, then the isomorphism
$\varphi_{\alpha \beta}: {\cal S}_{\alpha} |_{U_{\alpha} \cap U_{\beta}}
\rightarrow {\cal S}_{\beta} |_{U_{\alpha} \cap U_{\beta}}$,
where $U_{\alpha} = U$ and $U_{\beta} = V$, is the same as an isomorphism
$\varphi_{\rho}: {\cal S}_U \rightarrow \rho^* {\cal S}_V$,
since $U \cap V = U$, and the triple compatibility condition follows similarly.

To show that the second implies the first, consider sheaves
${\cal S}_{\alpha}$, ${\cal S}_{\alpha \beta}$, and
${\cal S}_{\alpha \beta \gamma}$ defined on elements of a cover
$\{ U_{\alpha} \}$ and its intersections.  Given an isomorphism
\begin{displaymath}
\alpha^{\alpha}_{\alpha \beta}: \: {\cal S}_{\alpha \beta}
\: \stackrel{\sim}{\longrightarrow} \: {\cal S}_{\alpha} |_{
U_{\alpha} \cap U_{\beta}}
\end{displaymath}
one can define an isomorphism $\varphi_{\alpha \beta}: {\cal S}_{\alpha}
|_{U_{\alpha} \cap U_{\beta}} \stackrel{\sim}{\longrightarrow}
{\cal S}_{\beta} |_{U_{\alpha \beta}}$ as the composition
\begin{displaymath}
\varphi_{\alpha \beta} \: \equiv \: \alpha^{\beta}_{\alpha \beta}
\circ \left( \alpha^{\alpha}_{\alpha \beta} \right)^{-1}:
\: {\cal S}_{\alpha}|_{U_{\alpha} \cap U_{\beta}}
\: \stackrel{\sim}{\longrightarrow} \: {\cal S}_{\beta} |_{U_{\alpha}
\cap U_{\beta}}
\end{displaymath}
and it is straightforward to check that these $\varphi_{\alpha \beta}$
satisfy the requisite condition on triple intersections.

From the second definition, we have a third equivalent definition,
which finally does make sense for generalized spaces.
A sheaf ${\cal S}$ on $X$ is defined by the following data:
\begin{enumerate}
\item For all spaces $Y$ and continuous maps $f: Y \rightarrow X$,
a sheaf ${\cal S}_f$ on $Y$.
\item For all commuting diagrams
\begin{displaymath}
\xymatrix{
Y_1 \ar[rr]^{\rho} \ar[dr]_{f_1} & & Y_2 \ar[dl]^{f_2} \\
& X &
}
\end{displaymath}
an isomorphism $\varphi_{\rho}: {\cal S}_{f_1} \stackrel{\sim}{\longrightarrow}
\rho^* {\cal S}_{f_2}$
which obeys the consistency condition that for all commuting triples
\begin{displaymath}
\xymatrix{
Y_1 \ar[r]^{\rho_1} \ar[dr]_{f_1} &
Y_2 \ar[r]^{\rho_2} \ar[d]^{f_2} &
Y_3 \ar[dl]^{f_3} \\
& X &
}
\end{displaymath}
the isomorphisms obey
\begin{displaymath}
\varphi_{\rho_2 \circ \rho_1} \: = \:
\rho_1^* \varphi_{\rho_2} \circ \varphi_{\rho_1}: \:
{\cal S}_{f_1} \: \stackrel{\sim}{\longrightarrow} \:
(\rho_2 \circ \rho_1)^* {\cal S}_{f_3}
\end{displaymath}
\end{enumerate}

To see that the definition above implies the second definition is clear;
just restrict the possible $Y$'s to open subsets of $X$.
Conversely, since the second definition gave rise to an ordinary
sheaf ${\cal S}$ on $X$, we can define ${\cal S}_{f} \equiv
f^* {\cal S}$, and it is straightforward to check that with this
definition, $\varphi_{\rho} = \mbox{Id}$ for all $\rho$, which automatically
satisfies the condition on triples.

So far all we have done is wander through a succession of equivalent
definitions of a sheaf on an ordinary space.  The reader may well wonder
why we bothered.  The answer is that we have, finally, generated a definition
of sheaf which makes sense on generalized spaces as well as ordinary spaces.

As the reader may have guessed,
we  
define a sheaf of sets on a groupoid ${\cal F}$ by the
following data:
\begin{enumerate}
\item For all spaces $Y$ and maps $f: Y \rightarrow {\cal F}$, 
a sheaf ${\cal S}_f$ on $Y$.
\item For each commutative diagram
\begin{displaymath}
\xymatrix{
Y_1 \ar[rr]^{\rho} \ar[dr]_{f_1} & & Y_2 \ar[dl]^{f_2} \\
& {\cal F} &
}
\end{displaymath}
an isomorphism $\varphi_{\rho}: {\cal S}_{f_1} \stackrel{\sim}{
\longrightarrow} \rho^* S_{f_2}$.
\end{enumerate}
The isomorphisms $\varphi$ are required to obey a consistency condition,
namely for each commutative diagram
\begin{displaymath}
\xymatrix{
Y_1 \ar[r]^{\rho_1} \ar[dr]_{f_1} & Y_2 \ar[r]^{\rho_2} \ar[d]^{f_2}
& Y_3 \ar[dl]^{f_3} \\
& {\cal F} &
}
\end{displaymath}
the associated isomorphisms must obey
\begin{displaymath}
\varphi_{\rho_2 \circ \rho_1} \: = \: 
\rho_1^* \varphi_{\rho_2} \circ  \varphi_{\rho_1}: \:
{\cal S}_{f_1} \: \stackrel{\sim}{\longrightarrow} \:
(\rho_2 \circ \rho_1)^* {\cal S}_{f_3}
\: = \: \rho_1^* \left( \rho_2^* {\cal S}_{f_3} \right)
\end{displaymath}
As should be clear from the preceding analysis, this definition
agrees with the notion of ``sheaf'' for ordinary topological spaces.
There is a technical point we should mention -- the notion
``$\varphi_{\rho}$'' omits some information.  For a diagram of functors
to commute means merely that the two compositions must agree up
to an invertible natural transformation, and in fact the
isomorphisms $\varphi_{\rho}$ depend upon that natural transformation
in addition to $\rho$.  However, it seems customary to omit mention
of this dependence from the notation.

Now that we have defined sheaves on stacks ${\cal F}$, we
shall show that a sheaf on the quotient stack $[X/\Gamma]$
is the same thing as a $\Gamma$-equivariant sheaf on $X$.

\subsection{Sheaves on $[X/\Gamma]$ define $\Gamma$-equivariant sheaves on $X$}

First, we shall show that a sheaf on $[X/\Gamma]$
defines a $\Gamma$-equivariant sheaf on $X$.

Earlier we described a natural map $X \rightarrow [X/\Gamma]$,
that also describes $X$ as a principal $\Gamma$-bundle over
$[X/\Gamma]$.  By composing that map with actions of $g \in \Gamma$,
which map $X \stackrel{g}{\longrightarrow} X$, we can form
several maps $X \rightarrow [X/\Gamma]$, which we shall denote by
$\pi_g$.

Now, from the definition of sheaf on a stack given in the previous
subsection, for each map $\pi_g: X \rightarrow [X/\Gamma]$,
one associates a sheaf ${\cal S}_g$ on $X$.  Also, since
$\pi_g = \pi_1 \circ g$, there are isomorphisms $\varphi_g: {\cal S}_g
\stackrel{\sim}{\longrightarrow} g^* {\cal S}_1$.  
%Finally, since
%$gh = g \circ h$, there are isomorphisms
%$\varphi_h: {\cal S}_{gh} \stackrel{\sim}{\longrightarrow} h^* {\cal S}_g$.

Now, a $\Gamma$-equivariant structure on ${\cal S}_1$ consists of
isomorphisms ${\cal S}_1 \stackrel{\sim}{\longrightarrow} g^* {\cal S}_1$, 
respecting
the group law.  Although we have some isomorphisms, they are between
{\it a priori} distinct sheaves, and are not sufficient.

The remaining isomorphisms emerge from the fact that $\pi_1 = \pi_g$,
in the sense that they define isomorphic functors, and so the diagram
\begin{displaymath}
\xymatrix{
X \ar@/_1pc/[d]_{\pi_1}  \ar@/^1pc/[d]^{\pi_g} \\
 [X/\Gamma] 
}
\end{displaymath}
commutes.  (This was discussed earlier when the canonical map
$X \rightarrow [X/\Gamma]$ was originally introduced.)
As a result, we have additional isomorphisms $\phi_g: {\cal S}_1
\rightarrow {\cal S}_g$.

Given these additional isomorphisms, we can now define an isomorphism
$\alpha_g: {\cal S}_1 \stackrel{\sim}{\longrightarrow} g^* {\cal S}_1$, for each $g \in \Gamma$.
Specifically, define $\alpha_g$ to be the composition
\begin{displaymath}
\alpha_g: \:
{\cal S}_1 \: \stackrel{\phi_g}{\longrightarrow} \: {\cal S}_g \:
\stackrel{ \varphi_g }{\longrightarrow} \: g^* {\cal S}_1
\end{displaymath}

From the consistency condition for isomorphisms, it is
straightforward to check that
$\alpha_{gh} = h^*\alpha_g \circ \alpha_h$ for all $g, h \in \Gamma$,
and so our isomorphisms obey the group law.

Thus, given a sheaf on $[X/\Gamma]$, we can construct a
$\Gamma$-equivariant sheaf on $X$.

For an example, consider the principal $\Gamma$-bundle
$X \rightarrow [X/\Gamma]$ described earlier.  For any map
$Y \rightarrow [X/\Gamma]$, there is a principal $\Gamma$-bundle
over $Y$, namely $Y \times_{[X/\Gamma]} X$, so we can explicitly
recover a description of the bundle $X \rightarrow [X/\Gamma]$ in
terms of data over spaces $Y$.  The corresponding $\Gamma$-equivariant
sheaf on $X$ itself is then $X \times_{[X/\Gamma]} X =
X \times \Gamma$.  So, the $\Gamma$-equivariant sheaf on $X$
corresponding to the principal $\Gamma$-bundle $X \rightarrow
[X/\Gamma]$ is the trivial bundle $X \times \Gamma$ on $X$.

Next, we shall show that a $\Gamma$-equivariant sheaf
on $X$ defines a sheaf on $[X/\Gamma]$.

\subsection{$\Gamma$-equivariant sheaves on $X$ define sheaves on $[X/\Gamma]$}

Let ${\cal S}$ be a $\Gamma$-equivariant sheaf on $X$,
and let $\alpha_g: {\cal S} \stackrel{\sim}{\longrightarrow} g^* {\cal S}$
be the isomorphisms defining a $\Gamma$-equivariant structure on ${\cal S}$.
We shall outline how to construct a sheaf on $[X/\Gamma]$.
(The details are quite technical, so we have opted to only outline
how this is done.)
This means, for each map $\lambda: Y \rightarrow [X/\Gamma]$, we shall
outline how to
construct a sheaf ${\cal S}_{\lambda}$ on $Y$, as well as a set of
consistent isomorphisms.

Let $\lambda: Y \rightarrow [X/\Gamma]$ be a continuous map, defined by some
pair
\begin{displaymath}
\left( \, E \: \stackrel{\pi}{\longrightarrow} \: Y, \:
E \: \stackrel{f}{\longrightarrow} \: X \, \right)
\end{displaymath}
Since $f$ and ${\cal S}$ is $\Gamma$-equivariant, $f^* {\cal S}$ is
a $\Gamma$-equivariant sheaf on $E$, and so there is a sheaf
${\cal S}_{\lambda}$ such that $\pi^* {\cal S}_{\lambda} = f^* {\cal S}$.
Take the sheaf on $Y$ associated to $\lambda$ to be ${\cal S}_{\lambda}$.

Now, for every commutative diagram
\begin{equation}    \label{Gshftmp1}
\xymatrix{
Y_1 \ar[rr]^{\rho} \ar[dr]_{\lambda_1} & & Y_2 \ar[dl]^{\lambda_2} \\
& [X/\Gamma] &
}
\end{equation}
we need an isomorphism $\varphi_{\rho}: {\cal S}_{\lambda_1} 
\stackrel{\sim}{\longrightarrow} \rho^* {\cal S}_{\lambda_2}$.
Let $\lambda_i$ be defined by a pair
\begin{displaymath}
\left( \, E_i \: \stackrel{\pi_i}{\longrightarrow} \: Y_i, \:
E_i \: \stackrel{f_i}{\longrightarrow} \: X \, \right)
\end{displaymath}
For diagram~(\ref{Gshftmp1}) to commute means the maps
$\lambda_1$, $\lambda_2$ are isomorphic, i.e., there
exists a morphism 
\begin{displaymath}
(\rho, \gamma): \:
\left( \, E_1 \: \longrightarrow \: Y_1, \:
E_1 \: \stackrel{f_1}{\longrightarrow} \: X \, \right) \:
\longrightarrow \:
\left( \, E_2 \: \longrightarrow \: Y_2, \:
E_2 \: \stackrel{f_2}{\longrightarrow} \: X \, \right)
\end{displaymath}
which is to say, $\gamma: E_1 \rightarrow E_2$ is a map of principal
$\Gamma$-bundles making the diagrams
\begin{displaymath}
%\begin{array}{ccccccc}
%E_1 & \stackrel{\gamma}{\longrightarrow} & E_2 & &
%E_1 & \stackrel{\gamma}{\longrightarrow} & E_2 \\
%\downarrow & & \downarrow &
%\: \mbox{   ,   } \: &
%\makebox[0pt][r]{ $\scriptstyle{ f_1 }$ } \downarrow & &
%\downarrow \makebox[0pt][l]{ $\scriptstyle{ f_2 }$ } \\
%Y_1 & \stackrel{\rho}{\longrightarrow} & Y_2 & &
%X & = & X
%\end{array}
\xymatrix{
E_1 \ar[r]^{\gamma} \ar[d] & E_2 \ar[d] & \mbox{  } & 
E_1 \ar[rr]^{\gamma} \ar[dr]_{f_1}& & 
E_2 \ar[dl]^{f_2} \\
Y_1 \ar[r]^{\rho} & Y_2 & \mbox{  } &
& X &
}
\end{displaymath}
commute.  
Recall that $\pi_1^* {\cal S}_{\lambda_1} = f_1^*
{\cal S}$ and $\pi_2^* {\cal S}_{\lambda_2} = f_2^* {\cal S}$.
Now, since $f_1 = f_2 \circ \gamma$, $f_1^* {\cal S} \cong \gamma^* f_2^* 
{\cal S}$.  It can be shown that this isomorphism
descends to an isomorphism $\varphi_{\rho}: {\cal S}_{\lambda_1}
\stackrel{\sim}{\longrightarrow} \rho^* {\cal S}_{\lambda_2}$,
and moreover these isomorphisms satisfy the requisite properties.

A simple example may help make this a bit more clear.
Consider the principal $\Gamma$-bundle $X \rightarrow [X/\Gamma]$.
We pointed out in the last section that this could be understood
as the trivial $\Gamma$-equivariant bundle $X \times \Gamma$ on 
$X$.  Now, let us recover bundles over $Y$, for each map
$Y \rightarrow [X/\Gamma]$.  Let such a map be defined by the object
\begin{displaymath}
\left( \, E \: \stackrel{\pi}{\longrightarrow} \: Y, \:
E \: \stackrel{f}{\longrightarrow} \: X \, \right)
\end{displaymath}
Use the fact that
\begin{eqnarray*}
f^* ( X \times \Gamma) & = & E \times \Gamma \\
 & = & \pi^* E
\end{eqnarray*}
and so we take the associated bundle on $Y$ to be $E$.
Indeed, earlier we began by describing $X \rightarrow [X/\Gamma]$
in terms of bundles $Y \times_{[X/\Gamma]} X \rightarrow Y$,
and recall $Y \times_{[X/\Gamma]} X \cong E$.
So, in this example, we have precisely inverted our previous
construction, as one would have hoped.

\section{$[X/\Gamma]$ is smooth}    \label{smooth}

One of the most famous properties of quotient stacks is that,
so long as $X$ and $\Gamma$ are smooth, and $\Gamma$ acts by
diffeomorphisms, the quotient stack
$[X/\Gamma]$ is smooth, regardless of whether or not the
quotient space $X/\Gamma$ is smooth.

In this section, we shall review how this works.
We shall show that, in the case
$X$ is a smooth manifold, $\Gamma$ is discrete\footnote{The
smoothness result commonly quoted does not require $\Gamma$
to be discrete.  However, this is all we require for physics
purposes, and assuming that $\Gamma$ is discrete simplifies certain
technical aspects of the discussion.}, 
and
$\Gamma$ acts on $X$ by diffeomorphisms, 
the quotient stack
$[X/\Gamma]$ naturally admits a smooth structure.
We first show that $[X/\Gamma]$ is a topological manifold
(that is, a space which admits coordinate charts, but those charts
are not required to differ by diffeomorphisms on overlaps),
and then show how it is furthermore a smooth manifold, not just
a topological manifold.

Perhaps the fastest way to do this is to simply note that
if we change the previous discussion in this paper by replacing
all occurrences of ``topological space'' with ``manifold''
and, ``continuous map'' with ``smooth map,'' then most of the
discussion is unchanged.  If $\Gamma$ does not act freely,
then of course the quotient space no longer makes sense in this
context, but we could still define everything as groupoids over the category
of manifolds rather than ${\it Top}$ or $X/\Gamma$.
However, the reader may find such an approach unsatisfactory, as 
it  
would not directly explain how to put a smooth structure on
a topological quotient stack.  Therefore, we shall work through
direct constructions of smooth structures on topological
quotient stacks in detail.

\subsection{$[X/\Gamma]$ is a topological manifold}

To begin, we shall first show that $[X/\Gamma]$ is
a topological manifold.  Specifically, this is a topological space that is
locally homeomorphic to ${\bf R}^n$, though the coordinate charts are
not required to be related by diffeomorphism on overlaps
(though they are automatically homeomorphisms).

How should one define a notion of ``topological manifold'' for
a stack?  As usual, the correct procedure is to rewrite the definition
for ordinary spaces in a manner that makes sense for stacks.

Here is one way to describe a topological manifold.
Let $N$ be a topological space.  Then $N$ can be given the
structure of a topological manifold
if and only if there exists another topological manifold $M$
and a surjective local homeomorphism $f: M \rightarrow N$.

To check this statement, first note that
given such a pair $(M, f)$, it is straightforward to show that $N$
is locally homeomorphic to ${\bf R}^n$.  Let $\{ V_{\alpha} \}$
be an open cover of $M$, such that for each $\alpha$, 
the restriction of $f$ to $V_{\alpha}$ is a homeomorphism,
and without loss of generality assume furthermore that each
$V_{\alpha}$ is contained within some coordinate neighborhood.
Let $\psi_{\alpha}: V_{\alpha} \longrightarrow \psi_{\alpha}(
V_{\alpha} ) \subseteq {\bf R}^n$ be homeomorphisms onto open subsets
of ${\bf R}^n$, defining $M$ as a topological manifold.
Then $\{ f(V_{\alpha}) \}$ is an open cover of $N$, and
\begin{displaymath}
\psi_{\alpha} \circ \left( f|_{V_{\alpha}} \right)^{-1}: \:
f(V_{\alpha}) \: \longrightarrow \: {\bf R}^n
\end{displaymath}
is a homeomorphism, defining $N$ as a topological manifold.

Conversely, if $N$ is a topological manifold, let $\{ U_{\alpha} \}$
be an open cover such that each $U_{\alpha}$ is contained within a
coordinate patch.  Then, define $M$ to be the disjoint union of the
$U_{\alpha}$, and $f$ to be the amalgamated inclusions.

So, if we want to describe a topological space $N$ as a topological
manifold, one way to proceed is to use a topological manifold $M$
together with a surjective local homeomorphism $f: M \rightarrow N$.

Now, as the reader may have guessed, this particular definition
generalizes immediately to stacks.
We say a stack ${\cal F}$ is a topological (generalized) manifold
provided that there exists a topological manifold $M$ and
a surjective local homeomorphism $f: M \rightarrow {\cal F}$ (assumed
to be representable, of course).  Such a pair $(M, f)$ is known as
a ``topological atlas'' for ${\cal F}$.
(Recall from section~\ref{mapprops}
that we can make sense out of notions such as ``surjective''
and ``local homeomorphism'' for representable maps into stacks by saying
that such a map is surjective or a local homeomorphism if and only if
for all continuous maps $g: Y \rightarrow {\cal F}$, the projection
map $Y \times_{{\cal F}} M \stackrel{p_1}{\longrightarrow} Y$ is
surjective or a local homeomorphism.)

Note that if ${\cal F}$ is a stack with topological atlas $(M, f)$,
then for any topological manifold $Y$ and continuous map 
$Y \rightarrow {\cal F}$, the product $Y \times_{ {\cal F} } M$
is a topological manifold, since $p_1: Y \times_{ {\cal F} } M
\rightarrow Y$ is a local homeomorphism.

Now that we have defined what it means for a stack to be a topological
manifold, and shown that this definition makes sense for ordinary
spaces, let us check whether the quotient stack $[X/\Gamma]$ is a
topological manifold.

For $[X/\Gamma]$ to be a topological manifold, we must give
a topological manifold $M$ together with a surjective local homeomorphism.
For simplicity, assume that $\Gamma$ is discrete.
In that case, take $M = X$, which we assume is a topological manifold,
and take $f$ to be the canonical map $X \rightarrow [X/\Gamma]$.
As observed earlier, the canonical map is always surjective,
and in the special case that $\Gamma$ is discrete, it is also
a local homeomorphism.  Thus, we have given a topological manifold
$M$ together with a surjective local homeomorphism $f: M \rightarrow
[X/\Gamma]$, and so, so long as $X$ is a topological manifold,
$[X/\Gamma]$ is also a topological manifold.
In other words, $X$ together with the canonical map $X \rightarrow [X/\Gamma]$
form a topological atlas for $[X/\Gamma]$.

To check that this is not vacuous, let us try to do the same thing
for the quotient space $X/\Gamma$, in the case that $\Gamma$ is not
freely acting.  In this case, the canonical projection map
$X \rightarrow X/\Gamma$ fails to be a local homeomorphism over 
the fixed points of $\Gamma$.  In a nutshell, the difficulty
is that ${\bf R}^n/\Gamma \not\sim {\bf R}^n$.  (For example,
consider the special case that the only fixed point of $\Gamma$
is at the origin of ${\bf R}^n$.  Consider the space obtained by
omitting that point, so $\Gamma$ acts freely on what remains.
Now, ${\bf R}^n - \{ 0 \}$ is homotopic to $S^{n_1}$, and
(for $n > 2$) is simply-connected.  However, $({\bf R}^n - \{ 0 \})/\Gamma$
is homotopic to $S^{n-1}/\Gamma$, and is not simply-connected,
but has $\pi_1 = \Gamma$.) 
Thus, if $\Gamma$ does not act freely, then $X/\Gamma$ is not a topological
manifold.

\subsection{$[X/\Gamma]$ is a smooth manifold}

The difference between a smooth manifold and a topological
manifold is that in a smooth manifold, the coordinate charts
on overlapping open sets must be related by diffeomorphism.
How can we set up that condition in the language above?
As before, we shall first find a definition of smooth manifold
that can be directly applied to stacks, and then check that the
quotient stack $[X/\Gamma]$ can be given the relevant structure.

\subsubsection{Rephrase smoothness for spaces}

Let us first rephrase the usual definition of smooth manifold.
Before we can do this, we need a short lemma.

The lemma is as follows: let $M$, $N$ be topological spaces, 
$f: M \rightarrow N$ a surjective local homeomorphism, 
$Y$ a smooth manifold, and $g: Y \rightarrow N$ any continuous map.
Then $Y \times_N M$ is naturally a smooth manifold, with smooth structure
induced by $Y$.  To prove this, let $\{ V_{\alpha} \}$ be a cover of $M$
such that $f$ restricted to any $V_{\alpha}$ is a homeomorphism.
Then, $Y \times_N M$ has an open cover of the form
\begin{displaymath}
\left\{ \, g^{-1}( f(V_{\alpha}) ) \times_N V_{\alpha} \:
\cong \: g^{-1}( f(V_{\alpha})) \, \right\}
\end{displaymath}
Since $g$ is continuous, and an open subset of a smooth manifold is itself
a smooth manifold, we have an obvious smooth structure.
(More concretely, if $\psi$ is a coordinate chart mapping 
an open subset of $g^{-1}( f(V_{\alpha}))$
to an open subset of ${\bf R}^n$, define $\phi$, a coordinate
chart on an open subset of $g^{-1}(f(V_{\alpha})) \times_N M$,
by $\phi(a,b) = \psi(a)$.  This is not only a well-defined chart,
but also differs from other charts by diffeomorphisms on overlaps.)

Now we are ready to rephrase the usual definition of smooth manifold.
Let $N$ be a topological space.  We claim that $N$ can be
naturally given the structure of a smooth manifold, if and only
if there exists a smooth manifold $M$ and a surjective local
homeomorphism $f: M \rightarrow N$, such that 
both of the projection maps $M \times_N M \longrightarrow M, M$ are
smooth.  (Note that $M \times_N M$ is itself a smooth manifold by virtue
of the lemma above, so it is indeed sensible to speak of
smooth maps $M \times_N M \rightarrow M$.)
%for all
%continuous maps $g: Y \rightarrow N$ with $Y$ a smooth manifold,
%both projection maps $Y \times_N M \longrightarrow Y, M$ are smooth.
%%\begin{enumerate}
%%\item $Y \times_N M$ is a smooth manifold, and  %%% faulty b/c needed
%%                                                %%% to say how $Y \times_N M$
%%                                                %%% could be smooth.
%%\item the projection map $Y \times_N M \stackrel{p_1}{\longrightarrow}
%%Y$ is smooth
%%\end{enumerate}

To show that this is true, first consider the case that $N$ is
a smooth manifold.  Let $\{ U_{\alpha} \}$ be an open cover of $N$,
such that each $U_{\alpha}$ is contained within a coordinate chart.
Let $\psi_{\alpha}: U_{\alpha} \rightarrow {\bf R}^n$ be those coordinate
charts.

Define $M$ to be the disjoint union of the $U_{\alpha}$, and
$f: M \rightarrow N$ to be the amalgamated inclusion.
It is straightforward to check that smoothness of the projection
maps
\begin{displaymath}
M \times_N M \: \left( \, = \: \coprod_{\alpha, \beta} \, U_{\alpha}
\cap U_{\beta} \, \right) \: \stackrel{p_{1,2}}{\longrightarrow}
M, \: M
\end{displaymath}
is equivalent to demanding that the coordinate charts be related
by diffeomorphisms on overlaps, guaranteed by the fact that $N$ is a smooth
manifold.

%Let $Y$ be a smooth manifold, and $g: Y \rightarrow N$ any continuous
%map.  We need to check that the two projection maps $Y \times_N M
%\longrightarrow Y, M$ are both smooth.  Now, note that
%\begin{displaymath}
%Y \times_N M \: = \: \coprod_{\alpha} \, \left( \, Y \times_N U_{\alpha}
%\, \right) \: = \: \coprod_{\alpha} \, g^{-1}\left( \, U_{\alpha} \,
%\right)
%\end{displaymath}
%The projection map $Y \times_N M \stackrel{p_1}{\longrightarrow} Y$
%is just the amalgamated inclusion $\coprod_{\alpha} g^{-1}(U_{\alpha})
%\rightarrow Y$.  Since smoothness is local, this is trivially smooth.
%**** WHAT ABOUT PROJECTION TO $M$ ?

Thus, one direction of the claim above is proven.

Conversely, let $N$ be a topological space, with $M$ and $f: M \rightarrow
N$ as stated above.  We shall construct a smooth structure on $N$.

To do this, let $\{ V_{\alpha} \}$ be an open cover of $M$,
such that each $V_{\alpha}$ is contained within a coordinate chart,
and such that the restriction of $f$ to each $V_{\alpha}$ is a homeomorphism
onto its image.  Let $\psi_{\alpha}: V_{\alpha} \rightarrow {\bf R}^n$
be the homeomorphisms defining the smooth structure on $M$.
For notational convenience, define $f_{\alpha} = f|_{V_{\alpha}}$,
and $U_{\alpha} = f_{\alpha}(V_{\alpha})$.

Now, $\{ U_{\alpha} \}$ is an open cover of $N$, and the maps
$\psi_{\alpha} \circ f_{\alpha}^{-1}: U_{\alpha} \rightarrow {\bf R}^n$
define coordinate charts on $N$, as for topological manifolds.
To show that $N$ has the structure of a smooth manifold, not just
a topological manifold, we need to show that these coordinate charts
are related by diffeomorphism on overlaps, i.e., that the map between
open subsets of ${\bf R}^n$ defined by the composition
\begin{equation}    \label{overlapdiffeo}
\psi_{\beta} \left( \, V_{\beta} |_{ f_{\beta}^{-1}(U_{\alpha} \cap U_{\beta}) } \, \right) \:
\stackrel{ \psi_{\beta}^{-1} }{\longrightarrow} \:
V_{\beta}|_{ f_{\beta}^{-1}(U_{\alpha} \cap U_{\beta} ) } \:
\stackrel{f_{\beta} }{\longrightarrow } \: U_{\alpha} \cap U_{\beta} \:
\stackrel{f_{\alpha}^{-1}}{\longrightarrow} \: V_{\alpha}|_{f_{\alpha}^{-1}(
U_{\alpha} \cap U_{\beta} )} \:
\stackrel{\psi_{\alpha} }{\longrightarrow} \:
\psi_{\alpha}\left(  \, V_{\alpha}|_{f_{\alpha}^{-1}(
U_{\alpha} \cap U_{\beta} )} \, \right) 
\end{equation}
is a diffeomorphism.

This overlap condition ultimately follows from the constraint
that the projection map $M \times_N M \stackrel{p_2}{\longrightarrow} M$
is smooth.  The topological space $M \times_N M$ has an open cover of
the form
\begin{displaymath}
\left\{ \, V_{\alpha}|_{ f_{\alpha}^{-1}( U_{\alpha} \cap U_{\beta} ) }
\times_N V_{\beta} |_{ f_{\beta}^{-1}( U_{\alpha} \cap U_{\beta} ) }
\, \right\}
\end{displaymath}
These open sets naturally form coordinate charts on $M \times_N M$,
with the maps 
\begin{displaymath}
\phi_{\alpha \beta}: \:
V_{\alpha} |_{ f_{\alpha}^{-1}( U_{\alpha} \cap U_{\beta} ) } \times_N
V_{\beta} |_{ f_{\beta}^{-1}( U_{\alpha} \cap U_{\beta} ) } \:
\longrightarrow \: {\bf R}^n
\end{displaymath}
defined by $\psi_{\beta}: V_{\beta}
\rightarrow {\bf R}^n$ by, $\phi_{\alpha \beta}(a, b) = \psi_{\beta}(b)$
for $(a, b) \in V_{\alpha} \times V_{\beta}$.

The projection map $p_1: M \times_N M \rightarrow M$ acts by mapping
$(a, b) \mapsto a$, so the constraint that the projection map be smooth
simply means that the map between open sets of ${\bf R}^n$ defined by
equation~(\ref{overlapdiffeo}) must be smooth.  Doing this for all
$\alpha$, $\beta$, we see that both the homeomorphism in
equation~(\ref{overlapdiffeo}) and its inverse must be smooth,
hence it is a diffeomorphism, as required.

Thus, the coordinate charts on the topological manifold $N$ are related
by diffeomorphism on overlaps, and so $N$ is a smooth manifold.

\subsubsection{Smoothness for stacks}

We are almost ready to rephrase the results above to apply to stacks.
Before we do so, we need an analogue of the lemme that we introduced
at the beginning of the last subsection.

Let $Y$ be a smooth manifold, $M$ a topological
space, $f: M \rightarrow {\cal F}$ a surjective local homeomorphism,
and $g: Y \rightarrow {\cal F}$ any continuous map.
Then the smooth structure on $Y$ induces a smooth structure
on the topological manifold $Y \times_{ {\cal F} } M$.

To prove this, let $p: Y \times_{ {\cal F} } M \rightarrow Y$ denote
the projection, and let $\{ W_{\alpha} \}$ denote an open cover of
$Y \times_{ {\cal F} }M$ such that for all $\alpha$,
$p|_{W_{\alpha}}$ is a homeomorphism.
Let $\psi_{\alpha}: p( W_{\alpha} ) \rightarrow {\bf R}^n$ be charts
on $Y$.  We claim that $\psi_{\alpha} \circ p|_{W_{\alpha}}: W_{\alpha}
\rightarrow {\bf R}^n$ are good charts on $Y \times_{ {\cal F} } M$.
To check this is very easy:  after all, if $W_{\alpha} \cup W_{\beta}
\neq \emptyset$, then 
\begin{displaymath}
\left( \psi_{\alpha} \circ p|_{W_{\alpha}} \right) 
\circ
\left( \psi_{\beta} \circ p|_{W_{\beta}} \right)^{-1}
\: = \:
\psi_{\alpha} \circ \psi_{\beta}^{-1}
\end{displaymath}
we see immediately that our charts on $Y \times_{ {\cal F} }M$ behave
well on overlaps, hence $Y \times_{{\cal F}}M$ is a smooth manifold.

Now, given this lemma, we can now use the results of the previous
subsection to give a definition of smoothness applicable to stacks.
We say that a stack ${\cal F}$ is
smooth if there exists a smooth manifold $M$ together with
a (representable) surjective local homeomorphism $f: M \rightarrow {\cal F}$,
with the property that the projection maps $M \times_{{\cal F}} M
\stackrel{p_{1,2}}{\longrightarrow} M, M$ are smooth.
Such a pair $(M, f)$ is known as a ``smooth atlas'' for ${\cal F}$.
(By virtue of the lemma, $M \times_{ {\cal F} } M$ is itself
a smooth manifold, so it makes sense to speak of a map
$M \times_{ {\cal F}}M \rightarrow M$ as being smooth.)

With this definition, it is straightforward to check that
the quotient stack $[X/\Gamma]$ is smooth, so long as
$X$ is smooth, $\Gamma$ is discrete, and $\Gamma$ acts by diffeomorphisms.
In this case, take $M = X$, and take $f$ to be the canonical map
$X \rightarrow [X/\Gamma]$.  We showed in section~\ref{mapprops}
that the canonical map is a surjective local homeomorphism (when
$\Gamma$ is discrete).  Also, from section~\ref{Xprinbundle}
we know that $X \times_{[X/\Gamma]} X = X \times \Gamma$,
and that the projection maps to $X$ are, respectively, the 
obvious projection $X \times \Gamma \rightarrow X$,
and the evaluation map $X \times \Gamma \rightarrow X$.
The first is always smooth, and the second is smooth so long as
$\Gamma$ acts by diffeomorphisms.  Thus, we have explicitly
constructed a smooth structure on $[X/\Gamma]$.

Next, what does it mean for a continuous map into $[X/\Gamma]$ to
be smooth?  As before, we shall first examine what this means for
ordinary manifolds, in a context that easily generalizes to stacks.

Let $N$ be a topological space, $M$ a smooth manifold, and
$f: M \rightarrow N$ a surjective local homeomorphism as
before defining a smooth structure on $N$.
Let $Y$ be another smooth manifold and $g: Y \rightarrow N$ a continuous map.
Then, $g$ is smooth if and only if the projection maps
$Y \times_N M \longrightarrow Y, M$ are both smooth.
(This is easy to check -- smoothness of the second projection map
$Y \times_N M \longrightarrow M$ is equivalent to smoothness of the
map $g: Y \rightarrow N$ between manifolds, with smooth structure defined
on $N$ by $M$.)

Thus, given a smooth manifold $Y$ and a continuous map $Y \rightarrow
{\cal F}$ into some stack with a smooth structure defined by a pair
$(M, f)$ as above, we say that the map $g$ is smooth if both of the
projection maps $Y \times_{{\cal F}} M \longrightarrow Y, M$ are smooth.
(By virtue of the lemma given at the beginning of this subsection,
the smooth structure on $Y$ induces a smooth structure on
$Y \times_{{\cal F}} M$, hence it makes sense to speak of maps from
$Y \times_{{\cal F}} M$ as being smooth.)

For example, the surjective local homeomorphism $f: M \rightarrow {\cal F}$
that defines a smooth structure on ${\cal F}$,
is itself smooth, by definition of smooth map.  After all,
by definition of smooth structure on ${\cal F}$, we assume that
both of the projections $M \times_{{\cal F}} M \longrightarrow M, M$
are smooth.

\subsection{Differential forms and metrics on stacks}

We can define differential forms on stacks in a manner closely analogous
to functions (as described in section~\ref{funcstack}).
Let ${\cal F}$ be a stack with a smooth atlas $(M, f)$.
A differential $n$-form $\omega$ on ${\cal F}$
is an assignment of a differential $n$-form $\omega_Y$ to each
smooth map $Y \rightarrow {\cal F}$ (for $Y$ a smooth manifold),
such that for any commuting diagram
\begin{equation}    \label{nform1}
\xymatrix{
Y_1 \ar[rr]^{ g } \ar[dr] & & Y_2 \ar[dl] \\
 & {\cal F} &
}
\end{equation}
one has $\omega_2 \circ g = \omega_1$.

As usual, note that if ${\cal F}$ is itself an honest space,
then this definition uniquely specifies an $n$-form on ${\cal F}$
in the usual sense.

In the special case that ${\cal F} = [X/\Gamma]$,
where $X$ is smooth, and $\Gamma$ is a discrete group acting
by diffeomorphisms, it is straightforward to check that an
$n$-form on $[X/\Gamma]$ is the same thing as a $\Gamma$-invariant
$n$-form on $X$.  As the argument is identical to that for
functions on stacks, we shall omit the details.

At the risk of beating a dead horse, one can also define
metrics on stacks.  To each map $Y \rightarrow {\cal F}$,
one assigns a symmetric two-tensor $g_Y$, obeying the same
constraint on commutative diagrams as above for differential forms.
If ${\cal F}$ is itself an honest space, this is equivalent to
the usual notion of a metric, and if ${\cal F}$ is the
quotient stack $[X/\Gamma]$, as above, then this description
is equivalent to a $\Gamma$-invariant metric on $X$.

\section{Sigma models on stacks, redux}     \label{redux}

In section~\ref{sigmamodel} we pointed out that a sum
over (equivalence classes of) maps into $[X/\Gamma]$
duplicated both the twisted sector sum and the functional integral
in each twisted sector of a string orbifold, a smoking gun for
an interpretation of a string orbifold as a sigma model on a stack.
In this section, we shall complete this picture by describing
the classical action for a sigma model on a stack (with an atlas),
and show how this generalizes both ordinary sigma models and
string orbifolds.

Let ${\cal F}$ be a stack with smooth atlas $X$,
and let $\Sigma$ be the worldsheet (more generally, worldspace)
of the sigma model.
For any map $\phi: \Sigma \rightarrow {\cal F}$,
let $\Phi: \Sigma \times_{ {\cal F} } X \rightarrow X$
denote the projection map (which encodes part of $\phi$).
ote that both $X$
and $\Sigma \times_{ {\cal F} } X$ are smooth spaces,
so $\Phi$ is a map in the usual sense.
Let $\phi^* G$ denote the pullback of the metric on ${\cal F}$
to $\Sigma$.  (Recall that we defined metrics on stacks in terms
of their pullbacks, so we are guaranteed to know $\phi^* G$
as an immediate consequence of the definition of $G$,
and $\phi^* G$ is an honest symmetric two-tensor on $\Sigma$.)
Then, the natural proposal for the
basic bosonic terms in the classical action for
a sigma model on ${\cal F}$ are given by
\begin{equation}
\int d^2 \sigma \, \left( \phi^* G_{\mu \nu} \right)
h^{ \alpha \beta }
\left( \frac{ \partial \Phi^{\mu} }{ \partial \sigma^{\alpha} } \right)
\left( \frac{ \partial \Phi^{\nu} }{ \partial \sigma^{\beta} } \right)
\end{equation}
where this is integrated over a lift\footnote{Sensible essentially
because the (projection) map $\Sigma \times_{ {\cal F} } X
\rightarrow \Sigma$ is a surjective local homeomorphism.} of $\Sigma$
to $\Sigma \times_{ {\cal F} } X$,
and where $h^{\alpha \beta}$ denotes the worldsheet metric.
(Working out analogues of the other terms in a sigma model
is completely straightforward, given the form of the bosonic terms;
we leave them as an exercise for the dedicated reader.)
Then, the path integral sums over (equivalence classes\footnote{After
all, we wish to perform a sum, hence we must take equivalence
classes.} of) maps
$\Sigma \rightarrow {\cal F}$.

First, let us compare to ordinary sigma models.
Suppose ${\cal F}$ is an ordinary space, rather than a stack.
Then we can\footnote{Other choices of atlases work similarly.
For example, if we take $X$ to be a disjoint union of elements of an
open cover of the target space, then we recover an expression for
the sigma model explicitly in terms of elements of a cover.} 
take $X = {\cal F}$.  As a consequence,
$\Sigma \times_{ {\cal F} } X = \Sigma$, so the classical action
is integrated over $\Sigma$, and the path integral is the usual sum over
maps.  Also, in this case, $\phi$ coincides with the
`projection' map $\Phi$, so the classical action above
completely agrees with the usual classical action for a sigma model
on a space.  In other words, if ${\cal F}$ is a space,
then we recover the standard sigma model.

Now, suppose that ${\cal F}$ is a global quotient stack,
{\it i.e.}, ${\cal F} = [X/\Gamma]$ where $\Gamma$ is discrete
and acts by diffeomorphisms.  For simplicity, we shall also
assume that $X$ is smooth.  Then, an atlas for this stack
is simply $X$ itself.  A map $\phi: \Sigma \rightarrow {\cal F}$
is determined by two pieces of data, namely
\begin{itemize}
\item A principal $\Gamma$-bundle on $\Sigma$
\item A $\Gamma$-equivariant map from the total space of
the bundle into $X$
\end{itemize}
(Less formally, a twisted sector and a twisted sector map,
respectively.)
The fibered product $\Sigma \times_{ {\cal F} } X$
is given by the total space of the principal $\Gamma$ bundle on
$\Sigma$, and the projection map $\Sigma \times_{ {\cal F} } X
\rightarrow X$ is given by the $\Gamma$-equivariant map into $X$.
In any event, it is now straightforward to check that the
classical action for a sigma model on ${\cal F}$ described
above coincides with the action appearing in string orbifolds
(including branch cuts induced by the lift), and the path
integral sum over maps $\phi: \Sigma \rightarrow {\cal F}$
duplicates both the twisted sector sum as well as the
sum over maps within each twisted sector, as described earlier.

So far we have only recovered known results; let us now try 
something new.  Suppose the target ${\cal F}$ is a gerbe.
For simplicity, we shall assume that ${\cal F}$ is the
canonical trivial $G$-gerbe on a space $X$.
Such a gerbe is described by the quotient stack $[X/G]$,
where the action of $G$ on $X$ is trivial.
Using the notion of sigma model on a stack as above,
one quickly finds that the path integral for this target space
is the same as the path integral for a sigma model on $X$,
up to an overall multiplicative factor (equal to the number of
equivalence classes of principal $G$-bundles on $Y$).
As overall factors are irrelevant in path integrals,
the result appears to be that a string on the canonical trivial
gerbe is the same as a string on the underlying space.
More generally, it is natural to
conjecture that strings on flat gerbes should be equivalent to
strings on underlying spaces, but with flat $B$ fields.  Such
a result which would nicely dovetail with the well-known
fact that a coherent sheaf on a flat gerbe is equivalent to a `twisted' sheaf
on the underlying space, the same twisting that occurs in the presence
of a $B$ field.  For physicists, this is an alternative to the
description in terms of modules over Azumaya algebras that has
recently been popularized \cite{kapustin}.

So far we have only discussed classical actions for sigma models
on stacks, but there is much more that must be done before 
one can verify that the notion of a sigma model on a stack
is necessarily sensible.  In effect, we have only considered
local behavior, but in order to be sure this notion is sensible
after quantization, one also needs to consider global phenomena.
Such considerations were the source of much hand-wringing when
nonlinear sigma models on ordinary spaces were first introduced
(see for example \cite{mmn}), and must be repeated for stacks.

\section{Well-behavedness of sigma models on stacks}    
\label{wellbehave}

Assuming that our proposal for a sigma model on a stack
makes sense, under what circumstances might one expect 
the corresponding CFT to be well-behaved?

In an ordinary sigma model on a space, the CFT will become
badly behaved if the target space metric degenerates,
making the kinetic terms in the sigma model action poorly-behaved.
In other words, one usually expects an ordinary sigma model
to be well-behaved so long as the target space is smooth.

In the proposal of the last section, for a sigma model on
stack ${\cal F}$ with atlas $X$, the kinetic terms in the action
are formulated in terms of maps into $X$, hence one requirement
for a well-behaved CFT (if the CFT actually exists) would be
that $X$ is smooth.

If our stack ${\cal F}$ were an honest space, we would now be
done.  However, for a general stack ${\cal F}$,
our proposal was more complicated than just a sum over
maps into the atlas -- the action was formulated on a lift
of the worldsheet $Y$ to $Y \times_{ {\cal F} } X$,
which introduces branch cuts.  We shall not attempt to analyze
what additional constraints the presence of such branch cuts
introduce into requirements for a well-behaved CFT,
but notice that surely any such constraints would be constraints
on the map $X \rightarrow {\cal F}$. 

So far we have briefly (and loosely) argued that for
the (hypothetical) CFT for a sigma model on a stack ${\cal F}$ to be
well-behaved, one needs the atlas $X$ to be smooth,
in addition to some (undetermined) criteria on the map
$X \rightarrow {\cal F}$.
Recall from earlier that the same criteria are needed 
for the stack ${\cal F}$ to be smooth -- one needs the atlas $X$
to be smooth, and the map $X \rightarrow {\cal F}$ must
satisfy additional criteria.
Now, without knowing precisely which criteria the map $X \rightarrow 
{\cal F}$ must satisfy, we cannot claim that the criteria for
a well-behaved CFT necessarily completely agree with the
criteria for the target stack ${\cal F}$ to be smooth,
but we are at least in the right ballpark, and some of the criteria
({\it i.e.}, $X$ smooth) are indeed identical.

In other words, conditions needed physically for a sigma model
into ${\cal F}$ to be well-behaved, are at least in the same
ballpark as the criteria for ${\cal F}$ to be smooth.
At least very naively, it looks very likely that
the notion of smoothness for stacks is the physically-relevant
notion, {\it i.e.}, the conditions for a stack to be smooth
appear to agree with the conditions for the corresponding
CFT to be well-behaved.

\section{Twist fields, associated stacks, and orbifold
Euler characteristics}     \label{twistfield}

In this section, we shall explain how twist fields arise
in this picture of string orbifolds as sigma models on
certain stacks.
We will be naturally led to a description of twist fields 
in terms of cohomology of stacks associated to quotient stacks,
a description which is actually known to certain mathematicians,
though its form will be somewhat different from what many
physicists have expected.

To understand how twist fields arise in a sigma model on a quotient
stack, we need to recall for a moment why massless states of
a sigma model are ever
described in terms of the cohomology of the target.
Somewhat loosely, in general terms one expects the circumference of
a closed string to be roughly proportional to the energy of the string,
so massless modes should be described by strings that have shrunk to
points.  So, if one is only interested in massless modes, then it suffices
to consider maps of points into the target space, and from the usual
reasoning (see for example \cite{edmorse}), one recovers the cohomology of the
target space.

Now, consider a string propagating\footnote{Here, we are taking a more
Hamiltonian approach to matters than before.  Previously we have considered
path integrals for sigma models on quotient stacks, and so considered
maps from the worldsheet into the quotient stack.  By contrast,
here we are interested in string states, not path integrals,
so we consider maps from
$S^1$ into the quotient stack.} on a quotient stack.
As emphasized earlier, a map from a closed string into the quotient
stack $[X/\Gamma]$ is described by a principal $\Gamma$-bundle over
$S^1$, together with a $\Gamma$-equivariant map from the total space
of the bundle into $X$.

When the string shrinks to zero size, this data can be described
more compactly.  Naively, one would say that for a zero-size string,
this data is the same as a $\Gamma$-equivariant map from 
a principal $\Gamma$-bundle over $X$, i.e., $\Gamma$ itself, into $X$,
which is the same as a point of $[X/\Gamma]$.

\begin{figure}
\centerline{\psfig{file=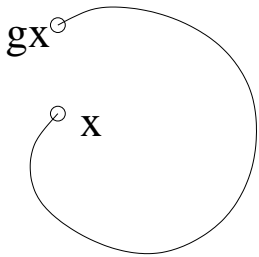,width=1.5in}}
\caption{\label{figtw1} A twist sector state; $g \in \Gamma$} 
\end{figure}

However, that is not quite the whole story -- in particular, it ignores
the possibility that the principal $\Gamma$-bundle on the shrinking 
$S^1$ is nontrivial.  (Equivalently, it ignores the twisted sectors,
illustrated in figure~\ref{figtw1}.)
We need to include information about that bundle.
Now, a principal bundle on $S^1$ can be described by a single transition 
function.  As the $S^1$ shrinks, a map into $[X/\Gamma]$ is trying to look
over each point of $S^1$ like a single map $\Gamma \rightarrow X$, 
together with a single transition function that should look increasingly
like an automorphism of that single map $\Gamma \rightarrow X$.
So, intuitively, in the limit that the $S^1$ shrinks to a point,
a map will be described by a point of $[X/\Gamma]$, together
with an automorphism of that point.

For the points of $[X/\Gamma]$ that project down to smooth points
on $X/\Gamma$, the only automorphism is the identity, so there is no
new information.  However, over singularities of $X/\Gamma$,
the points of $[X/\Gamma]$ have extra automorphisms, in addition
to the identity.  Thus, we get extra maps.
Thus, the low-energy spectrum of a string orbifold is not precisely
the cohomology of the quotient stack $[X/\Gamma]$, but rather some
associated space which has extra points.  

We can make this discussion somewhat more formal.
For any stack ${\cal F}$
the inertia group stack\footnote{In the francophone literature, this is
also known as the ramification stack.  Also, another way to describe
the inertia group stack $I_{{\cal F}}$ associated to any stack ${\cal F}$
is as
\begin{displaymath}
I_{{\cal F}} \: = \: {\cal F} \times_{{\cal F} \times {\cal F}} {\cal F}
\end{displaymath}
where both the maps ${\cal F} \rightarrow {\cal F} \times {\cal F}$
are the diagonal.} $I_{{\cal F}}$ of ${\cal F}$
\cite[def'n 1.12]{vistoli}
is defined by saying that a map $Y \rightarrow I_{{\cal F}}$
is given by a map $Y \rightarrow {\cal F}$ together with an
automorphism of that map. 
(Phrased another way, the category $I_{{\cal F}}$ has as objects,
pairs consisting of an object of ${\cal F}$ together with an
automorphism of that object.  Morphisms in the category are
morphisms in ${\cal F}$ which are compatible with the
automorphisms of the objects.)
More to the point, from the discussion above we see that our
shrunken $S^1$'s are mapped into the inertia group stack
$I_{[X/\Gamma]}$ associated to the quotient stack, not the
quotient stack itself.

In principle, twist fields are the cohomology of this auxiliary
space, the inertia group stack $I_{[X/\Gamma]}$.
To put this matter in better perspective,
the inertia group stack $I_{[X/\Gamma]}$ 
associated to a quotient stack has the form
\cite[p. 38]{toen}
\begin{equation}   \label{toenresult}
I_{[X/\Gamma]} \: \cong \:
%\amalg_{[g]} [ X^g / C(g) ]
\coprod_{[g]} \, \left[ X^g / C(g) \right]
\end{equation}
(where $[g]$ denotes the conjugacy class of $g \in \Gamma$ and
$C(g)$ is the centralizer of $g$ in $\Gamma$)
which compares very closely to the usual expression for the
stringy Euler characteristic of a string orbifold
\cite{hirzhofer}:
\begin{displaymath}
\chi_{orb}(X, \Gamma) \: = \: \sum_{[g]}
\, \chi\left( X^g / C(g) \right)
\end{displaymath}
Expression~(\ref{toenresult}) is proven in \cite[p. 38]{toen},
as well as in appendix~\ref{mytoenproof} in a different fashion.
Hopefully the obvious relationship between the form of the
inertia group stack and orbifold Euler characteristics should
help convince skeptical physicists.

Now, although this particular description of twist fields does
not seem to be widely known to physicists, it is known to
some mathematicians -- by thinking about string orbifolds in terms
of sigma models, we have recovered a description of twist fields.
Also, although this form is quite obscure, it can be
shown to be equivalent to a much simpler description in terms of
group actions on covers.

Another point we should speak to is the fact that this particular
description of twist fields is rather different from what
many physicists have traditionally expected.  In some corners
of the physics community, it has been conjectured that twist fields
should be understandable in terms of some cohomology of
the quotient {\it space}, and not only is the expression above
not a cohomology of the quotient space, it is not even a cohomology
of the quotient stack!  In fact, after considerable work,
it can be shown that the description above is essentially equivalent
to a description in terms of group actions on covers, something that
many physicists would not ordinarily consider a `final' description.
So, one way of rephrasing our argument is that, if a string orbifold
truly is a sigma model on a quotient stack, then a description of
twist fields in terms of group actions on covers is perhaps the
best one can hope for -- a description in terms of a cohomology
of the quotient space need not exist.

Another point we should speak to is the naive contradiction
that has arisen, due to the face that the massless spectrum
of our proposed sigma model is not the cohomology of the target.
We implicitly resolved this paradox earlier
when we noted that the `zero-momentum' part of the loop space
need no longer agree with the original stack, a fact which
invalidates the standard result.
Phrased another way, although a string orbifold describes strings
propagating on a quotient stack $[X/\Gamma]$, it appears that
twist fields appear via cohomology calculations in
the associated inertia group stack $I_{[X/\Gamma]}$.
Ordinarily, massless modes would come from cohomology of the
space the string propagates on, but stacks are a little
more subtle.  If the stack ${\cal F}$ can be represented by
an ordinary space, then all points have only the trivial
automorphism, and so the inertia group stack $I_{{\cal F}}$ and
${\cal F}$ are the same -- so, this subtlety that the space strings
propagate on is distinct from the space that the massless modes are associated
to, can only crop up on stacks, not ordinary spaces.

As our discussion so far has been rather abstract, let us take
a moment to consider some concrete examples.
For example, consider the orbifold of ${\bf C}^2$ by
${\bf Z}_2$.  Most points of $[{\bf C}^2/{\bf Z}_2]$ do not have
automorphisms beyond the identity.  However, the point of
$[{\bf C}^2/{\bf Z}_2]$ lying over the singularity of
${\bf C}^2/{\bf Z}_2$ does have a nontrivial automorphism.
After all, that point is the map ${\bf Z}_2 \rightarrow {\bf C}^2$
whose image is the fixed point, and one can compose this map
with a nontrivial action of ${\bf Z}_2$ to recover the same map again.
Thus, the cohomology of $I_{[X/\Gamma]}$ should look like the cohomology
of $[X/\Gamma]$, together with an extra generator reflecting the
cohomology of an extra point.  This extra point is the object of
$[{\bf C}^2/{\bf Z}_2]$ lying over the singularity of ${\bf C}^2/{\bf Z}_2$,
together with a nontrivial automorphism.  The reader should recognize
that nontrivial automorphism as precisely the ``twist'' in the twist
field appearing in the low-energy spectrum of the string orbifold of
${\bf C}^2$ by ${\bf Z}_2$.  In the next section we shall check
more formally that the inertia group stack $I_{[X/\Gamma]}$ correctly
reproduces the general formula for the orbifold Euler characteristic,
but for the moment, the reader should pause to consider that we
have directly computed a twist field in which the ``twist''
is manifest.

More generally, it should be clear that quotient stacks
$[{\bf C}^n/{\bf Z}_k]$ should have extra maps over the singularities,
precisely duplicating the appropriate twist fields.

For most of this paper, we have argued that standard
properties of string orbifolds (twisted sectors,
well-behaved CFT, and B fields,
for example) are really properties of stacks, and have nothing
to do with string theory {\it per se}.  However, in order to properly
understand twist fields, we have had to use strings -- we have used
the fact that the extended objects are one-dimensional crucially
in determining that the massless fields should be described
by cohomology of $I_{[X/\Gamma]}$.  For other, higher-dimensional, extended
objects, the analogous analysis would give a different result.
See section~\ref{membranetwist} for more details on higher-dimensional
analogues of twist fields.

\section{Non-effective group actions}    \label{noneff}

Sometimes in string theory we consider orbifolds whose group
only acts on the fields of the theory, and leaves the underlying
space invariant.  For example, one sometimes orbifolds by $(-)^F$,
commonly in situations where one wishes to break supersymmetry at
string scale.  In such cases, one still has the structure of twisted
sectors, even though the group acts trivially on the underlying space.
Is this consistent with the interpretation of string orbifolds
as sigma models on quotient stacks?

Happily, this is completely consistent with the interpretation
of strings as sigma models on quotient stacks.
For example, if $\Gamma$ acts trivially on $X$, then
$[X/\Gamma]$ and $X$ are not homeomorphic.  In other words, even when $\Gamma$
acts trivially on $X$, in the sense that $g \cdot x = x$ for
all $g \in \Gamma$ and points $x \in X$, a sigma model on the quotient
stack $[X/\Gamma]$ still has the structure of twisted sectors and
so forth.

On a slightly different note, non-effective group actions
can also be used to construct other interesting stacks.
For example, the trivial gerbe over a space (the formal structure
corresponding to B fields, analogous to bundles for gauge fields)
can be described as the quotient stack of a $U(1)$ action that acts
trivially on the space.  (Although we have been primarily interested
in quotients by discrete groups in this paper, most of our discussion
also holds for the case that the orbifold group is continuous.)

In particular, the quotient stack $[\mbox{pt}/\Gamma]$, 
where $\mbox{pt}$ denotes
some fixed point with a trivial $\Gamma$-action, is known as the
classifying stack for $\Gamma$ and denoted $B \Gamma$.
The trivial gerbe over a space $X$ is simply $X \times B \Gamma$,
and other gerbes are twisted versions of this (i.e., other gerbes
look locally, but not necessarily globally, like $X \times B \Gamma$).

\section{B fields}    \label{Bfield}

The standard lore concerning string orbifolds nowadays says that
in string orbifolds, the B field associated with blowup modes is
naturally nonzero.  (Such a B field could be rotated to take on other
values; but there is a specific natural value associated with orbifold
points.)  Technically, for ${\bf C}^2/{\bf Z}_2$, by resolving
the singularity one is led to believe that for a string orbifold,
$B = \frac{1}{2}$ \cite{paulz2}, and for ${\bf C}^2/{\bf Z}_n$,
there are weaker arguments\footnote{For some rigorous results,
see instead \cite{sarah}; also, \cite{katrin1}.} 
suggesting that $B = \frac{1}{n}$
\cite{douglaszn}.  Few other cases have been analyzed in detail,
though it has been suggested that there exist other examples
in which the B field is naturally zero in a string orbifold.

Witten has observed \cite{edstrings95} that if the B field associated
to a blowup mode is naturally nonzero, then one should believe that the
corresponding conformal field theory is nonsingular, judging by 
results on theta angles in linear sigma models \cite{edphases}.

How does this tie into the present work?
If string orbifolds are sigma models on quotient stacks,
then how can we see the B field?  For that matter, what does
it mean to have a B field on a shrunken divisor?  How can one talk
about holonomy on a zero-size object?

Surprisingly enough, one can get answers to these questions
with comparatively little work.  One way of describing a quotient
stack $[X/\Gamma]$ is as a quotient space $X/\Gamma$ with
some ``extra structure'' at the singularities.  In particular,
recall that when $\Gamma$ acts freely, $[X/\Gamma] \cong X/\Gamma$,
so the difference between the quotient stack and the quotient space
arises at the fixed points.  Also recall that, in the category
of points $[X/\Gamma]_{pt}$ of the quotient stack $[X/\Gamma]$,
objects lying over singularities of $X/\Gamma$ have extra automorphisms
 -- these extra automorphisms are precisely the ``extra structure''
referred to in comparing $[X/\Gamma]$ to $X/\Gamma$.

How does the B field fit into this picture?
A very quick way to see a connection is as follows.
The extra structure possessed by the quotient stack $[X/\Gamma]$
over singularities of the quotient space $X/\Gamma$ is precisely
a gerbe at a point.  In the literature on stacks, this seems to
be known as the ``residual gerbe.''  (More generally,
recall that gerbes are merely special types of stacks, so one
should not be at all surprised to find gerbes lurking in the present
context.)
Moreover, for $[{\bf C}^2/{\bf Z}_n]$
singularities, for example, the gerbe corresponds to a B field 
with holonomy quantized by $1/n$.  Phrased another way, if one
resolves the quotient space, and pulls back the quotient stack to the
resolution of the quotient space, then the restriction of the pullback
to the exceptional divisor will be a gerbe describing a B field with
holonomy lying in $\{0, 1/n, 2/n, 3/n, \cdots, (n-1)/n \}$.

% Dave Ben-Zvi says, that Laumon has proven that ANY algebraic stack
% of finite type is a union of pieces which are gerbes for different
% groups.

Now, this analysis was very quick, and with a bit of effort,
one can do a much better job, by considerations of D-brane probes,
described as coherent sheaves.  We will need two important facts
in this analysis:
\begin{itemize}
\item First, a coherent sheaf on a quotient stack $[X/\Gamma]$ is precisely
a $\Gamma$-equivariant coherent sheaf on $X$ (which is not quite the
same as a sheaf on the quotient space $X/\Gamma$).  Recall that the
Douglas-Moore construction \cite{dougmoore} of D-branes
on string orbifolds describes $\Gamma$-equivariant
objects on the covering space, so in other words, the Douglas-Moore
construction of branes on orbifolds precisely corresponds to
coherent sheaves on quotient stacks.
\item Second, a coherent sheaf on a flat gerbe is the same
thing as a `twisted' sheaf on the underlying space, {\it i.e.},
twisted in the sense of `bundles' on D-branes with B fields.
Put another way, sheaves on gerbes are an alternative to 
modules over Azumaya algebras as popularized in \cite{kapustin}.
A thoroughly ham-handed attempt to describe this phenomenon is
presented in \cite{stacks}.
\end{itemize}
Given that all stacks look locally like either orbifolds or gerbes,
these two cases justify using coherent sheaves to describe
D-branes on more general stacks.

Now, in order to see what quotients stacks have to do with B fields,
let us consider a naive `blowup' of the stack $[{\bf C}^2 / {\bf Z}_2 ]$.
In particular, the minimal resolution of the quotient
singularity ${\bf C}^2/{\bf Z}_2$ is the same as the quotient
$( \mbox{Bl}_1 {\bf C}^2 ) / {\bf Z}_2$, where the ${\bf Z}_2$
action has been extended trivially over the exceptional divisor
of the blouwp.  Thus, the quotient stack 
$[ ( \mbox{Bl}_1 {\bf C}^2 ) / {\bf Z}_2 ]$
is a naive stacky analogue of the resolution of the quotient space
${\bf C}^2/{\bf Z}_2$, and among other things, is a stack over
the resolution.

Finally, consider D-brane probes of this stack, viewed
as coherent sheaves.
Away from the exceptional divisor, this stack looks like the 
corresponding space, so a D-brane away from the exceptional divisor
thinks it is propagating on the underlying space.
A coherent sheaf over the exceptional divisor, on the other hand,
is a sheaf on a gerbe, and so describes a D-brane in the presence
of a B field.  Thus, we see the advertised B field.
In fact, we can read off even more -- the gerbe over the exceptional
divisor is a ${\bf Z}_2$-gerbe, so the corresponding B field holonomy
must be either $0$ or $1/2$.  
This certainly puts us in the right ballpark to recover the
results in \cite{paulz2,sarah,katrin1,douglaszn}.  In order to find out
precisely which of these values the B field on the exceptional divisor
would take on, one would have to do a more careful study of the
deformation theory of the quotient stack, which we shall
not attempt here.  
However, the fact that one naturally gets a B field associated with
singularities, with, up to an as-yet-undetermined discrete choice,
the right value, is enough for the moment.

%In passing, the reader might wonder at the fact that, in our naive
%pullback argument, the generalized space our string propagated on
%was a gerbe.  This may sound somewhat unusual, but in fact is consistent.
%For example, in \cite{stacks} we pointed out that the twisted
%``bundles'' appearing on D-brane worldvolumes could naturally be understood
%as sheaves on stacks (the stack describing the gerbe that the B field
%couples to).  Put another way, in \cite{stacks} we argued that
%one should think of D-branes as propagating on stacks, not spaces,
%although we did not phrase our result in this fashion at the time.

More generally, 
%based on the work in \cite{stacks} and the appearance
%of the B field in quotient stacks, 
it seems likely that one may be
able to understand B fields in standard Calabi-Yau compactifications
through a stack argument.  It seems not unreasonable that the
business of ``analytically continuing around singularities'' may now
be understood as equivalently compactifying on a stack describing
the B field\footnote{An analogy may make this more clear.
For a line bundle with connection on a given space, there are often
two very different looking representatives of the corresponding
equivalence class of bundles with connection:  one representative
in which the transition functions of the bundle are all identically 1,
and the connection is a globally-defined, closed (but not exact) form;
another in which the connection is identically zero, and the information
about Wilson lines is encoded in trivializable-but-not-trivial
transition functions.}; presumably, one would find that the stacks in question
are only singular at special points, hence one could have a smooth
``analytic continuation.''

\section{M theory}     \label{Mthy}

Most of this paper has been devoted to giving a solid
interpretation of string orbifolds in terms of quotient stacks.
However, although we are providing a great deal of insight into
the deep underpinnings of string orbifolds, we have not yet told
the reader any new physics.

We shall now point out some possible new physics originating from
these insights.

\subsection{Generalities}

It has sometimes been stated that orbifolds are not well-understood
in the context of M-theory\footnote{We are here using ``M-theory'' to
refer to the hypothetical quantum theory underlying eleven dimensional
supergravity, in accordance with its original usage in the
literature, as opposed to some unifying master theory.}.
After all, in the past string orbifold
constructions have seemed to rely deeply on having strings,
which of course is not appropriate for M-theory.

However, in light of the arguments given in this paper, we can
finally shed light on M-theory orbifolds.  In essence, we have argued
in this paper that string orbifolds can be understood without strings,
so we can now apply these methods to M-theory.

If we think of M-theory as a limit of IIA string theory, then it is
natural to believe that the limit of a IIA string theory on
the generalized space $[X/\Gamma]$ (i.e., a string orbifold)
will be M-theory on $[X/\Gamma]$.  

Since the generalized
space $[X/\Gamma]$ is smooth, regardless of whether or not $\Gamma$
acts freely, M-theory on this generalized space should be well-behaved.

Just as strings propagating on $[X/\Gamma]$ possess twisted sectors,
a membrane propagating on $[X/\Gamma]$ should also have twisted sectors,
in the obvious sense (see \cite{cdt} for a more pictorial explanation).

\subsection{Hypothetical membrane twist fields}   \label{membranetwist}

Now that we understand twist fields geometrically in strings,
we can now speculate about their analogues in membranes.
Such things may play an important role.  For example,
in the Horava-Witten \cite{petred}
description of heterotic $E_8 \times E_8$ theory
from M-theory, they were forced to add $E_8$ gauge multiplets to the
boundary by hand.  Their appearance was enforced by various anomaly
cancellations, but still, they had to add them manually.
If we understood the Horava-Witten picture in terms of stacks,
then just as string twist fields naturally emerge, so might
the $E_8$ multiplets from considerations of the M-theory three-form
potential.

In this subsection, we shall not attack the problem of understanding
the occurrence of $E_8$ boundary multiplets in Horava-Witten theory,
but rather we shall attack a much simpler problem.  We shall merely
examine the twist fields that crop up on membranes if we formally
follow the same argument from which we produced string twist fields.
We will not claim to understand the physics, but the mathematics
is clear. 

One might naively guess that, if one could make sense out
of the effective action of a membrane orbifold, that it would have
the same twist fields as strings, but that does not seem to be correct.
If one takes a Hamiltonian approach as above, and considers, for example,
membranes whose spatial cross-section looks like $T^2$,
then to take into account the possibility of a nontrivial
principal $\Gamma$-bundle on the shrinking $T^2$, one must include
not a single automorphism, but a pair of commuting automorphisms.
Thus, if we naively follow the same logic as for strings,
we see that membranes would have different numbers of twist fields,
i.e., the number of twist fields seen by an extended object
on an orbifold depends upon the dimension and topology of its
spatial cross-section.

Unfortunately there do not seem to be any examples in which
the number of twist fields has any correlation with blowups
in some natural family, unlike the case of strings.
For example, for a $T^3$ membrane on $[{\bf C}^2/{\bf Z}_2]$,
one would expect either 2 or 3 twist fields (depending upon an ordering
issue we have not been careful about).  The closest one can come 
is a 4-real-parameter family of blowups, describing a blowup of
the Calabi-Yau resolution of ${\bf C}^2/{\bf Z}_2$.
(One real parameter is the K\"ahler modulus of the resolution;
two real parameters are the location of the second blowup on the first;
and the last is the K\"ahler modulus of the second blowup.)
The resulting blown-up space is not Calabi-Yau, but perhaps for
membranes this is not unnatural.  If there were some natural way to eliminate
some of those four real parameters, perhaps by some sort of overall
scaling or modding out an action of $GL(2, {\bf C})$ on the first exceptional
divisor, then we would get the desired number of moduli, but we cannot
see any natural reason to do such a thing.

The fact that, unlike strings, there do not seem to be any
examples in which the number of ``membrane twist fields'' correlates
with the number of parameters in a family of resolutions or deformations 
of the quotient space
might be an indication that our purely formal motivation
for ``membrane twist fields,'' given above, has no physical basis,
and so should be ignored.  Alternatively, it might be an indication
that the story is more interesting for membranes than previously
assumed.

\section{New compactifications}    \label{newcomp}

%Although we have not emphasized this point in this paper,
%we have implicitly uncovered a new family of string compactifications.
%Quotient stacks are examples of ``generalized spaces,'' which cannot
%be understood within old-fashioned point-set topology.  Given that
%we now have an explicit example of string propagation on a stack
%(furnished by good old-fashioned string orbifolds, no less),
%one can now study string propagation on other stacks and generalized
%spaces.
%
%This may seem somewhat radical, but related examples are known.
%For example, in \cite{stacks} we pointed out that the
%twisted ``bundles'' on a D-brane worldvolume in the presence of
%a nontrivial B-field are naturally understood in terms of sheaves
%on stacks (in particular, stacks describing the B field, i.e., gerbes).
%Thus, one can think of D-branes as living on stacks instead of spaces.

Although the emphasis in this paper has been on string
orbifolds, part of the point of this paper is to lay the
groundwork for a new class of string compactifications:
compactification on stacks. 
One way of thinking about stacks is as slight generalizations
of spaces, on which one can perform differential geometry,
hence it is natural to ask whether one can compactify a
string on such objects.  
Indeed, understanding whether string compactification on stacks
even makes sense is a prerequisite to being able to fully
understand whether, as suggested here, string orbifolds really
are the same as strings compactified on quotient stacks.

Towards this goal of understanding string compactification
on stacks, we have not only described in detail how one
makes sense of differential geometry on stacks,
but also, for example, described a proposal for a classical
action for a sigma model on a stack (section~\ref{redux}),
which generalizes both ordinary sigma models and string orbifolds.
Of course, possessing a proposal for a classical action is not
sufficient; for example, when sigma models on ordinary spaces
were first being considered, it was noted that global effects
can lead to obstructions to quantization \cite{mmn}.

One of the most important ways to check whether
the notion of string compactification on stacks is sensible,
is to examine some nontrivial examples.  Hence, we are faced
with a chicken-and-egg problem:  it is difficult to check whether
the notion of compactification on stacks is truly sensible without
nontrivial examples, yet until one understands whether this is a sensible
notion, it is difficult to tell whether one has any nontrivial
examples.  String orbifolds appear to offer the first nontrivial
examples, but we may have been misled by formalism.

In any event, although in principle one can now consider
string compactification on stacks, it is not clear how useful
this notion actually can hope to be.  Not only are stacks closely
related to spaces, but in fact, all stacks look locally
either like spaces, orbifolds, or gerbes, a fact which greatly
cuts down on possible compactifications.  Also, in order
to be consistent with standard results on string orbifolds,
it seems very likely that compactification on many stacks
is identical to compactification on some underlying space with
a flat B field turned on.
At the end of the day, we can at present only imagine
two general directions along these lines that need to
be pursued:
\begin{itemize}
\item Local orbifolds.  Local orbifolds can be described
in terms of stacks, in addition to global orbifolds.
By considering string compactification on stacks,
one may be able to give local orbifolds a solid physical
foundation.
\item Deformation theory of string orbifolds.
If string orbifolds really are the same as strings compactified
on quotient stacks, then their deformation theory must be
understood in terms of deformation theory of the
underlying quotient stack, not a quotient space.
Hopefully this coincides with a quotient space ``with B fields,''
and we have given indirect evidence to support such
a conclusion, but much work remains to check whether this
is reasonable.
\end{itemize}

\section{Noncommutative geometry}   \label{ncg}

On a slightly different topic, the reader might be curious if
there is any relationship between stacks and noncommutative geometry.
After all, both are defined indirectly, by deforming some structure
that characterizes spaces:
\begin{enumerate}
\item Noncommutative geometry is defined via rings of functions.
The ring of functions on a space characterizes the space,
so one formally gets ``noncommutative spaces'' by deforming the ring
structure.
\item We have defined stacks (generalized spaces) via the category
of maps into the space.  The category of maps into a space characterizes
the space, so one can formally define new ``generalized spaces''
by deforming the category.
\end{enumerate}

Now, stacks and noncommutative spaces are {\it not} the same.
However, there is lore in the mathematics community that they are
related, that they can both be used to attack the same sorts of problems.

One example of this is well-known, and was given a ham-handed
treatment in \cite{stacks}.
There, we pointed out that the twisted ``bundles'' on D-brane worldvolumes
(twisted because of a nontrivial B field) can be understood either
in terms of noncommutative geometry (as modules on Azumaya algebras,
at least for torsion $H$, as seems to now be well-known),
or in terms of sheaves on stacks.

Other examples also exist.  For example, one of the original
motivating examples for noncommutative geometry in physics 
\cite{douglashull} consisted of noting that understanding T-duality
in certain B field backgrounds seemed to involve understanding quotients
of tori by irrational rotations, which was made more nearly sensible
via noncommutative geometry.  However, one could also describe
such a quotient in terms of a quotient stack -- the basic properties
we have discussed in this paper continue to hold regardless of the details
of the group action on the space (just so long as the group acts by
diffeomorphisms).

We shall not pursue possible connections further here,
but hope to return to this matter in future work.

\section{Conclusions}     \label{conclusions}

In this paper we have argued that string orbifolds appear to
literally be sigma models on quotient stacks.
We argued this first by unraveling
definitions to see that a sum over maps into a quotient stack
duplicated both the twisted sector sum as well as the
path integral sum in each twisted sector,
after which we pointed out that a natural ansatz for a 
classical action for a sigma model on a stack generalized both
classical actions for ordinary sigma models and 
string orbifold classical actions.
If all we had accomplished was to use quotient stacks as a highly
overcomplicated mechanism for describing group actions on covers,
then there would hardly have been a point;
instead, we go on to point
out that this new understanding of the target space geometry sheds new
light
a number of physical features of string orbifolds.  For example, the fact
that string orbifold CFT's behave as though they were sigma models
on smooth spaces is an immediate consequence of the fact that quotient
stacks are smooth spaces.  The old lore concerning B fields and quotient
singularities has a natural understanding here, as do twist fields and
orbifold Euler characteristics.

However, we feel we have barely scratched the surface of what can
be done with quotient stacks in particular and stacks in general.
There are a number of unanswered questions and natural followups
that beg to be pursued:
\begin{enumerate}
\item New compactifications.  Quotient stacks are examples of generalized
spaces, which have not been considered previously in the context
of string compactifications.  In principle, given that we now
realize we have an example of strings on stacks (i.e., string orbifolds),
one could now study compactification on other stacks.
We have only considered quotient stacks (in this paper) and
gerbes (in \cite{stacks}, where we gave a ham-handed treatment of
the well-known fact that the twisted ``bundles''
on D-brane worldvolumes were naturally understood as bundles on stacks).
What is needed is some new family of stacks, Calabi-Yau in some appropriate
sense, that one could study string compactification on.
\item Deformation theory of stacks.  We know of no mathematical work on
deformation theory of stacks, which would be very useful if one
wished to seriously pursue studies of string compactifications on stacks,
for example, or if one wished to completely and rigorously understand
$B$ fields on exceptional divisors of resolutions.  
Therefore, we encourage mathematicians reading this paper
to embark on such studies, or at least send us references to existing work.
%\item V-manifolds.  It is clear that quotient stacks should be
%very closely related to the V-manifolds described in,
%for example, \cite{satake1,satake2}, though we have not yet had the
%time to work out a mathematically precise statement of the relationship.
%This would be interesting to do.
%\item Cohomology of stacks.  Although we have argued that twist fields
%should, in principle, be computable as cohomology of the associated
%inertia group stack $I_{[X/\Gamma]}$, and furthermore shown how this
%naturally leads to the standard formula for the orbifold Euler
%characteristic, we have not written down the specific cohomology
%theory that string theory must be computing \cite{zaslow1}.  This would be
%interesting to do.  It may turn out to be equivalent to the
%cohomology group defined in \cite{ruan1}.
%\item Geometric reproduction of orbifold CFT results.
%Now that we have an honest geometric understanding of string orbifolds,
%it should be possible to understand many more features of orbifold CFT's
%in terms of the geometry of quotient stacks.  We have only had the time
%to consider the largest features, but many more features of
%string orbifold CFT's, such as the orbifold symmetry of correlation
%functions, OPE's of twist fields, and more, may all also be
%understandable directly in terms of the geometry of quotient stacks.
\item Relation between twist fields and blow-up modes.  In the
special cases of orbifolds with a lot of supersymmetry, there is a famous
connection between twist fields and blow-up modes.  Unfortunately,
although in principle quotient stacks should surely shed
light on this, we only have results at present that apply
to all quotient stacks and string orbifolds -- results that
apply to special cases require a degree of refinement not possessed
by the coarse analysis methods used in this paper.
Using stacks should surely shed light on this matter, and may even
give yet another way of understanding the McKay correspondence,
and so should be pursued vigorously by interested mathematicians.
\item Asymmetric orbifolds.  We have considered only geometric orbifolds
in this paper; we do not yet understand asymmetric orbifolds.  These
are presently under investigation.
\item B fields in standard Calabi-Yau compactifications.
At the end of section~\ref{Bfield} we conjectured that it might be
possible to understand the old business about ``analytically continuing
around singularities'' in terms of, equivalently compactifying on a stack
encoding the B field.  As we elaborated on this notion at some length
there, we will not say more here, aside from pointing out that this
would be interesting to follow up.
\item Horava-Witten $E_8$ multiplets.  In the Horava-Witten picture
of heterotic $E_8 \times E_8$ theory from M-theory \cite{petred},
the $E_8$ gauge multiplets are added by hand.  Ideally, one would like
to understand them as emerging somewhat more naturally, as some analogue
of twist fields.  Now that we have a better understanding of twist fields,
perhaps this could be worked out.  Since the M-theory three-form,
modulo the gravitational correction, is currently believed to be 
the Chern-Simons form of some underlying $E_8$ gauge theory, it seems
not unreasonable that if one were to study possible twist fields
in the same spirit as our discussion of discrete torsion and twist fields,
one might find $E_8$ gauge multiplets appearing naturally.
\end{enumerate}

\section{Acknowledgements}

We would like to thank P.~Aspinwall, D.~Ben-Zvi, S.~Dean, J.~Distler, T.~Gomez,
R.~Hain,
S.~Katz, A.~Knutson, D.~Morrison,
B.~Pardon, R.~Plesser, and M.~Stern for useful conversations.
In particular, we would especially like to thank T.~Gomez for extensive
communications concerning quotient stacks in particular, and A.~Knutson for
numerous communications regarding related general mathematics.

\appendix

\section{A space is determined by the maps into it}   \label{mapsgivespace}

In this section, we shall describe how a topological space
is completely determined by the continuous
maps into it.

How does this work?  Before giving a thorough category-theoretic
argument, there is some basic intuition that can be gained.

First, if we know all of the continuous maps from all topological
spaces into $X$, then the points are easy to pick off.
The points are just the maps, $\mbox{point} \rightarrow X$.

Now, knowing the points of a topological space is not sufficient
to determine that space; one must also know the open sets.
This also is relatively straightforward.  For example,
in the special case that $X$ is known in advance to be
compact and Hausdorff, we can use the fact that 
\cite[chapter 3.5]{munkres} every closed subset
of a compact Hausdorff space is precisely the image of a compact space under
some continuous map (and, conversely, the image of any compact
space under a continuous map is a closed subset).  
So, in the special case that $X$ is known
to be compact and Hausdorff, the closed subsets of $X$
are precisely the images of compact spaces.  Knowing all of the closed
sets is, of course, equivalent to knowing all of the open sets,
and so this gives us a topology on $X$.

A much more efficient way to proceed is to use category theory.
Here is the mathematically precise statement:  topological spaces
$X$ and $Y$ are homeomorphic if and only
if there is a groupoid morphism\footnote{Defined over the category ${\it Top}$
of topological spaces, whose objects are topological spaces and
morphisms are continuous maps.  In other words, the map $(X)_{map}
\rightarrow (Y)_{map}$ is required to preserve the projection
to ${\it Top}$, so that a map $( Z \rightarrow X )$ (an object
of $(X)_{map}$) is mapped to a map of the form $(Z \rightarrow Y)$.}
$(X)_{map} \rightarrow (Y)_{map}$
over ${\it Top}$ which defines an equivalence of categories.

If $X$ and $Y$ are homeomorphic, then it is easy to construct
such an equivalence of categories.  Let $f: X \rightarrow Y$
be a homeomorphism, then define a functor $F: (X)_{map}
\rightarrow (Y)_{map}$ as, the functor that maps
\begin{displaymath}
\left( \, X \: \stackrel{Id}{\longrightarrow} \: X \, \right)
\: \mapsto \:
\left( \, X \: \stackrel{f}{\longrightarrow} \: Y \, \right)
\end{displaymath}
The fact that such a functor is determined by the image
of $( \mbox{Id}: X \rightarrow X)$ is a result of the
Yoneda lemma \cite{stacks, gomez, vistoli}.
In more detail, an object $( Z \rightarrow X)$ maps to
$( Z \longrightarrow X \stackrel{f}{\longrightarrow} Y )$,
and morphisms are mapped trivially.
To show that this defines an equivalence of categories,
define a functor $G: (Y)_{map} \rightarrow (X)_{map}$ as the functor
that maps
\begin{displaymath}
\left( \, Y \: \stackrel{Id}{\longrightarrow} \: Y \, \right)
\: \mapsto \:
\left( \, Y \: \stackrel{f^{-1}}{\longrightarrow} \: X \, \right)
\end{displaymath}
It is straightforward to check that $F \circ G = \mbox{Id}$ and
$G \circ F = \mbox{Id}$.

Conversely, suppose that $(X)_{map}$ and $(Y)_{map}$ are equivalent
as categories.\footnote{With the equivalence preserving the projection
map to ${\it Top}$.}  Let $F: (X)_{map} \rightarrow (Y)_{map}$
be a functor defining the equivalence, itself defined by the
object $( X \stackrel{f}{\longrightarrow} Y)$ of $(Y)_{map}$, as
above.
Let 
$G: (Y)_{map} \rightarrow (X)_{map}$ be the corresponding functor,
defined by the object $( Y \stackrel{g}{\longrightarrow} X )$
of $(X)_{map}$.
Let $\eta: F \circ G \stackrel{\sim}{\Longrightarrow} \mbox{Id}$,
$\gamma: G \circ F \stackrel{\sim}{\Longrightarrow} \mbox{Id}$
invertible natural transformations.

Now, $F \circ G$ maps
\begin{displaymath}
\left( \, Y \: \stackrel{Id}{\longrightarrow} \: Y \, \right) \:
\mapsto \:
\left( \, Y \: \stackrel{g}{\longrightarrow} \: X \:
\stackrel{f}{\longrightarrow} \: Y \, \right)
\end{displaymath}
and $G \circ F$ maps
\begin{displaymath}
\left( \, X \: \stackrel{Id}{\longrightarrow} \: X \, \right) \:
\mapsto \:
\left( \, X \: \stackrel{f}{\longrightarrow} \: Y \:
\stackrel{g}{\longrightarrow} \: X \, \right)
\end{displaymath}

Since there is an invertible natural transformation $\eta:
F \circ G \Rightarrow \mbox{Id}$, associated to the object
$( Y \stackrel{Id}{\longrightarrow} Y )$ there is a morphism
\begin{displaymath}
\eta: \: \left( \, Y \: \stackrel{f \circ g }{\longrightarrow} \: Y \, \right)
\: \longrightarrow \:
\left( \, Y \: \stackrel{Id}{\longrightarrow} \: Y \, \right)
\end{displaymath}
For $\eta$ to be a morphism means that the diagram
\begin{displaymath}
%\begin{array}{ccc}
%Y & \stackrel{\eta}{\longrightarrow} & Y \\
%\makebox[0pt][r]{ $\scriptstyle{ f \circ g}$ } \downarrow &
%& \parallel \\
%Y & = & Y
%\end{array}
\xymatrix{
Y \ar[rr]^{\eta} \ar[dr]_{  f \circ g } & & Y \ar[dl]^{=} \\
& Y & 
}
\end{displaymath}
commutes.
For the natural transformation $\eta: F \circ G \Rightarrow \mbox{Id}$
to be invertible means that the map $\eta$ is a homeomorphism,
hence $f \circ g$ is a homeomorphism.
Similarly, using the existence of an invertible natural transformation
$\gamma: G \circ F \Rightarrow \mbox{Id}$, one can show that
$g \circ f$ is a homeomorphism.

So far we have argued that $f \circ g: Y \rightarrow Y$
and $g \circ f: X \rightarrow X$ are homeomorphisms.
For $f \circ g$ to be a homeomorphism means that $f$ is surjective
and $g$ is injective; similarly, for $g \circ f$ to be a homeomorphism
means that $g$ is surjective and $f$ is injective.  Thus, $f$ and $g$
are bijections.  By assumption, they are also continuous.
Also, since $f^{-1} = g \circ (f \circ g)^{-1}$, i.e., since $f^{-1}$
can be written as the composition of continuous maps, $f^{-1}$ is continuous.
Thus, $f: X \rightarrow Y$ is a homeomorphism, and we are done.

\section{$Y \times_{[X/\Gamma]} X$ is a principal $\Gamma$-bundle
over $Y$}

In the main text, we made several claims regarding the fiber
product $Y \times_{[X/\Gamma]} X$; in this appendix, we shall work
out the technical details.

Recall a map $Y \rightarrow [X/\Gamma]$ is defined by 
a pair
\begin{displaymath}
\left( \, E \: \stackrel{\pi}{\longrightarrow} \: Y, \:
E \: \stackrel{f}{\longrightarrow} \: X \, \right)
\end{displaymath}
We shall show that the fiber product $Y \times_{[X/\Gamma]} X$
is naturally identified with $E$.
Specifically, we shall show that the categories
\begin{displaymath}
( Y \times_{[X/\Gamma]} X )_{map} \: \cong \:
(E)_{map}
\end{displaymath}
are equivalent, and that the functors
\begin{eqnarray*}
\left( \, Y \times_{[X/\Gamma]} X \: \stackrel{p_1}{\longrightarrow} \: Y \, 
\right) & \cong & \left( \, E \: \stackrel{\pi}{\longrightarrow} \: Y \, 
\right) \\
\left( \, Y \times_{[X/\Gamma]} X \: \stackrel{p_2}{\longrightarrow} \: X \,
\right) & \cong & \left( \, E \: \stackrel{f}{\longrightarrow} \: X \, 
\right) 
\end{eqnarray*}
are equivalent, in each of the next subsections.

\subsection{Technical setup}

Before giving the technical definition of the fiber product
$( Y \times_{[X/\Gamma]} X)_{map}$, we shall first recall
the form of functors $(Y)_{map} \rightarrow [X/\Gamma]_{map}$.

Recall that, given the pair $(E \rightarrow Y, E \rightarrow X)$
defining a map $Y \rightarrow [X/\Gamma]$, the corresponding
functor $h_Y: (Y)_{map} \rightarrow [X/\Gamma]_{map}$ is defined
as follows:
\begin{enumerate}
\item Objects:  This functor maps
\begin{displaymath}
\left( \, W \: \stackrel{g}{\longrightarrow} \: Y \, \right)
\end{displaymath}
to the object
\begin{displaymath}
\left( \, g^* E \: \longrightarrow \: W, \:
g^* E \: \longrightarrow \: E \: \stackrel{f}{\longrightarrow} X 
\, \right)
\end{displaymath}
where the map $g^* E \rightarrow E$ is canonical.
\item Morphisms:  Let
\begin{displaymath}
\lambda: \: \left( \, W_1 \: \stackrel{g_1}{\longrightarrow} \: Y \,
\right) \: \longrightarrow \: \left( \, W_2 \: \stackrel{g_2}{
\longrightarrow} \: Y \, \right)
\end{displaymath}
be a morphism, i.e., the diagram
\begin{displaymath}
\xymatrix{
W_1 \ar[rr]^{\lambda} \ar[dr]_{g_1} & & W_2 \ar[dl]^{g_2} \\
& Y &
}
\end{displaymath}
commutes.
The functor maps $\lambda$ to $(\lambda, \lambda^{\#})$,
where
$\lambda^{\#}: \lambda^* g_2^* E ( = 
g_1^* E) \longrightarrow g_2^* E$
is the canonical map.
\end{enumerate}

The functor $h_X: (X)_{map} \rightarrow [X/\Gamma]_{map}$
is defined similarly.  In this case, one takes the trivial bundle
$X \times \Gamma$ in place of $E$, and the $\Gamma$-equivariant
map is the evaluation map.

Now that we have recalled the definitions of the functors $h_Y$ and $h_X$,
we shall review the technical definition of the fiber product
$( Y \times_{[X/\Gamma]} X)_{map}$, folliwng for example \cite{stacks}
and references therein.

The category $(Y \times_{[X/\Gamma]} X)_{map}$ consists of
\begin{enumerate}
\item Objects are triples
\begin{displaymath}
\left( \, \gamma_1: W \longrightarrow Y, \:
\gamma_2: W \longrightarrow X,
\alpha: h_Y(\gamma_1) \stackrel{\sim}{\longrightarrow} h_X(\gamma_2)
\, \right)
\end{displaymath}
Note that since $\alpha$ is a morphism in $[X/\Gamma]_{map}$,
$\alpha$ is not only a bundle isomorphism making
\begin{displaymath}
\xymatrix{
\gamma_1^* E \ar[rrr]^(.4){\alpha} \ar[drr] & & &
\gamma_2^*(X \times \Gamma) = W \times \Gamma \ar[dl] \\
& & W &
}
\end{displaymath}
commute,
but in addition the diagram
\begin{displaymath}
\xymatrix{
\gamma_1^* E \ar[rr]^{\alpha} \ar[dr] & &  W \times \Gamma  \ar[dl] \\
& X &
}
\end{displaymath}
is also required to commute.
\item Morphisms
\begin{eqnarray*}
\lefteqn{ \left( \, \gamma_1: W_1 \longrightarrow Y, \:
\gamma_2: W_1 \longrightarrow X, \:
\alpha_1:  h_Y(\gamma_1) \stackrel{\sim}{\longrightarrow} h_X(\gamma_2) \,
\right) } \\
& & \longrightarrow \: \left( \,
\delta_1: W_2 \longrightarrow Y, \: \delta_2: W_2 \longrightarrow X, \:
\alpha_2: h_Y(\delta_1) \stackrel{\sim}{\longrightarrow} h_X(\delta_2)
\, \right)
\end{eqnarray*}
are given by pairs $(\rho_Y, \rho_X)$, where
$\rho_Y, \rho_X: W_1 \rightarrow W_2$ are continuous maps such that
the diagrams
\begin{displaymath}
%\begin{array}{ccccccc}
%W_1 & \stackrel{\rho_Y}{\longrightarrow} & W_2 & \: \: 
%& W_1 & \stackrel{\rho_X}{
%\longrightarrow} & W_2 \\
%\makebox[0pt][r]{ $\scriptstyle{ \gamma_1 }$ } \downarrow & &
%\downarrow \makebox[0pt][l]{ $\scriptstyle{ \delta_1 }$ } 
%& \mbox{,  } & \makebox[0pt][r]{ $\scriptstyle{ \gamma_2}$ } \downarrow & &
%\downarrow \makebox[0pt][l]{ $\scriptstyle{ \delta_2 }$ } \\
%Y & = & Y & \: \: & X & = & X \\
%\end{array}
\xymatrix{
W_1 \ar[rr]^{\rho_Y} \ar[dr]_{\gamma_1} & & W_2 \ar[dl]^{\delta_1} 
& \mbox{  } & 
W_1 \ar[rr]^{\rho_X} \ar[dr]_{\gamma_2} & & W_2 \ar[dl]^{\delta_2} \\
& Y & & \mbox{  } &
& X &
}
\end{displaymath}
and
\begin{displaymath}
%\begin{array}{ccc}
%\gamma_1^* E & \stackrel{\alpha_1}{\longrightarrow} &
%\gamma_2^*(X \times \Gamma) = W_1 \times \Gamma \\
%\makebox[0pt][r]{ $\scriptstyle{ h_Y(\rho_Y) }$ } \downarrow & &
%\downarrow \makebox[0pt][l]{ $\scriptstyle{ h_X(\rho_X) }$ } \\
%\delta_1^* E & \stackrel{\alpha_2}{\longrightarrow} & 
%\delta_2^* (X \times \Gamma) = W_2 \times \Gamma 
%\end{array}
\xymatrix{
\gamma_1^* E \ar[r]^(.3){\alpha_1} \ar[d]_{ h_Y(\rho_Y) } &
\gamma_2^*(X \times \Gamma) = W_1 \times \Gamma 
\ar[d]^{ h_X(\rho_X) } \\
\delta_1^* E \ar[r]^(.3){\alpha_2} &
\delta_2^* (X \times \Gamma) = W_2 \times \Gamma
}
\end{displaymath}
commute.
In passing, we should mention that it is straightforward to show that
the constraint that $h_X(\rho_X) \circ \alpha_1 = \alpha_2 \circ
h_Y(\rho_Y)$ implies that $\rho_X = \rho_Y$; however, we will typically
still refer to a morphism in this category as a pair
$(\rho_Y, \rho_X)$. 
\end{enumerate}

The projection maps $p_{1,2}: Y \times_{[X/\Gamma]} X \longrightarrow
Y, X$ induce projection functors $P_{1,2}: ( Y \times_{[X/\Gamma]} X )_{map}
\longrightarrow (Y)_{map}, (X)_{map}$; we shall discuss these in more
detail later.

%\subsection{ $Y \times_{[X/\Gamma]} X \: \cong \: E$ }
% Curiously enough, the line above caused a TeX error.
% As far as I can tell, there must be some limit on the number
% of occurrences of `` \xyz '' in a header;
% too many caused an ``Illegal parameter number in reserved@a''
% or some such error message.

\subsection{ $Y \times_{[X/\Gamma]} X$ versus $E$ }

In this subsection we shall argue that the categories
$(Y \times_{[X/\Gamma]} X)_{map}$ and $(E)_{map}$ are equivalent,
by explicitly constructing a functor $F: (Y \times_{[X/\Gamma]} X)_{map}
\rightarrow (E)_{map}$ and showing that it defines an equivalence
of categories.

Define a functor $F: (Y \times_{[X/\Gamma]} X)_{map}
\rightarrow (E)_{map}$ as follows:
\begin{enumerate}
\item Objects:  Let $( \gamma_1: W \rightarrow Y,
\gamma_2: W \rightarrow X, \alpha: h_Y(\gamma_1) \stackrel{\sim}{
\longrightarrow} h_X(\gamma_2) )$ be an object of
$(Y \times_{[X/\Gamma]} X)_{map}$.  Let $s: W \rightarrow
\gamma_2^* (X \times \Gamma) ( = W \times \Gamma)$ be the
identity section.  The image under $F$ is the map $W \rightarrow E$ 
defined by
\begin{displaymath}
W \: \stackrel{s}{\longrightarrow} \: \gamma_2^*( X \times \Gamma )
= W \times \Gamma \: \stackrel{\alpha^{-1}}{\longrightarrow} \:
\gamma_1^* E \: \stackrel{canonical}{\longrightarrow} \: E
\end{displaymath}
\item Morphisms:
Let 
\begin{displaymath}
\begin{array}{cl}
(\rho_Y, \rho_X): & 
\left( \gamma_1: W_1 \rightarrow Y, \gamma_2: W_1 \rightarrow X,
\alpha_1: h_Y(\gamma_1) \stackrel{\sim}{\longrightarrow} h_X(\gamma_2)
\right)\\
& \: \longrightarrow \:
\left( \delta_1: W_2 \rightarrow Y, \delta_2: W_2 \rightarrow X,
\alpha_2: h_Y(\delta_1) \stackrel{\sim}{\longrightarrow} h_X(\delta_2)
\right)
\end{array}
\end{displaymath}
be a morphism in $(Y \times_{[X/\Gamma]} X)_{map}$.
The image of $(\rho_Y, \rho_X)$ under $F$ is defined to be $\rho_X$.
\end{enumerate}
With this definition, $F$ is a well-defined covariant functor
$(Y \times_{[X/\Gamma]} X)_{map}
\rightarrow (E)_{map}$.

To show that the functor $F: (Y \times_{[X/\Gamma]} X)_{map}
\rightarrow (E)_{map}$ defines an equivalence of categories, we shall
next construct a functor $G: (E)_{map} \rightarrow (Y \times_{[X/\Gamma]}
X)_{map}$ and show that $F \circ G \cong \mbox{Id}$ and
$G \circ F \cong \mbox{Id}$.

Define a functor $G: (E)_{map} \rightarrow (Y \times_{[X/\Gamma]} X)_{map}$
as follows:
\begin{enumerate}
\item Objects:  Let $( m: W \rightarrow E )$ be an object of 
$(E)_{map}$, i.e., a continuous map into $E$.
The image under $G$ is defined to be the object
\begin{displaymath}
\left( \, \gamma_1: W \longrightarrow Y, \:
\gamma_2: W \longrightarrow X, \:
\alpha: \gamma_1^* E \longrightarrow W \times \Gamma \, \right)
\end{displaymath}
where
\begin{eqnarray*}
\gamma_1 & \equiv & \pi \circ m \\
\gamma_2 & \equiv & f \circ m \\
\alpha^{-1}(w,g) & \equiv & \left( w, g \cdot m(w) \right)
\: \in \: \gamma_1^* E \: \subseteq \: W \times E
\end{eqnarray*}
\item Morphisms:  Let $\lambda: ( W_1 \stackrel{m_1}{\longrightarrow}
E) \longrightarrow ( W_2 \stackrel{m_2}{\longrightarrow} E)$
be a morphism, i.e.,
the diagram
\begin{displaymath}
\xymatrix{
W_1 \ar[rr]^{\lambda} \ar[dr]_{m_1} & & W_2 \ar[dl]^{m_2} \\
& E & 
}
\end{displaymath}
commutes.
Define the image of $G$ to be the morphism $(\lambda, \lambda)$
in $( Y \times_{[X/\Gamma]} X)_{map}$.
\end{enumerate}
With this definition, $G$ is a well-defined covariant functor
$(E)_{map} \rightarrow (Y \times_{[X/\Gamma]} X)_{map}$.

Finally, it is straightforward to check that
$F \circ G = \mbox{Id}_{(E)_{map}}$ and $G \circ F
= \mbox{Id}_{(Y \times_{[X/\Gamma]} X)_{map}}$.
(This is actually a stronger statement than needed; one only
requires $F \circ G \cong \mbox{Id}$, for example.)

Thus, the categories $(Y \times_{[X/\Gamma]} X)_{map}$
and $(E)_{map}$ are equivalent, from which we conclude that
$Y \times_{[X/\Gamma]} X \cong E$.

\subsection{ $( Y \times_{[X/\Gamma]} X \rightarrow Y)$ 
versus $( E \rightarrow Y )$ }

Given that $Y \times_{[X/\Gamma]} X \cong E$, one would naturally
suspect that the map $Y \times_{[X/\Gamma]} X \stackrel{p_1}{\longrightarrow} Y$
is equivalent to the map $E \stackrel{\pi}{\longrightarrow} Y$,
and indeed this is the case.

We shall verify this by showing that the natural functor
$P_1: (Y \times_{[X/\Gamma]} X)_{map} \rightarrow (Y)_{map}$ is
equivalent to the functor $\Pi: (E)_{map} \rightarrow (Y)_{map}$.

First, we shall take a moment to carefully define these functors.

By definition of fiber product, 
the functor $P_1: (Y \times_{[X/\Gamma]} X)_{map} \rightarrow (Y)_{map}$
is defined as follows:
\begin{enumerate}
\item Objects:  Let $(\gamma_1: W \rightarrow Y, \gamma_2: W \rightarrow X,
\alpha: h_Y(\gamma_1) \stackrel{\sim}{\longrightarrow} h_X(\gamma_2) )$
be an object of $(Y \times_{[X/\Gamma]} X)_{map}$.
The functor $P_1$ maps $(\gamma_1, \gamma_2, \alpha)$ to $\gamma_1$.
\item Morphisms:  Let
\begin{displaymath}
\begin{array}{cl}
(\rho_Y, \rho_X): & \left( \, \gamma_1: W_1 \longrightarrow Y, \:
\gamma_2: W_1 \longrightarrow X, \: \alpha_1: h_Y(\gamma_1) 
\stackrel{\sim}{\longrightarrow} h_X(\gamma_2) \, \right) \\
 & \: \longrightarrow \: 
\left( \, \delta_1: W_2 \longrightarrow Y, \: \delta_2: W_2 \longrightarrow
X, \: \alpha_2:  h_Y(\delta_1) \stackrel{\sim}{\longrightarrow} h_X(\delta_2)
\, \right)
\end{array}
\end{displaymath}
be a morphism in $(Y \times_{[X/\Gamma]} X)_{map}$.
The functor $P_1$ maps $(\rho_Y, \rho_X)$ to $\rho_Y$.
\end{enumerate}

The functor $\Pi:  (E)_{map} \rightarrow (Y)_{map}$ is defined as follows:
\begin{enumerate}
\item Objects:  Let $(\gamma: W \rightarrow E)$ be an object of $(E)_{map}$.
The functor $\Pi$ maps this object to $(\pi \circ \gamma: W \rightarrow Y)$.
\item Morphisms:  Let $\lambda: ( \gamma_1: W_1 \rightarrow E)
\longrightarrow ( \gamma_2: W_2 \rightarrow E)$ be a morphism, i.e.,
the diagram
\begin{displaymath}
\xymatrix{
W_1 \ar[rr]^{\lambda} \ar[dr]_{\gamma_1} & & W_2 \ar[dl]^{\gamma_2} \\
& E &
}
\end{displaymath}
commutes.
The functor $\Pi$ maps $\lambda$ to itself.
\end{enumerate}

Now, in what sense can these two functors be isomorphic?  After all,
they map between distinct categories!  The answer is that
one uses the functors $F: (Y \times_{[X/\Gamma]} X)_{map}
\rightarrow (E)_{map}$ and $G: (E)_{map}
\rightarrow (Y \times_{[X/\Gamma]} X)_{map}$, defined in the previous
subsection, to relate the categories in question.

More specifically, it is straightforward to check that
\begin{displaymath}
P_1 \circ G \: = \: \Pi: \: (E)_{map} \: \longrightarrow \:
(Y)_{map}
\end{displaymath}
and 
\begin{displaymath}
\Pi \circ F \: = \: P_1: \: ( Y \times_{[X/\Gamma]} X)_{map}
\: \longrightarrow \: (Y)_{map}
\end{displaymath}

(In fact, this result is stronger than necessary.  One would merely need to
require that $P_1 \circ G \cong \Pi$ and $\Pi \circ F \cong P_1$.)

Thus, we conclude that, as anticipated, the map
$Y \times_{[X/\Gamma]} X \stackrel{p_1}{\longrightarrow} Y$
is indeed the same as the map $E \stackrel{\pi}{\longrightarrow} Y$.

\subsection{ $( Y \times_{[X/\Gamma]} X \rightarrow X )$ versus
$( E \rightarrow X )$ }

Considering that $Y \times_{[X/\Gamma]} X \cong E$ and the projection
map $Y \times_{[X/\Gamma]} X \stackrel{p_1}{\longrightarrow} Y$
is isomorphic to the map $E \stackrel{\pi}{\longrightarrow} Y$,
one would naturally suspect that the other projection map
$Y \times_{[X/\Gamma]} X \stackrel{p_2}{\longrightarrow} X$
is isomorphic to the $\Gamma$-equivariant map 
$E \stackrel{f}{\longrightarrow} X$.  

In this subsection, we shall
show that this suspicion is correct.
Specifically, we shall show that the natural functor
$P_2: (Y \times_{[X/\Gamma]} X)_{map} \rightarrow (X)_{map}$
is isomorphic to the functor $\tilde{f}: (E)_{map} \rightarrow
(X)_{map}$.

By definition of fiber product,
the functor $P_2: ( Y \times_{[X/\Gamma]} X )_{map}
\rightarrow (X)_{map}$ is defined as follows:
\begin{enumerate}
\item Objects:  Let $(\gamma_1: W \rightarrow Y, \gamma_2: W \rightarrow X,
\alpha: h_Y(\gamma_1) \stackrel{\sim}{\longrightarrow} h_X(\gamma_2) )$
be an object of $(Y \times_{[X/\Gamma]} X)_{map}$
The functor $P_2$ maps $(\gamma_1, \gamma_2, \alpha)$ to $\gamma_2$.
\item Morphisms:  Let
\begin{displaymath}
\begin{array}{cl}
(\rho_Y, \rho_X): & \left( \, \gamma_1: W_1 \longrightarrow Y, \:
\gamma_2: W_1 \longrightarrow X, \: \alpha_1: h_Y(\gamma_1)
\stackrel{\sim}{\longrightarrow} h_X(\gamma_2) \, \right) \\
& \: \longrightarrow \:
\left( \, \delta_1: W_2 \rightarrow Y, \: \delta_2: W_2 \rightarrow X, \:
\alpha_2: h_Y(\delta_1) \stackrel{\sim}{\longrightarrow} h_X(\delta_2) \,
\right)
\end{array}
\end{displaymath}
be a morphism in $( Y \times_{[X/\Gamma]} X)_{map}$.
The functor $P_2$ maps $(\rho_Y, \rho_X)$ to $\rho_X$.
\end{enumerate}

Define the functor $\tilde{f}: (E)_{map} \rightarrow (X)_{map}$ as follows:
\begin{enumerate}
\item Objects:  Let $( m: W \rightarrow E )$ be an object of $(E)_{map}$.
The functor $\tilde{f}$ maps $m$ to $( f \circ m: W \rightarrow X)$.
\item Morphisms:  Let $\lambda: (m_1: W_1 \rightarrow E) \longrightarrow
(m_2: W_2 \rightarrow E)$ be a morphism, i.e., the diagram
\begin{displaymath}
\xymatrix{
W_1 \ar[rr]^{\lambda} \ar[dr]_{ m_1 } & & W_2 \ar[dl]^{m_2} \\
& E &
}
\end{displaymath}
The functor $\tilde{f}$ maps $\lambda$ to itself.
\end{enumerate}

Just as in the last subsection, we run into an apparent problem
understanding how these functors can be isomorphic, as they map
between distinct categories.  The answer to this dilemma is to use
the functors  $F: (Y \times_{[X/\Gamma]} X)_{map}
\rightarrow (E)_{map}$ and $G: (E)_{map}
\rightarrow (Y \times_{[X/\Gamma]} X)_{map}$, defined earlier, to
relate the categories in question.

It is straightforward to check that
\begin{displaymath}
\tilde{f} \circ F \: = \: P_2: \: ( Y \times_{[X/\Gamma]} X)_{map}
\: \longrightarrow \: (X)_{map} 
\end{displaymath}
and
\begin{displaymath}
P_2 \circ G \: = \: \tilde{f}: \: (E)_{map} \: \longrightarrow \:
(X)_{map}
\end{displaymath}
As before, this result is stronger than strictly necessary -- one
merely requires $\tilde{f} \circ F \cong P_2$ and $P_2 \circ G \cong 
\tilde{f}$.

Thus, we conclude that the map $Y \times_{[X/\Gamma]} X \stackrel{p_2}{
\longrightarrow} X$ is indeed the same as the map
$E \stackrel{f}{\longrightarrow} Y$.

\section{A theorem on representable morphisms}  \label{represent}

In this section we shall present a standard theorem regarding representable
morphisms.

Specifically, let ${\cal F}$ be a groupoid over some category ${\cal C}$.
Then the following are equivalent \cite[section 4]{dm}:
\begin{enumerate}
\item The diagonal map $\Delta: {\cal F} \rightarrow {\cal F} \times
{\cal F}$ is representable.
\item For all objects $X, Y \in \mbox{Ob } {\cal C}$, and maps
$f: X \rightarrow {\cal F}$, $f: Y \rightarrow {\cal F}$,
the fiber product $X \times_{{\cal F}} Y$ is representable.
\item All morphisms $f: X \rightarrow {\cal F}$ are representable.
\end{enumerate}

The fact that the second and third items are equivalent follows
directly from the definition of representable morphism.

Let us show that the first statement implies the second.
Suppose $\Delta$ is representable.  Let $f: X \rightarrow
{\cal F}$, $g: Y \rightarrow {\cal F}$ be any two morphisms into ${\cal F}$.
Now, for $\Delta$ to be representable means that for all objects
$Z \in \mbox{Ob } {\cal C}$ and for all morphisms
$h: Z \rightarrow {\cal F}$, the fiber product
$Z \times_{{\cal F} \times {\cal F}} {\cal F}$ is a representable stack.
Since $\Delta$ is representable, and $f \times g: X \times Y \rightarrow
{\cal F} \times {\cal F}$ is a morphism, we have that
$( X \times Y) \times_{{\cal F} \times {\cal F}} {\cal F}$ is representable.
However, from unraveling definitions one can quickly show
\begin{displaymath}
( X \times Y ) \times_{{\cal F} \times {\cal F}} {\cal F}
\: \cong \: X \times_{{\cal F}} Y
\end{displaymath}
and so we have the second statement.

Now, we shall show that the second statement implies the first.
Suppose for all $X, Y \in \mbox{Ob } {\cal C}$, and for all
maps $f \times g: X \times Y \rightarrow {\cal F} \times {\cal F}$,
the fiber product $X \times_{{\cal F}} Y$ is representable.
From the fact that
\begin{displaymath}
X \times_{{\cal F}} Y
\: \cong \: 
( X \times Y ) \times_{{\cal F} \times {\cal F}} {\cal F}
\end{displaymath}
it should be immediately clear that $\Delta$ is also representable.

\section{Proof of form of the associated inertia group stack}
\label{mytoenproof}

In this section we shall prove that
\begin{displaymath}
I_{[X/\Gamma]} \: \cong \:
\coprod_{[g]} \, \left[ X^g / C(g) \right]
\end{displaymath}
which we used to understand orbifold Euler characteristics.
This result is also proven in \cite[p. 38]{toen}, but we
shall use somewhat different methods to arrive at this result,
and we felt it important for completeness to include our methods
here.

Very roughly, the idea is as follows.  An object of $(I_{[X/\Gamma]})_{map}$
is a triple
\begin{displaymath}
\left( \, E \: \stackrel{\pi}{\longrightarrow} \: Y, \:
E \: \stackrel{f}{\longrightarrow} \: X, \:
E \: \stackrel{\lambda}{\longrightarrow} \: E \, \right)
\end{displaymath}
where $E$ is a principal $\Gamma$-bundle over $Y$,
$f: E \rightarrow X$ is $\Gamma$-equivariant, and
$\lambda: E \rightarrow E$ is a base-preserving automorphism
compatible with $f$.
Now, the grading by conjugacy classes of $\Gamma$ comes from
the fact that equivalence classes of automorphisms $\lambda$
are classified by conjugacy classes.  The fact that the quotient
stacks in each grading are quotients of $X^g$ follows from the fact
that if $\lambda$ is an automorphism determined by $[g]$,
then the image of $f$ lies in $X^g$ (or a naturally homeomorphic set).
The appearance of the centralizer $C(g)$ reflects the fact that
forcing morphisms in the category to be compatible with the
automorphism $\lambda$ removes all gauge transformations except those
determined by $C(g) \subseteq G$.

First, we shall prove the statement at the level of points, and
then we shall study continuous maps.

For future reference, we shall assume to have fixed a set
of representatives $g$ of the conjugacy classes of $\Gamma$.

\subsection{Points}

Define a functor 
\begin{displaymath}
F: \: \coprod_{[g]} \, \left[ X^g / C(g) \right]_{pt} \: 
\longrightarrow \left( I_{[X/\Gamma]} \right)_{pt}
\end{displaymath} 
as follows:
\begin{enumerate}
\item Objects:  Let $( f: C(g) \rightarrow X^g) \in \mbox{Ob }
[ X^g / C(g) ]_{pt}$.  The functor $F$ maps this object
to the object
\begin{displaymath}
\left( \, f: \: \Gamma \: \longrightarrow \: X, \:
g: \: \Gamma \: \longrightarrow \: \Gamma \, \right)
\end{displaymath}
where the $\Gamma$-equivariant map $f: \Gamma \rightarrow X$
is determined by extending $f: C(g) \rightarrow X$ to all of $\Gamma$
(after all, a $\Gamma$-equivariant map from any subset of $\Gamma$
is completely determined by the image of the identity),
and the map $g: \Gamma \rightarrow \Gamma$ is given by,
$h \mapsto g \cdot h$ for $x \in \Gamma$.
\item Morphisms:  Let 
\begin{displaymath}
\lambda: \: \left( \, f_1: \: C(g) \: \longrightarrow \: X^g \, \right)
\: \longrightarrow \: \left( \, f_2: \: C(g) \: \longrightarrow \: X^g
\, \right)
\end{displaymath}
be a morphism in $[X^g/C(g)]_{pt}$.  Since $\lambda$ is completely
determined by $\lambda(1)$, it extends to a map $\Gamma \rightarrow \Gamma$,
which commutes both with the extensions of $f_1$ and $f_2$, as well
as with $g$.  Hence, the functor $F$ maps $\lambda$ to its
extension.
\end{enumerate}

We claim that $F$ defines an equivalence of categories.
In order to prove this, we shall construct a functor 
\begin{displaymath}
G: \: \left( I_{[X/\Gamma]} \right)_{pt} \: \longrightarrow \:
\coprod_{[g]} \, \left[ X^g / C(g) \right]_{pt}
\end{displaymath}
such that $F \circ G \cong \mbox{Id}$ and $G \circ F \cong \mbox{Id}$.

Define a functor
\begin{displaymath}
G: \: \left( I_{[X/\Gamma]} \right)_{pt} \: \longrightarrow \:
\coprod_{[g]} \, \left[ X^g / C(g) \right]_{pt}
\end{displaymath}
as follows:
\begin{enumerate}
\item Objects:  Let the pair
\begin{displaymath}
\left( \, f: \: \Gamma \: \longrightarrow \: X, \:
\overline{g}: \: \Gamma \: \longrightarrow \: \Gamma \, \right)
\end{displaymath}
be an object of $(I_{[X/\Gamma]})_{pt}$, meaning that the diagram
\begin{displaymath}
\xymatrix{
\Gamma \ar[rr]^{ \overline{g} } \ar[dr]_{f} & & \Gamma \ar[dl]^{f} \\
& X &
}
\end{displaymath}
commutes.
Clearly, the map $\overline{g}$ is determined by some element of
$\Gamma$, which we shall also denote $\overline{g}$.  Let
$g$ denote the fixed representative of the conjugacy class of
$\Gamma$ to which $\overline{g}$ belongs, and define $h \in \Gamma$
by, $\overline{g} = h g h^{-1}$.  Then the following diagram commutes:
\begin{displaymath}
\xymatrix{
\Gamma \ar[rr]^{ g } \ar[dr]_{  f \circ h } & & \Gamma \ar[dl]^{  f \circ h} \\
& X &
}
\end{displaymath}
Commutivity of the diagram above clearly implies that
$\mbox{im } f \circ h \subseteq X^g$.  Since $f \circ h$ is
completely determined by the image of the identity, we can
restrict it to arbitrary subgroups; define $f'$ to be
the restriction of $f \circ h$ to $C(g) \subseteq \Gamma$.
Then, $G$ maps the object given to $( f': C(g) \rightarrow X^g)
\in \mbox{Ob } [X^g/C(g)]_{pt}$.
\item Morphisms:  Let
\begin{displaymath}
k: \: \left( \, f_1: \: \Gamma \: \longrightarrow \: X, \:
\overline{g}_1: \: \Gamma \: \longrightarrow \: \Gamma \, \right)
\: \longrightarrow \:
\left( \, f_2: \: \Gamma \: \longrightarrow \: X, \:
\overline{g}_2: \: \Gamma \: \longrightarrow \: \Gamma \, \right)
\end{displaymath}
be a morphism in $(I_{[X/\Gamma]})_{pt}$, which means that the 
following two diagrams commute:
\begin{displaymath}
%\begin{array}{ccccccc}
%\Gamma & \stackrel{k}{\longrightarrow} & \Gamma & & 
%\Gamma & \stackrel{k}{\longrightarrow} & \Gamma \\
%\makebox[0pt][r]{ $\scriptstyle{ f_1 }$ } \downarrow & &
%\downarrow \makebox[0pt][l]{ $\scriptstyle{ f_2}$ } &
%\: \mbox{   ,   } \: & 
%\makebox[0pt][r]{ $\scriptstyle{ \overline{g}_1 }$ } \downarrow & &
%\downarrow \makebox[0pt][l]{ $\scriptstyle{ \overline{g}_2 }$ } \\
%X & = & X & &
%\Gamma & \stackrel{k}{\longrightarrow} & \Gamma
%\end{array}
\xymatrix{
\Gamma \ar[rr]^{ k } \ar[dr]_{f_1} & & \Gamma \ar[dl]^{f_2} & \mbox{  } &
\Gamma \ar[r]^{k} \ar[d]_{ \overline{g}_1 } & \Gamma \ar[d]^{\overline{g}_2}\\
& X & & \mbox{  } &
\Gamma \ar[r]^{ k } & \Gamma
}
\end{displaymath}
We shall use $k$ to denote the morphism and the corresponding
group element (the image of the identity) interchangeably.
From commutivity of the right diagram, we see that $\overline{g}_1$
and $\overline{g}_2$ are in the same conjugacy class.
Let $g$ denote the fixed representative of that conjugacy class,
and define $h_1, h_2 \in \Gamma$ by,
$\overline{g}_1 = h_1 g h_1^{-1}$ and $\overline{g}_2 = h_2 g h_2^{-1}$.
Define $f'_1$ to be the restriction of $f_1 \circ h_1$ to $C(g)$,
define $f'_2$ to be the restriction of $f_2 \circ h_2$ to $C(g)$,
and define $k' = h_2^{-1} k h_1$.  Using the right diagram it
is straightforward to check that $k' \in C(g)$, and hence
defines a $C(g)$-equivariant map $C(g) \rightarrow C(g)$.
Also, using the left diagram it is straightforward to check
that $k'$ is compatible with $f'_1$ and $f'_2$.  Hence,
define $G$ to map $k$ to the morphism
\begin{displaymath}
k': \: \left( \, f'_1: \: C(g) \: \longrightarrow \: X^g \, \right)
\: \longrightarrow \:
\left( \, f'_2: \: C(g) \: \longrightarrow \: X^g \, \right)
\end{displaymath}
in $[X^g/C(g)]_{pt}$.
\end{enumerate}

Now, it is straightforward to check that $G \circ F = \mbox{Id}$
on the category $\coprod_{[g]} [X^g/C(g)]_{pt}$.
We shall also show that $F \circ G \cong \mbox{Id}$ 
on the category $( I_{[X/\Gamma]} )_{pt}$, by constructing
an invertible natural transformation $\eta: F \circ G \Rightarrow
\mbox{Id}$.

Construct the invertible natural transformation $\eta: F \circ G
\Rightarrow \mbox{Id}$ as follows.
For any object
\begin{displaymath}
\left( \, f: \: \Gamma \: \longrightarrow \: X, \:
\overline{g}: \: \Gamma \: \longrightarrow \: \Gamma \, \right)
\end{displaymath}
in $(I_{[X/\Gamma]})_{pt}$, define $\eta \equiv h$, where
$h \in \Gamma$ is such that $\overline{g} = h g h^{-1}$.
It is straightforward to check that this is a well-defined morphism
in $(I_{[X/\Gamma]})_{pt}$, and furthermore that $\eta$ is an
invertible natural transformation.

Hence, $G \circ F = \mbox{Id}$ and $F \circ G \cong \mbox{Id}$,
so $F$ is an invertible natural transformation, which means that
\begin{displaymath}
\left( \, I_{[X/\Gamma]} \, \right)_{pt} \: \cong \:
\coprod_{[g]} \, \left[ X^g / C(g) \right]_{pt}
\end{displaymath}

In the next subsection we shall check the statement completely,
by showing that the categories of continuous maps into each are
equivalent.

\subsection{Continuous maps}

In this section we shall show that
\begin{displaymath}
\left( \, I_{[X/\Gamma]} \, \right)_{map} \: \cong \:
\coprod_{[g]} \, \left[ X^g / C(g) \right]_{map}
\end{displaymath}
in the special case that $\Gamma$ is abelian.
(As we just showed that the desired statement is true at the
level of points for all $\Gamma$, not just those which are abelian,
and reference~\cite{toen} contains an independent proof, we are
highly confident of the result.)

Define a functor 
\begin{displaymath}
F: \: \coprod_{[g]} \, \left[ X^g / C(g) \right]_{map} \:
\longrightarrow \: \left( \, I_{[X/\Gamma]} \, \right)_{map}
\end{displaymath}
as follows:
\begin{enumerate}
\item Objects:  Let $( E \rightarrow Y, f: E \rightarrow X^g)$
be an object of $[X^g/C(g)]_{map}$ for some conjugacy-class-representative
$g$.   Define $F$ to map this object to the triple
\begin{displaymath}
\left( \, E \: \longrightarrow \: Y, \:
E \: \stackrel{f}{\longrightarrow} \: X^g \: \hookrightarrow \: X, \:
E \: \stackrel{g}{\longrightarrow} \: E \, \right)
\end{displaymath}
where $g: E \rightarrow E$ is defined by sending any $e \in E$
to $g \cdot E$. 
\item Morphisms:  Let 
\begin{displaymath}
(\rho, \psi): \:
\left( \, E_1 \: \longrightarrow \: Y_1, \: 
f_1: \: E_1 \: \longrightarrow \: X^g \, \right) \: \longrightarrow \:
\left( \, E_2 \: \longrightarrow \: Y_2, \:
f_2: \: E_2 \: \longrightarrow \: X^g \, \right)
\end{displaymath}
be a morphism in $\coprod_{[g]} [ X^g / C(g) ]_{map}$,
i.e., $\rho: Y_1 \rightarrow Y_2$ and $\psi: E_1 \rightarrow E_2$
are morphisms making appropriate diagrams commute.
Define $F$ to send $(\rho, \psi)$ to $(\rho, \psi)$.
\end{enumerate}

We claim that $F$ defines an equivalence of categories.  To prove this,
we shall construct a functor 
\begin{displaymath}
G: \:  \left( \, I_{[X/\Gamma]} \, \right)_{map} \: \longrightarrow \:
 \coprod_{[g]} \, \left[ X^g / C(g) \right]_{map} 
\end{displaymath}
and check that $F \circ G \cong \mbox{Id}$ and $G \circ F \cong \mbox{Id}$.

Define a functor
\begin{displaymath}
G: \:  \left( \, I_{[X/\Gamma]} \, \right)_{map} \: \longrightarrow \:
 \coprod_{[g]} \, \left[ X^g / C(g) \right]_{map} 
\end{displaymath}
as follows:
\begin{enumerate}
\item Objects:  Let
\begin{displaymath}
\left( \, E \: \longrightarrow \: Y, \:
f: \: E \: \longrightarrow \: X, \:
\overline{g}: \: E \: \longrightarrow \: E \, \right)
\end{displaymath}
be an object of $(I_{[X/\Gamma]})_{map}$, which means that the diagram
\begin{displaymath}
\xymatrix{
E \ar[rr]^{ \overline{g} } \ar[dr]_{f} & & E \ar[dl]^{f} \\
& X &
}
\end{displaymath}
commutes.
Without loss of generality, assume that $Y$ is connected.
Then, the base-preserving map $\overline{g}: E \rightarrow E$ is determined by
some $\overline{g} \in \Gamma$, as $e \in E \mapsto \overline{g} \cdot e$.
Let $g$ denote the fixed representative of the conjugacy class of
$\overline{g} \in \Gamma$, and define $h$ by, $\overline{g} = 
h g h^{-1}$.  Define $G$ to map the object above to the pair
\begin{displaymath}
\left( \, E \: \longrightarrow \: Y, \:
f \circ h: \: E \: \longrightarrow \: X \, \right)
\end{displaymath}
which define an object of $\coprod_{[g]} [X^g/C(g)]_{map}$.
\item Morphisms:  Let
\begin{eqnarray*}
\lefteqn{ (\rho, \psi): \:
\left( \, E_1 \rightarrow Y_1, \:
f_1: \: E_1 \rightarrow X, \: \overline{g}_1: \: E_1 \rightarrow E_1 \, 
\right) } \\
 & & \: \longrightarrow \:
\left( \, E_2 \rightarrow Y_2, \:
f_2: \: E_2 \rightarrow X, \: \overline{g}_2: \: E_2 \rightarrow E_2 \, 
\right)
\end{eqnarray*}
be a morphism in $(I_{[X/\Gamma]})_{map}$, i.e., $\rho: Y_1 \rightarrow
Y_2$ and $\psi: E_1 \rightarrow E_2$ are morphisms making the diagrams
\begin{displaymath}
%\begin{array}{ccccccccccc}
%E_1 & \stackrel{\psi}{\longrightarrow} & E_2 & & 
%E_1 & \stackrel{\psi}{\longrightarrow} & E_2 & &
%E_1 & \stackrel{\psi}{\longrightarrow} & E_2 \\
%\downarrow & & \downarrow & \: \mbox{   ,   } \: &
%\makebox[0pt][r]{ $\scriptstyle{ f_1}$ } \downarrow & &
%\downarrow \makebox[0pt][l]{ $\scriptstyle{f_2}$ } & \: \mbox{   ,   } \: &
%\makebox[0pt][r]{ $\scriptstyle{ \overline{g}_1 }$ } \downarrow & &
%\downarrow \makebox[0pt][l]{ $\scriptstyle{ \overline{g}_2 }$ } \\
%Y_1 & \stackrel{\rho}{\longrightarrow} & Y_2 & &
%X & = & X & &
%E_1 & \stackrel{\psi}{\longrightarrow} & E_2 
%\end{array}
\xymatrix{
E_1 \ar[r]^{\psi} \ar[d] & E_2 \ar[d] & \mbox{  } &
E_1 \ar[rr]^{\psi} \ar[dr]_{ f_1 } & & E_2 \ar[dl]^{f_2} & \mbox{  } &
E_1 \ar[r]^{\psi} \ar[d]_{ \overline{g}_1 } & E_2 \ar[d]^{\overline{g}_2}\\
Y_1 \ar[r]^{\rho} & Y_2 & \mbox{  } &
& X & & \mbox{  } &
E_1 \ar[r]^{\psi} & E_2 
}
\end{displaymath}
commute.  Now, it is straightforward to check that the rightmost
diagram implies that $\overline{g}_1$ and $\overline{g}_2$ are conjugate.
Define $h_1$ and $h_2$ by, $\overline{g}_1 = h_1 g h_1^{-1}$,
$\overline{g}_2 = h_2 g h_2^{-1}$ for some fixed conjugacy-class
representative $g \in \Gamma$.
Define $\psi' \equiv h_2^{-1} \circ \psi \circ h_1$.
Define $G$ to map the morphism $(\rho, \psi)$ to $(\rho, \psi')$.
\end{enumerate}

Next, in order to show that $F$ is an equivalence of categories,
we must show that $F \circ G \cong \mbox{Id}$ and
$G \circ F \cong \mbox{Id}$.  Now, is it straightforward to check
that $G \circ F = \mbox{Id}$ on the category $\coprod_{[g]} [X^g/C(g)]_{map}$.
We shall show that $F \circ G \cong \mbox{Id}$ on the category
$(I_{[X/\Gamma]})_{map}$ by constructing an invertible
natural transformation $\eta: F \circ G \Rightarrow \mbox{Id}$.

Construct the invertible natural transformation $\eta: F \circ G 
\Rightarrow \mbox{Id}$ as follows.  For any object
\begin{displaymath}
\left( \, E \: \longrightarrow \: Y, \:
E \: \stackrel{f}{\longrightarrow} \: X, \:
E \: \stackrel{ \overline{g} }{\longrightarrow} \: E \, \right)
\end{displaymath}
(where we have assumed, for simplicity, that $Y$ is connected)
define $\eta \equiv h$, where $\overline{g} = h g h^{-1}$,
for some fixed conjugacy-class representative $g \in \Gamma$.
It is straightforward to check that this makes $\eta$ a well-defined
morphism in $(I_{[X/\Gamma]})_{map}$,
and in fact makes $\eta$ an invertible natural transformation.

Hence, $G \circ F = \mbox{Id}$ and $F \circ G \cong \mbox{Id}$,
so $F$ is an invertible natural transformation, which means that
\begin{displaymath}
\left( \, I_{[X/\Gamma]} \, \right)_{map} \: \cong \:
\coprod_{[g]} \, \left[ X^g / C(g) \right]_{map}
\end{displaymath}
as advertised.

Again, in this subsection we have assumed that $\Gamma$ is abelian.
However, our proof that the categories of points are equivalent
worked for all $\Gamma$, not just $\Gamma$ abelian,
and reference~\cite{toen} contains an independent proof, so we
are confident of the result.

\end{document}